\documentclass[12pt,a4paper]{article}
\usepackage{amsmath,amssymb,amsthm}
\usepackage[margin=1.0in]{geometry}
\usepackage{cite}
\allowdisplaybreaks
\usepackage[colorlinks=true
,urlcolor=blue
,anchorcolor=blue
,citecolor=blue
,filecolor=blue
,linkcolor=blue
,menucolor=blue
,pagecolor=blue
,linktocpage=true
,pdfproducer=medialab
,pdfa=true
]{hyperref}
\numberwithin{equation}{section}

\def\e{\epsilon}

\def\l{\lambda}

\def\be{\begin{equation}}
\def\ee{\end{equation}}
\def\bea{\begin{eqnarray}}
\def\eea{\end{eqnarray}}

\def\pa{\partial}

\def\lp{\left(}
\def\rp{\right)}

\def\nn{\nonumber}
\def\ie{{\it i.e., }}

\makeatletter
\renewcommand\section{\@startsection {section}{1}{\z@}%
	{-3.5ex \@plus -1ex \@minus -.2ex}
	{2.3ex \@plus.2ex}%
	{\normalfont\large\bfseries}}
\renewcommand\subsection{\@startsection{subsection}{2}{\z@}%
	{-3.25ex\@plus -1ex \@minus -.2ex}%
	{1.5ex \@plus .2ex}%
	{\normalfont\bfseries}}
\makeatother

\begin{document}

\begin{center}
\addtolength{\baselineskip}{.5mm}
\thispagestyle{empty}
\begin{flushright}
\end{flushright}

\vspace{20mm}

{\Large  \bf On effective field theory of F-theory beyond leading order}
\\[15mm]
{Hamid R. Bakhtiarizadeh\footnote{bakhtiarizadeh@sirjantech.ac.ir}}
\\[5mm]
{\it Department of Physics, Sirjan University of Technology, Sirjan, Iran}

\vspace{20mm}

{\bf  Abstract}
\end{center}

We construct a proposal for effective  bosonic field theory at order $ \alpha'^3 $ in twelve dimensions, whose compactification on a circle and on a torus respectively yields eleven-dimensional and type IIB supergravity theories at eight-derivative level. The couplings $ ({\partial {G_5}})^2 R^2 $, $ ({\partial {F_4}})^2 R^2 $, $ ({\partial {F_4}})^4 $, $ ({\partial {G_5}})^4 $ and $ ({\partial {G_5}})^2({\partial {F_4}})^2 $ in twelve-dimensional supergravity are determined with this requirement that an ansatz of these couplings should admits a consistent truncation to the eleven-dimensional and type IIB supergravity theories. The self-duality condition of the five-form field strength in twelve dimensions is also understood by considering the RR five-form field strength of type IIB theory at linear order.

\vfill
\newpage


\section{Introduction}\label{int}

Type IIA and most of supergravity theories in lower dimensions can be obtained directly from eleven-dimensional supergravity \cite{Cremmer:1978km}, which is a low-energy description of the M-theory \cite{Witten:1995ex}, by applying the Kaluza-Klein (KK) reduction \cite{KK} on a compact circle. Unlike type IIA supergravity, the type IIB theory cannot be obtained by reduction of eleven-dimensional supergravity on a manifold.

The $ SL(2,\mathbb{Z}) $ symmetry of type IIB superstring theory, which is the modular group of a torus, provide strong evidence that the type IIB superstring theory in ten dimensions may have its origin in a twelve-dimensional theory known as F-theory \cite{Vafa:1996xn}. It is shown that type IIB superstring theory is obtained by nonperturbative compactification of F-theory on a torus for which the complex structure identified by axio-dilaton. In the other words, if one starts with twelve dimensions and compactifies on $ T^2 $, then $ SL(2,\mathbb{Z}) $ gets interpreted as the symmetry of the torus in ten dimensions. The dilaton field is not constant in these compactifications, and since the value of the dilaton determines the string coupling constant, thus these solutions cannot be studied perturbatively except in orientifolds \cite{Vafa:1996xn}. The low-energy description of this theory is believed to be the twelve-dimensional supergravity \cite{Ferrara:1996wv,Tseytlin:1996ne,Kar:1997cx,Nishino:1997gq,Khviengia:1997rh,Choi:2014vya,Choi:2015gia,Berman:2015rcc}. Compactification of F-theory on various manifolds which leads to the theories in lower dimensions has also been studied in \cite{Grimm:2010ks,Bonetti:2011mw,Grimm:2011fx,Cvetic:2012xn,DelZotto:2014fia,Martucci:2014ska,Junghans:2014zla,Malmendier:2014uka,Morrison:1996na,Morrison:1996pp}.

Till now, there is no comprehensive theory of supergravity in twelve dimensions. Some proposals have been introduced, but it seems none of them is complete. Serious difficulties in writing the action such as the 12-dimensional minimal fermion for which should be a superpartner components with spin higher than $ 2 $ in four dimensions \cite{Nahm:1977tg} as well as extra field degrees of freedom than elven-dimensional supergravity \cite{Choi:2014vya,Choi:2015gia}, make a big obstruction in writing the low-energy effective action of F-theory. 

With the present understanding, it is believed that the $ 12 $-dimensional theory should contain type IIB field contents in $ D = 10 $ along with those of M-theory in $ D = 11 $. In the other words, there might exist a twelve-dimensional theory that could be compactified to ten dimensions on a torus, in such a way that type IIB supergravity could be extracted as a consistent truncation. The consistency of truncation is crucial, since the solutions of the type IIB supergravity will also be solutions of the equations of motion of the twelve-dimensional theory. The proposed twelve-dimensional action should also be able to consistently reproduce $ D = 11 $ supergravity action after compactification on a circle \cite{Khviengia:1997rh}.

The supergravity action in twelve dimensions also consists of the lowest-order action plus an infinite number of higher-derivative terms beyond the leading order \cite{Grimm:2013bha,Minasian:2015bxa}. The higher-order terms of supergraity actions have also a significant importance in the study of particle interactions. The role of $ \alpha' $ corrections to the supergravity actions is of crucial importance in the study of various physical phenomena. Imprints of string/M/F theory arise from corrections that are at higher order in $ \alpha' $. Upon reduction to four dimensions, such higher-order terms are of particular phenomenological interest \cite{Grimm:2010ks,Antoniadis:1997eg,Grimm:2013bha}. Our main focus in this paper is $ \alpha'^3 $ corrections to the effective action of F-theory. In \cite{Grimm:2012rg,GarciaEtxebarria:2012zm,Grimm:2013gma,Grimm:2013bha,Junghans:2014zla}, there have been some efforts in deriving $ \alpha'^3 $ corrections to low-energy effective action of F-theory. 

In this paper, in fact we use the idea of Vafa's paper \cite{Vafa:1996xn} and generalize it to the eight-derivative level. We use the KK procedure to find these corrections. We start with making an ansatz for various possible couplings in twelve dimensions and putting them under a consistent truncation to capture the possible ones in elven and ten dimensions. Then, we compare them with their counterparts in eleven-dimensional and type IIB supergravity to derive the eight-derivative couplings in twelve dimensions. To this end, we follow the approach introduced in \cite{Khviengia:1997rh} for a consistent truncation of the bosonic fields of lowest-order supergravity action in $ D=12 $ to $ D=11 $ and $ D=10 $. We have already tested the correctness of our method in \cite{Bakhtiarizadeh:2017ojz} where we have calculated the known gauge field corrections to eleven-dimensional supergravity \cite{Peeters:2005tb} by a circular compactification to ten dimensions. 

Let us now briefly review the bosonic sector of $ 12 $-dimensional supergravity action at leading order given in Ref. \cite{Khviengia:1997rh}. The twelve-dimensional Lagrangian density at leading order is given by 
\bea
{\cal L}_{12}= e R-\frac{1}{2} e (\pa \psi)^2-\frac{1}{48} e^{a\psi} {F_4}^2 -\frac{1}{240} e e^{b\psi} {G_5}^2 +\lambda B_4 \wedge dA_3 \wedge dA_3,\label{12dsugra}
\eea
where $ e=\sqrt{-g} $, $ R $ is the Ricci scalar, $ \psi $ is the dilaton, $ a = i/\sqrt{5} $ and $ b=-2i/\sqrt{5} $. The $ 4 $-form and $ 5 $-form field strengths in twelve dimensions are defined as: $ F_4=dA_3 $ and $ G_5=dB_4 $, respectively. The coefficient $ \lambda $ can be obtained by comparison of the dimensionally-reduced Lagrangian in $ D = 11 $ with the bosonic sector of $ 11 $-dimensional supergravity, and it is $ \sqrt{3}/4 $. Thus, the bosonic field content of supergavity in twelve dimensions now includes the metric $ g_{MN} $, the dilaton $ \psi $, and the $ 3 $-form and $ 4 $-form potentials $ A_3 $ and $ B_4 $. 

As we know, the theories in $ D < 11 $ need real dilaton couplings, the theory in $ D = 11 $ itself needs zero dilaton coupling, and the theories in $ D > 11 $ need imaginary dilaton couplings. The imaginary couplings, regardless of being undesirable, are needed to make a consistent truncation to the fields of type IIB supergravity possible. In this paper, we drop the imaginary couplings, \ie those contain the dilaton field in twelve dimensions. 

It has been shown in \cite{Khviengia:1997rh} that the truncation of the twelve-dimensional Lagrangian (\ref{12dsugra}) to the type IIB theory in $ D=10 $ is consistent only up to linear order when the $ 5 $-form field strength is involved. Since, as we will see in Sec. \ref{10d}, the $ 12 $-dimensional $ 5 $-form field strength $ G_5 $ in $ D = 10 $ is given simply by $ F_5 = dB_4 $ without any Chern-Simons correction whereas the corresponding one in type IIB supergravity has a Chern-Simons correction as $ F_5 = dB_4 + \l \e_{ij} dA_{2}^{(i)} \wedge dA_{2}^{(j)} $. Although, the equation of motion and Bianchi identity for $ G_5 $ makes we cannot in general, consistently impose the self-duality condition $ G_5 = \star G_5 $, but the self-dual $ 5 $-form $ (G_5 + \star G_5) $ satisfies precisely the same equation of motion and Bianchi identity as in type IIB supergravity. This is exactly the same condition we already have included manually in calculating the couplings containing the RR $ 5 $-form field strength $ F_5 $ in ten dimensions \cite{Bakhtiarizadeh:2013zia,Bakhtiarizadeh:2017bpl,Bakhtiarizadeh:2015exa,Bakhtiarizadeh:2017ojz}. As a result, one can expect that automatically obtain the self-dual couplings of $ 5 $-form field strength $ G_5 $ in twelve dimensions by comparing the truncated couplings with the corresponding ones containing the RR self-dual $ 5 $-form field strength in ten dimensions. 

The structure of the paper is arranged as follows. First, we prepare an ansatz with unknown coefficients made of all possible contractions of tensors for various eight-derivative couplings in twelve dimensions. These bases are given in appendix. In Sec. \ref{11d}, we provide a consistent truncation of these bases on a circle to obtain the couplings in $ D=11 $. In Sec. \ref{10d}, we make a toroidal compactification of the bases to find the couplings in $ D = 10 $. In the next sections we will find the $ 12 $-dimensional couplings $ ({\partial {G_5}})^2 R^2 $, $ ({\partial {F_4}})^2 R^2 $, $ ({\partial {F_4}})^4 $, $ ({\partial {G_5}})^4 $ and $ ({\partial {G_5}})^2({\partial {F_4}})^2 $, with this requirement that the couplings arising from reduction of the bases to $ D=10 $ and $ D=11 $, should be consistent with the corresponding known ones in type IIB and eleven-dimensional supergravity, respectively. Finally, Sec. \ref{dis} is devoted to discussion.

In this paper our notations on coordinate indices are: a,b,c,... for indices in twelve dimensions, {\sf a,b,c,...} for indices in eleven dimensions, and {\tt a,b,c,...} for indices in ten dimensions.

\section{Reduction to eleven-dimensional supergravity}\label{11d}

Let us first review the standard Kaluza-Klein procedure introduced in Ref. \cite{Khviengia:1997rh} to reduce the $ 12 $-dimensional theory, first to $ D = 11 $ and then, in the next section, to $ D = 10 $. In an obvious notation, $ F_4 $ reduces to $ F_4 $, $ F_3^{(i)} $, $ F_2^{(ij)} $,... after compactification on internal circles labeled by $ i, j $ and $ G_5 $ similarly reduces to $ G_5 $, $ G_4^{(i)} $, $ G_3^{(ij)} $,.... The dimensional reduction of the Riemann curvature also gives rise to $ R_4 $, $ R_3^{(i)} $, $ R_2^{(ij)} $,....

Since the dimensionally-reduced theory in eleven dimensions contains more fields than ones in $ D = 11 $ supergravity, it is obvious that some of them must be set to zero. The crucial point is that this truncation must be consistent, \ie setting the fields to zero must be consistent with their equations of motion. It has been shown that we may consistently set
\bea
G_5 = F_3^{(1)} = 0. \label{11drules}
\eea
Setting both field strengths in (\ref{11drules}) to zero simultaneously follows from their equations of motion derived from dimensionally-reduced action at leading order.

To obtain $ D = 11 $ supergravity, one should be able to reduce the remaining system of fields further, so that in particular we have only a single independent $ 4 $-form field strength, rather than two. In doing so, we take $ F_4 $ and $ G_4^{(1)} $ to be proportional, again to ensure that this truncation of the theory is consistent with the equations of motion. Thus one may define
\bea
G_4^{(1)} = \sqrt{\frac{2}{3}} {\sf {F}_4 },\, F_4 = \frac{1}{\sqrt{3}} {\sf  {F}_4 }, \label{11drules2}
\eea
where ${\sf {F}_4 = d {\sf A}_3 }$ is the ordinary $ 4 $-form field strength of $ 11 $-dimensional supergravity. By applying the definitions (\ref{11drules2}), one arrives at a consistent truncation of the dimensionally-reduced theory in eleven dimensions at leading order. In the following, we are going to extend the above idea to make a consistent truncation of the eight-derivative couplings in $ 12 $-dimensional supergravity to $ D=11 $.

Our starting-points are the $ 12 $-dimensional bases given in appendix. First, we make the ansatz (\ref{G5G5RR}) for $ (\pa {G_5})^2 R^2 $ terms in the effective Lagrangian of F-theory at order $ \alpha'^3 $. Then, we consistently truncate it on a circle upon the above compactification rules to capture the coupling $ (\pa {F_4})^2 R^2 $ in eleven dimensions. Consequently, it takes the following form
\bea
&& {\sf \frac{1}{3}\lp\right. 5 a_1 {F_4}_{efgh,i} {F_4}^{efgh,i} R_{abcd} \
	R^{abcd} + 4 a_2 {F_4}_{efgi,h} \
	{F_4}^{efgh,i} R_{abcd} R^{abcd} } \nonumber \\ && {\sf + 4 a_3 {F_4}_{d}{}^{fgh,i} {F_4}_{efgh,i} R_{abc}{}^{e} R^{abcd} + 3 a_4 \
	{F_4}_{d}{}^{fgh,i} {F_4}_{efgi,h} R_{abc}{}^{e} R^{abcd} } \nonumber \\ 
&&{\sf - 4 a_5 {F_4}_{d}{}^{fgh,i} {F_4}_{fghi,e} R_{abc}{}^{e} R^{abcd} + 5 a_6 \
	{F_4}_{fghi,e} {F_4}^{fghi}{}_{,d} R_{abc}{}^{e} R^{abcd} }\nonumber \\ 
&&{\sf + 3 a_7 {F_4}_{ce}{}^{gh,i} {F_4}_{dfgh,i} R_{ab}{}^{ef} R^{abcd} + 2 a_8 \
	{F_4}_{ce}{}^{gh,i} {F_4}_{dfgi,h} R_{ab}{}^{ef} R^{abcd} }\nonumber \\ 
&&{\sf - 3 a_9 {F_4}_{ce}{}^{gh,i} {F_4}_{dghi,f} R_{ab}{}^{ef} R^{abcd} + 4 a_{10} \
	{F_4}_{c}{}^{ghi}{}_{,e} {F_4}_{dghi,f} R_{ab}{}^{ef} \
	R^{abcd} }\nonumber \\ 
&&{\sf + 3 a_{11} {F_4}_{cd}{}^{gh,i} {F_4}_{efgh,i} R_{ab}{}^{ef} R^{abcd} + 2 a_{12} \
	{F_4}_{cd}{}^{gh,i} {F_4}_{efgi,h} R_{ab}{}^{ef} R^{abcd} }\nonumber \\ 
&&{\sf - 3 a_{13} {F_4}_{cd}{}^{gh,i} {F_4}_{eghi,f} R_{ab}{}^{ef} R^{abcd} + 4 a_{14} \
	{F_4}_{c}{}^{ghi}{}_{,d} {F_4}_{eghi,f} R_{ab}{}^{ef} \
	R^{abcd} }\nonumber \\ 
&&{\sf + 4 a_{15} {F_4}_{c}{}^{ghi}{}_{,e} {F_4}_{fghi,d} R_{ab}{}^{ef} R^{abcd} + 5 a_{16} \
	{F_4}_{efgh,i} {F_4}^{efgh,i} R_{acbd} R^{abcd} }\nonumber \\ 
&&{\sf + 4 a_{17} {F_4}_{efgi,h} {F_4}^{efgh,i} R_{acbd} R^{abcd} + 4 a_{18} \
	{F_4}_{d}{}^{fgh,i} {F_4}_{efgh,i} R_{acb}{}^{e} R^{abcd} }\nonumber \\ 
&&{\sf + 3 a_{19} {F_4}_{d}{}^{fgh,i} {F_4}_{efgi,h} R_{acb}{}^{e} R^{abcd} - 4 a_{20} \
	{F_4}_{d}{}^{fgh,i} {F_4}_{fghi,e} R_{acb}{}^{e} R^{abcd} }\nonumber \\ 
&&{\sf + 5 a_{21} {F_4}_{fghi,e} {F_4}^{fghi}{}_{,d} R_{acb}{}^{e} R^{abcd} + 3 a_{22} \
	{F_4}_{be}{}^{gh,i} {F_4}_{dfgh,i} R_{ac}{}^{ef} R^{abcd} }\nonumber \\ 
&&{\sf + 2 a_{23} {F_4}_{be}{}^{gh,i} {F_4}_{dfgi,h} R_{ac}{}^{ef} R^{abcd} - 3 a_{24} \
	{F_4}_{be}{}^{gh,i} {F_4}_{dghi,f} R_{ac}{}^{ef} R^{abcd} }\nonumber \\ 
&&{\sf + 4 a_{25} {F_4}_{b}{}^{ghi}{}_{,e} {F_4}_{dghi,f} R_{ac}{}^{ef} R^{abcd} + 3 a_{26} \
	{F_4}_{bd}{}^{gh,i} {F_4}_{efgh,i} R_{ac}{}^{ef} R^{abcd} }\nonumber \\ 
&&{\sf - 3 a_{27} {F_4}_{b}{}^{ghi}{}_{,d} {F_4}_{efgh,i} R_{ac}{}^{ef} R^{abcd} + 2 a_{28} \
	{F_4}_{bd}{}^{gh,i} {F_4}_{efgi,h} R_{ac}{}^{ef} R^{abcd} }\nonumber \\ 
&&{\sf - 3 a_{29} {F_4}_{bd}{}^{gh,i} {F_4}_{eghi,f} R_{ac}{}^{ef} R^{abcd} + 4 a_{30} \
	{F_4}_{b}{}^{ghi}{}_{,d} {F_4}_{eghi,f} R_{ac}{}^{ef} \
	R^{abcd} }\nonumber \\ 
&&{\sf - 3 a_{31} {F_4}_{be}{}^{gh,i} {F_4}_{fghi,d} R_{ac}{}^{ef} R^{abcd} + 4 a_{32} \
	{F_4}_{b}{}^{ghi}{}_{,e} {F_4}_{fghi,d} R_{ac}{}^{ef} \
	R^{abcd} }\nonumber \\ 
&&{\sf + 4 a_{33} {F_4}_{e}{}^{ghi}{}_{,b} {F_4}_{fghi,d} R_{ac}{}^{ef} R^{abcd} + 3 a_{34} \
	{F_4}_{bf}{}^{gh,i} {F_4}_{degh,i} R_{a}{}^{e}{}_{c}{}^{f} R^{abcd} }\nonumber \\ 
&&{\sf + 2 a_{35} {F_4}_{bf}{}^{gh,i} {F_4}_{degi,h} R_{a}{}^{e}{}_{c}{}^{f} R^{abcd} + 3 a_{36} {F_4}_{be}{}^{gh,i} {F_4}_{dfgh,i} R_{a}{}^{e}{}_{c}{}^{f} R^{abcd} }\nonumber \\ 
&&{\sf + 2 a_{37} {F_4}_{be}{}^{gh,i} {F_4}_{dfgi,h} R_{a}{}^{e}{}_{c}{}^{f} R^{abcd} - 3 a_{38} {F_4}_{bf}{}^{gh,i} {F_4}_{dghi,e} R_{a}{}^{e}{}_{c}{}^{f} R^{abcd} }\nonumber \\ 
&&{\sf + 4 a_{39} {F_4}_{b}{}^{ghi}{}_{,f} {F_4}_{dghi,e} R_{a}{}^{e}{}_{c}{}^{f} R^{abcd} - 3 a_{40} {F_4}_{be}{}^{gh,i} {F_4}_{dghi,f} R_{a}{}^{e}{}_{c}{}^{f} R^{abcd} }\nonumber \\ 
&&{\sf + 4 a_{41} {F_4}_{b}{}^{ghi}{}_{,e} {F_4}_{dghi,f} R_{a}{}^{e}{}_{c}{}^{f} R^{abcd} + 3 a_{42} {F_4}_{bd}{}^{gh,i} {F_4}_{efgh,i} R_{a}{}^{e}{}_{c}{}^{f} R^{abcd} }\nonumber \\ 
&&{\sf + 2 a_{43} {F_4}_{bd}{}^{gh,i} {F_4}_{efgi,h} R_{a}{}^{e}{}_{c}{}^{f} R^{abcd} + 4 a_{44} {F_4}_{b}{}^{ghi}{}_{,f} {F_4}_{eghi,d} R_{a}{}^{e}{}_{c}{}^{f} R^{abcd} }\nonumber \\ 
&&{\sf - 3 a_{45} {F_4}_{bd}{}^{gh,i} {F_4}_{eghi,f} R_{a}{}^{e}{}_{c}{}^{f} R^{abcd} + 4 a_{46} {F_4}_{b}{}^{ghi}{}_{,d} {F_4}_{eghi,f} R_{a}{}^{e}{}_{c}{}^{f} R^{abcd} }\nonumber \\ 
&&{\sf + 4 a_{47} {F_4}_{b}{}^{ghi}{}_{,e} {F_4}_{fghi,d} R_{a}{}^{e}{}_{c}{}^{f} R^{abcd} + 4 a_{48} {F_4}_{b}{}^{ghi}{}_{,d} {F_4}_{fghi,e} R_{a}{}^{e}{}_{c}{}^{f} R^{abcd} }\nonumber \\ 
&&{\sf + 2 a_{49} {F_4}_{bfg}{}^{h,i} {F_4}_{cdeh,i} R_{a}{}^{efg} R^{abcd} + a_{50} \
	{F_4}_{bfg}{}^{h,i} {F_4}_{cdei,h} R_{a}{}^{efg} R^{abcd} }\nonumber \\ 
&&{\sf - 2 a_{51} {F_4}_{bf}{}^{hi}{}_{,e} {F_4}_{cdgh,i} R_{a}{}^{efg} R^{abcd} - 2 a_{52} \
	{F_4}_{bfg}{}^{h,i} {F_4}_{cdhi,e} R_{a}{}^{efg} R^{abcd} }\nonumber \\ 
&&{\sf + 3 a_{53} {F_4}_{bf}{}^{hi}{}_{,g} {F_4}_{cdhi,e} R_{a}{}^{efg} R^{abcd} + 3 a_{54} \
	{F_4}_{bf}{}^{hi}{}_{,e} {F_4}_{cdhi,g} R_{a}{}^{efg} \
	R^{abcd} }\nonumber \\ 
&&{\sf - 2 a_{55} {F_4}_{bfg}{}^{h,i} {F_4}_{cehi,d} R_{a}{}^{efg} R^{abcd} + 3 a_{56} \
	{F_4}_{bf}{}^{hi}{}_{,g} {F_4}_{cehi,d} R_{a}{}^{efg} \
	R^{abcd} }\nonumber \\ 
&&{\sf + 3 a_{57} {F_4}_{bf}{}^{hi}{}_{,e} {F_4}_{cghi,d} R_{a}{}^{efg} R^{abcd} + 2 a_{58} \
	{F_4}_{bcf}{}^{h,i} {F_4}_{degh,i} R_{a}{}^{efg} R^{abcd} }\nonumber \\ 
&&{\sf + a_{59} {F_4}_{bcf}{}^{h,i} {F_4}_{degi,h} R_{a}{}^{efg} R^{abcd} - 2 a_{60} \
	{F_4}_{bcf}{}^{h,i} {F_4}_{dehi,g} R_{a}{}^{efg} R^{abcd} }\nonumber \\ 
&&{\sf + 3 a_{61} {F_4}_{bc}{}^{hi}{}_{,f} {F_4}_{dehi,g} R_{a}{}^{efg} R^{abcd} + 3 a_{62} \
	{F_4}_{bf}{}^{hi}{}_{,c} {F_4}_{dehi,g} R_{a}{}^{efg} \
	R^{abcd} }\nonumber \\ 
&&{\sf + 2 a_{63} {F_4}_{bce}{}^{h,i} {F_4}_{dfgh,i} R_{a}{}^{efg} R^{abcd} - 2 a_{64} \
	{F_4}_{bc}{}^{hi}{}_{,e} {F_4}_{dfgh,i} R_{a}{}^{efg} \
	R^{abcd} }\nonumber \\ 
&&{\sf - 2 a_{65} {F_4}_{be}{}^{hi}{}_{,c} {F_4}_{dfgh,i} R_{a}{}^{efg} R^{abcd} + a_{66} \
	{F_4}_{bce}{}^{h,i} {F_4}_{dfgi,h} R_{a}{}^{efg} R^{abcd} }\nonumber \\ 
&&{\sf - 2 a_{67} {F_4}_{bce}{}^{h,i} {F_4}_{dfhi,g} R_{a}{}^{efg} R^{abcd} + 3 a_{68} \
	{F_4}_{bc}{}^{hi}{}_{,e} {F_4}_{dfhi,g} R_{a}{}^{efg} \
	R^{abcd} }\nonumber \\ 
&&{\sf + 3 a_{69} {F_4}_{be}{}^{hi}{}_{,c} {F_4}_{dfhi,g} R_{a}{}^{efg} R^{abcd} - 2 a_{70} \
	{F_4}_{bcf}{}^{h,i} {F_4}_{dghi,e} R_{a}{}^{efg} R^{abcd} }\nonumber \\ 
&&{\sf + 3 a_{71} {F_4}_{bc}{}^{hi}{}_{,f} {F_4}_{dghi,e} R_{a}{}^{efg} R^{abcd} + 3 a_{72} \
	{F_4}_{bf}{}^{hi}{}_{,c} {F_4}_{dghi,e} R_{a}{}^{efg} \
	R^{abcd} }\nonumber \\ 
&&{\sf + 3 a_{73} {F_4}_{cf}{}^{hi}{}_{,b} {F_4}_{dghi,e} R_{a}{}^{efg} R^{abcd} + 2 a_{74} \
	{F_4}_{bcd}{}^{h,i} {F_4}_{efgh,i} R_{a}{}^{efg} R^{abcd} }\nonumber \\ 
&&{\sf + a_{75} {F_4}_{bcd}{}^{h,i} {F_4}_{efgi,h} R_{a}{}^{efg} R^{abcd} - 2 a_{76} \
	{F_4}_{bcd}{}^{h,i} {F_4}_{efhi,g} R_{a}{}^{efg} R^{abcd} }\nonumber \\ 
&&{\sf + 3 a_{77} {F_4}_{bc}{}^{hi}{}_{,d} {F_4}_{efhi,g} R_{a}{}^{efg} R^{abcd} - 2 a_{78} \
	{F_4}_{bcf}{}^{h,i} {F_4}_{eghi,d} R_{a}{}^{efg} R^{abcd} }\nonumber \\ 
&&{\sf + 3 a_{79} {F_4}_{bc}{}^{hi}{}_{,f} {F_4}_{eghi,d} R_{a}{}^{efg} R^{abcd} + 3 a_{80} \
	{F_4}_{cd}{}^{hi}{}_{,e} {F_4}_{fghi,b} R_{a}{}^{efg} \
	R^{abcd} }\nonumber \\ 
&&{\sf - 2 a_{81} {F_4}_{bce}{}^{h,i} {F_4}_{fghi,d} R_{a}{}^{efg} R^{abcd} + 3 a_{82} \
	{F_4}_{bc}{}^{hi}{}_{,e} {F_4}_{fghi,d} R_{a}{}^{efg} \
	R^{abcd} }\nonumber \\ 
&&{\sf + 3 a_{83} {F_4}_{be}{}^{hi}{}_{,c} {F_4}_{fghi,d} R_{a}{}^{efg} R^{abcd} - 2 a_{84} \
	{F_4}_{bcd}{}^{h,i} {F_4}_{fghi,e} R_{a}{}^{efg} R^{abcd} }\nonumber \\ 
&&{\sf + 3 a_{85} {F_4}_{bc}{}^{hi}{}_{,d} {F_4}_{fghi,e} R_{a}{}^{efg} R^{abcd} + 3 a_{86} \
	{F_4}_{cd}{}^{hi}{}_{,b} {F_4}_{fghi,e} R_{a}{}^{efg} \
	R^{abcd} }\nonumber \\ 
&&{\sf + a_{87} {F_4}_{aceg}{}^{,i} {F_4}_{bdfh,i} R^{abcd} R^{efgh} -  a_{89} \
	{F_4}_{aceg}{}^{,i} {F_4}_{bdfi,h} R^{abcd} R^{efgh} }\nonumber \\ 
&&{\sf + 2 a_{90} {F_4}_{ace}{}^{i}{}_{,g} {F_4}_{bdfi,h} R^{abcd} R^{efgh} + 2 a_{91} \
	{F_4}_{ace}{}^{i}{}_{,f} {F_4}_{bdgi,h} R^{abcd} R^{efgh} }\nonumber \\ 
&&{\sf + 2 a_{92} {F_4}_{ace}{}^{i}{}_{,g} {F_4}_{bdhi,f} R^{abcd} R^{efgh} + 2 a_{93} \
	{F_4}_{ace}{}^{i}{}_{,g} {F_4}_{bfhi,d} R^{abcd} R^{efgh} }\nonumber \\ 
&&{\sf + a_{94} {F_4}_{abeg}{}^{,i} {F_4}_{cdfh,i} R^{abcd} R^{efgh} -  a_{96} \
	{F_4}_{abeg}{}^{,i} {F_4}_{cdfi,h} R^{abcd} R^{efgh} }\nonumber \\ 
&&{\sf + 2 a_{97} {F_4}_{abe}{}^{i}{}_{,g} {F_4}_{cdfi,h} R^{abcd} R^{efgh} + a_{98} \
	{F_4}_{abef}{}^{,i} {F_4}_{cdgh,i} R^{abcd} R^{efgh} }\nonumber \\ 
&&{\sf -  a_{100} {F_4}_{abef}{}^{,i} {F_4}_{cdgi,h} R^{abcd} R^{efgh} + 2 a_{101} \
	{F_4}_{abe}{}^{i}{}_{,f} {F_4}_{cdgi,h} R^{abcd} R^{efgh} }\nonumber \\ 
&&{\sf + 2 a_{102} {F_4}_{abe}{}^{i}{}_{,g} {F_4}_{cdhi,f} R^{abcd} R^{efgh} -  a_{103} \
	{F_4}_{abeg}{}^{,i} {F_4}_{cfhi,d} R^{abcd} R^{efgh} }\nonumber \\ 
&&{\sf + 2 a_{104} {F_4}_{abe}{}^{i}{}_{,g} {F_4}_{cfhi,d} R^{abcd} R^{efgh} + 2 a_{105} \
	{F_4}_{abe}{}^{i}{}_{,f} {F_4}_{cghi,d} R^{abcd} R^{efgh} }\nonumber \\ 
&&{\sf + a_{106} {F_4}_{abce}{}^{,i} {F_4}_{dfgh,i} R^{abcd} R^{efgh} -  a_{108} \
	{F_4}_{abce}{}^{,i} {F_4}_{dfgi,h} R^{abcd} R^{efgh} }\nonumber \\ 
&&{\sf + 2 a_{109} {F_4}_{abc}{}^{i}{}_{,e} {F_4}_{dfgi,h} R^{abcd} R^{efgh} + 2 a_{110} \
	{F_4}_{abe}{}^{i}{}_{,c} {F_4}_{dfgi,h} R^{abcd} R^{efgh} }\nonumber \\ 
&&{\sf + 2 a_{111} {F_4}_{ace}{}^{i}{}_{,b} {F_4}_{dfgi,h} R^{abcd} R^{efgh} -  a_{112} \
	{F_4}_{abce}{}^{,i} {F_4}_{dghi,f} R^{abcd} R^{efgh} }\nonumber \\ 
&&{\sf + 2 a_{113} {F_4}_{abc}{}^{i}{}_{,e} {F_4}_{dghi,f} R^{abcd} R^{efgh} + 2 a_{114} \
	{F_4}_{abe}{}^{i}{}_{,c} {F_4}_{dghi,f} R^{abcd} R^{efgh} }\nonumber \\ 
&&{\sf + a_{115} {F_4}_{abcd}{}^{,i} {F_4}_{efgh,i} R^{abcd} R^{efgh} -  a_{117} \
	{F_4}_{abcd}{}^{,i} {F_4}_{efgi,h} R^{abcd} R^{efgh} }\nonumber \\ 
&&{\sf + 2 a_{118} {F_4}_{abc}{}^{i}{}_{,d} {F_4}_{efgi,h} R^{abcd} R^{efgh} -  a_{119} \
	{F_4}_{abce}{}^{,i} {F_4}_{fghi,d} R^{abcd} R^{efgh} }\nonumber \\ 
&&{\sf + 2 a_{120} {F_4}_{abc}{}^{i}{}_{,e} {F_4}_{fghi,d} R^{abcd} R^{efgh} \left.\rp}. \label{F4F4RR11}
\eea
In obtaining this result, we have taken the Riemann curvature in eleven dimensions as $ 12 $-dimensional one. In our notations this means $ R_4={\sf R_4} $. This is plausible at least for linearized Riemann curvature where we have considered in this article.    

Now, we consider reduction of the basis (\ref{F4F4RR}) constructed of $ (\pa {F_4})^2 R^2 $ terms in $ 12 $-dimensional supergravity action. Dimensional reduction of the ansatz to eleven dimensions by compactifying it on a circle leads to $ (\pa {F_4})^2 R^2 $ terms based on the above compactification rules. It can easily be seen from Eq. (\ref{11drules2}) that $ (\pa {F_4})^2 R^2 $ coupling in eleven dimensions has the same structure as $ 12 $-dimensional one but with an overall factor $ 2/3 $.

A basis for $ (\pa {F_4})^4 $ terms in twelve dimensions is given by (\ref{F4F4F4F4}). Putting it under a circular reduction gives rise to $ (\pa {F_4})^4 $ coupling in eleven dimensions. According to Eqs. (\ref{11drules}) and (\ref{11drules2}), we observe that this coupling has the same form as the ansatz (\ref{F4F4F4F4}) which is now multiplied by an overall factor $ 4/9 $ due to the rule (\ref{11drules2}). 
  
In the following, we apply the same procedure to the basis (\ref{G5G5G5G5}) to derive the couplings in dimensionally-reduced theory. Reduction on a circle yields the terms $ (\pa {F_4})^4 $ which are given by
\bea
&&{\sf \frac{1}{9}\lp\right.-4 (d_1 + d_2) {F_4}_{ae}{}^{fg,h} \
{F_4}^{abcd,e} {F_4}_{bfh}{}^{i,j} {F_4}_{cgij,d} }\nonumber \\ 
&&{\sf - 2 (3 d_{11} + 2 d_{15} + d_5) {F_4}_{ab}{}^{fg,h} \
{F_4}^{abcd,e} {F_4}_{ceh}{}^{i,j} {F_4}_{dfij,g} }\nonumber \\ 
&&{\sf + \bigl(9 d_{12} - 2 (d_3 + 2 d_9)\bigr) {F_4}_{ab}{}^{fg,h} {F_4}^{abcd,e} \
{F_4}_{ce}{}^{ij}{}_{,h} {F_4}_{dfij,g} }\nonumber \\ 
&&{\sf+ \bigl(6 d_{16} \
- 4 (d_{27}  + d_3)\bigr) {F_4}_{ab}{}^{fg,h} {F_4}^{abcd,e} \
{F_4}_{cef}{}^{i,j} {F_4}_{dghi,j} }\nonumber \\ 
&&{\sf+ (-2 d_{13} + 9 \
d_{17} - 4 d_8) {F_4}_{abe}{}^{f,g} \
{F_4}^{abcd,e} {F_4}_{cf}{}^{hi,j} {F_4}_{dghi,j} }\nonumber \\ 
&& {\sf+ (4 d_{14} + 9 d_{18}) \
{F_4}_{ab}{}^{fg}{}_{,e} {F_4}^{abcd,e} {F_4}_{cf}{}^{hi,j} \
{F_4}_{dghi,j} }\nonumber \\ 
&&{\sf+ \bigl(3 d_{21} - 4 (d_{32}  + d_4)\bigr) {F_4}_{ab}{}^{fg,h} {F_4}^{abcd,e} \
{F_4}_{cef}{}^{i,j} {F_4}_{dghj,i} }\nonumber \\ 
&&{\sf- 2 (2 d_{10} + \
d_{19} - 3 d_{22}) {F_4}_{abe}{}^{f,g} \
{F_4}^{abcd,e} {F_4}_{cf}{}^{hi,j} {F_4}_{dghj,i} }\nonumber \\ 
&&{\sf + (4 d_{20} + 6 d_{23}) \
{F_4}_{ab}{}^{fg}{}_{,e} {F_4}^{abcd,e} {F_4}_{cf}{}^{hi,j} \
{F_4}_{dghj,i} }\nonumber \\ 
&&{\sf + 4 (d_{24} + d_{28}) \
{F_4}_{ae}{}^{fg}{}_{,b} {F_4}^{abcd,e} {F_4}_{cf}{}^{hi,j} \
{F_4}_{dghj,i} }\nonumber \\ 
&&{\sf + (4 d_{13} - 4 d_{25} - 6 d_{29}) \
{F_4}_{ab}{}^{fg,h} {F_4}^{abcd,e} {F_4}_{cef}{}^{i,j} \
{F_4}_{dgij,h} }\nonumber \\ 
&&{\sf+ (-2 d_{25} + 9 d_{30}  + 4 d_8) {F_4}_{ab}{}^{fg,h} {F_4}^{abcd,e} \
{F_4}_{ce}{}^{ij}{}_{,f} {F_4}_{dgij,h} }\nonumber \\ 
&&{\sf+ (4 d_{14} + 8 \
d_{26} + 9 d_{31}) {F_4}_{ab}{}^{fg,h} \
{F_4}^{abcd,e} {F_4}_{cf}{}^{ij}{}_{,e} {F_4}_{dgij,h} }\nonumber \\ 
&&{\sf + 2 (d_{19} - 3 d_{33} - 2 d_{48}) \
{F_4}_{ab}{}^{fg,h} {F_4}^{abcd,e} {F_4}_{cef}{}^{i,j} \
{F_4}_{dhij,g} }\nonumber \\ 
&&{\sf+ (2 d_{10} + 9 d_{34}  - 2 d_{48}) {F_4}_{ab}{}^{fg,h} {F_4}^{abcd,e} \
{F_4}_{ce}{}^{ij}{}_{,f} {F_4}_{dhij,g} }\nonumber \\ 
&&{\sf+ (-2 d_{27} - 9 \
d_{35} - 4 d_9) {F_4}_{abe}{}^{f,g} \
{F_4}^{abcd,e} {F_4}_{cf}{}^{hi,j} {F_4}_{dhij,g} }\nonumber \\ 
&&{\sf + (2 d_{20} + 9 d_{36} + 8 d_{49}) \
{F_4}_{ab}{}^{fg,h} {F_4}^{abcd,e} {F_4}_{cf}{}^{ij}{}_{,e} \
{F_4}_{dhij,g} }\nonumber \\ 
&&{\sf+ (d_3 + 6 d_{37}  + 9 d_{53}) {F_4}_{ab}{}^{fg,h} {F_4}^{abcd,e} \
{F_4}_{cdh}{}^{i,j} {F_4}_{efgi,j} }\nonumber \\ 
&&{\sf+ (3 d_{38} + d_4 + 9 d_{56}) {F_4}_{ab}{}^{fg,h} {F_4}^{abcd,e} \
{F_4}_{cdh}{}^{i,j} {F_4}_{efgj,i} }\nonumber \\ 
&&{\sf + (d_{16} - 4 d_{37} + 12 d_{39}) \
{F_4}_{abc}{}^{f,g} {F_4}^{abcd,e} {F_4}_{dg}{}^{hi,j} \
{F_4}_{efhi,j} }\nonumber \\ 
&&{\sf+ (d_{21} - 4 d_{38}  + 8 d_{40}) {F_4}_{abc}{}^{f,g} {F_4}^{abcd,e} \
{F_4}_{dg}{}^{hi,j} {F_4}_{efhj,i} }\nonumber \\ 
&&{\sf+ (-6 d_{41} - 6 \
d_{43} + d_5) {F_4}_{ab}{}^{fg,h} \
{F_4}^{abcd,e} {F_4}_{cdh}{}^{i,j} {F_4}_{efij,g} }\nonumber \\ 
&&{\sf + (6 d_{42} - 6 d_{52} + 4 d_6) \
{F_4}_{ab}{}^{fg,h} {F_4}^{abcd,e} {F_4}_{cdf}{}^{i,j} \
{F_4}_{eghi,j} }\nonumber \\ 
&&{\sf+ 2 (d_{42}  + 6 d_{44}) {F_4}_{abc}{}^{f,g} {F_4}^{abcd,e} \
{F_4}_{df}{}^{hi,j} {F_4}_{eghi,j} }\nonumber \\ 
&&{\sf+ \bigl(- d_{11} + 2 \
(d_{43} - 6 d_{45})\bigr) {F_4}_{abc}{}^{f,g} \
{F_4}^{abcd,e} {F_4}_{d}{}^{hij}{}_{,f} {F_4}_{eghi,j} }\nonumber \\ 
&&{\sf + (3 d_{46} - 6 d_{55} + 4 d_7) \
{F_4}_{ab}{}^{fg,h} {F_4}^{abcd,e} {F_4}_{cdf}{}^{i,j} \
{F_4}_{eghj,i} }\nonumber \\ 
&&{\sf+ 2 (d_{46} + 4 d_{47}) {F_4}_{abc}{}^{f,g} {F_4}^{abcd,e} \
{F_4}_{df}{}^{hi,j} {F_4}_{eghj,i} }\nonumber \\ 
&&{\sf+ (d_{28} + 8 d_{50} + 9 d_{51}) {F_4}_{ab}{}^{fg,h} {F_4}^{abcd,e} \
{F_4}_{ch}{}^{ij}{}_{,f} {F_4}_{egij,d} }\nonumber \\ 
&&{\sf + (-2 d_{41} - 12 d_{54} + d_{95}) \
{F_4}_{abc}{}^{f,g} {F_4}^{abcd,e} {F_4}_{dg}{}^{hi,j} \
{F_4}_{ehij,f} }\nonumber \\ 
&&{\sf+ 2 (d_{100} + d_{52} - 6 d_{57}) {F_4}_{abc}{}^{f,g} {F_4}^{abcd,e} \
{F_4}_{df}{}^{hi,j} {F_4}_{ehij,g} }\nonumber \\ 
&&{\sf+ \bigl(- d_{12} + 2 \
(d_{53} + 8 d_{58})\bigr) {F_4}_{abc}{}^{f,g} \
{F_4}^{abcd,e} {F_4}_{d}{}^{hij}{}_{,f} {F_4}_{ehij,g} }\nonumber \\ 
&&{\sf + (- d_{13} + 9 d_{61} - 6 d_{80}) \
{F_4}_{abe}{}^{f,g} {F_4}^{abcd,e} {F_4}_{cd}{}^{hi,j} \
{F_4}_{fghi,j} }\nonumber \\ 
&&{\sf+ \bigl(d_{14} + 9 (d_{62}  + d_{81})\bigr) {F_4}_{ab}{}^{fg}{}_{,e} \
{F_4}^{abcd,e} {F_4}_{cd}{}^{hi,j} {F_4}_{fghi,j} }\nonumber \\ 
&&{\sf+ (- d_1 - 8 d_{59} - 9 d_{63}) \
{F_4}_{abe}{}^{f,g} {F_4}^{abcd,e} {F_4}_{c}{}^{hij}{}_{,d} \
{F_4}_{fghi,j} }\nonumber \\ 
&&{\sf + \bigl(- d_{16} + 4 (d_{60} + 3 d_{64})\bigr) {F_4}_{abc}{}^{f,g} {F_4}^{abcd,e} \
{F_4}_{de}{}^{hi,j} {F_4}_{fghi,j} }\nonumber \\ 
&&{\sf+ (- d_{17} - 2 \
d_{61}  + 16 d_{65}) {F_4}_{abce}{}^{,f} {F_4}^{abcd,e} \
{F_4}_{d}{}^{ghi,j} {F_4}_{fghi,j} }\nonumber \\ 
&&{\sf+ 2 (d_{18} + 2 \
d_{62} + 8 d_{66}) {F_4}_{abc}{}^{f}{}_{,e} \
{F_4}^{abcd,e} {F_4}_{d}{}^{ghi,j} {F_4}_{fghi,j} }\nonumber \\ 
&&{\sf + (- d_{39} + d_{64} + 20 d_{67}) \
{F_4}_{abcd}{}^{,f} {F_4}^{abcd,e} {F_4}_{e}{}^{ghi,j} \
{F_4}_{fghi,j} }\nonumber \\ 
&&{\sf+ (- d_{19} + 6 d_{69}  - 6 d_{85}) {F_4}_{abe}{}^{f,g} {F_4}^{abcd,e} \
{F_4}_{cd}{}^{hi,j} {F_4}_{fghj,i} }\nonumber \\ 
&&{\sf+ (d_{20} + 6 d_{70} + 9 d_{87}) {F_4}_{ab}{}^{fg}{}_{,e} \
{F_4}^{abcd,e} {F_4}_{cd}{}^{hi,j} {F_4}_{fghj,i} }\nonumber \\ 
&&{\sf + (- d_{21} + 4 d_{68} + 8 d_{72}) \
{F_4}_{abc}{}^{f,g} {F_4}^{abcd,e} {F_4}_{de}{}^{hi,j} \
{F_4}_{fghj,i} }\nonumber \\ 
&&{\sf+ (- d_{22} - 2 d_{69} + 12 d_{73}) {F_4}_{abce}{}^{,f} {F_4}^{abcd,e} \
{F_4}_{d}{}^{ghi,j} {F_4}_{fghj,i} }\nonumber \\ 
&&{\sf+ 2 (d_{23} + 2 \
d_{70} + 6 d_{74}) {F_4}_{abc}{}^{f}{}_{,e} \
{F_4}^{abcd,e} {F_4}_{d}{}^{ghi,j} {F_4}_{fghj,i} }\nonumber \\ 
&&{\sf + (2 d_{24} + 4 d_{71} + 9 d_{75}) \
{F_4}_{abe}{}^{f}{}_{,c} {F_4}^{abcd,e} {F_4}_{d}{}^{ghi,j} \
{F_4}_{fghj,i} }\nonumber \\ 
&&{\sf+ (- d_{40} + d_{72}  + 15 d_{76}) {F_4}_{abcd}{}^{,f} {F_4}^{abcd,e} \
{F_4}_{e}{}^{ghi,j} {F_4}_{fghj,i} }\nonumber \\ 
&&{\sf+ (9 d_{101} + \
d_{25} - 6 d_{78}) {F_4}_{ab}{}^{fg,h} \
{F_4}^{abcd,e} {F_4}_{cdh}{}^{i,j} {F_4}_{fgij,e} }\nonumber \\ 
&&{\sf + (d_{26} + 9 d_{79}) {F_4}_{ab}{}^{fg,h} \
{F_4}^{abcd,e} {F_4}_{cd}{}^{ij}{}_{,h} {F_4}_{fgij,e} }\nonumber \\ 
&&{\sf+ (- d_1 - 4 d_{77}  - 9 d_{82}) {F_4}_{abe}{}^{f,g} {F_4}^{abcd,e} \
{F_4}_{cg}{}^{hi,j} {F_4}_{fhij,d} }\nonumber \\ 
&&{\sf+ \bigl(- d_{29} + 4 \
(d_{78} - 3 d_{83})\bigr) {F_4}_{abc}{}^{f,g} \
{F_4}^{abcd,e} {F_4}_{dg}{}^{hi,j} {F_4}_{fhij,e} }\nonumber \\ 
&&{\sf + \bigl(d_{31} + 4 (d_{79} + 4 d_{84})\bigr) {F_4}_{abc}{}^{f,g} {F_4}^{abcd,e} \
{F_4}_{d}{}^{hij}{}_{,g} {F_4}_{fhij,e} }\nonumber \\ 
&&{\sf+ (- d_{27} - 6 \
d_{60} - 9 d_{86}) {F_4}_{abe}{}^{f,g} {F_4}^{abcd,e} \
{F_4}_{cd}{}^{hi,j} {F_4}_{fhij,g} }\nonumber \\ 
&&{\sf+ (d_{28} + 4 d_{71} + 9 d_{88}) {F_4}_{ab}{}^{fg,h} {F_4}^{abcd,e} \
{F_4}_{ce}{}^{ij}{}_{,d} {F_4}_{fhij,g} }\nonumber \\ 
&&{\sf + \bigl(- d_{29} + 4 (d_{80} - 3 d_{89})\bigr) {F_4}_{abc}{}^{f,g} {F_4}^{abcd,e} \
{F_4}_{de}{}^{hi,j} {F_4}_{fhij,g} }\nonumber \\ 
&&{\sf+ \bigl(d_{31} + 4 \
(d_{81}  + 4 d_{90})\bigr) {F_4}_{abc}{}^{f,g} {F_4}^{abcd,e} \
{F_4}_{d}{}^{hij}{}_{,e} {F_4}_{fhij,g} }\nonumber \\ 
&&{\sf + (d_{66} + 25 \
d_{91}) {F_4}_{abcd,e} {F_4}^{abcd,e} {F_4}_{fghi,j} \
{F_4}^{fghi,j} }\nonumber \\ 
&&{\sf + (d_{74} + 20 d_{92}) {F_4}_{abcd,e} \
{F_4}^{abcd,e} {F_4}_{fghj,i} {F_4}^{fghi,j} }\nonumber \\ 
&&{\sf+ (d_{75} + \
16 d_{93}) {F_4}_{abce,d} {F_4}^{abcd,e} {F_4}_{fghj,i} \
{F_4}^{fghi,j} }\nonumber \\ 
&&{\sf - 2 (d_5 + 2 d_{94} + 3 d_{95}) \
{F_4}_{ab}{}^{fg,h} {F_4}^{abcd,e} {F_4}_{cef}{}^{i,j} \
{F_4}_{ghij,d} }\nonumber \\ 
&&{\sf+ (-2 d_2 - 4 d_{77} + 9 d_{96}) {F_4}_{ab}{}^{fg,h} {F_4}^{abcd,e} \
{F_4}_{ce}{}^{ij}{}_{,f} {F_4}_{ghij,d} }\nonumber \\ 
&&{\sf + (2 d_{15} + 2 \
d_{94} - 9 d_{97}) {F_4}_{abe}{}^{f,g} \
{F_4}^{abcd,e} {F_4}_{cf}{}^{hi,j} {F_4}_{ghij,d} }\nonumber \\ 
&&{\sf + (- d_{96} + 16 d_{98}) {F_4}_{abc}{}^{f,g} \
{F_4}^{abcd,e} {F_4}_{e}{}^{hij}{}_{,f} {F_4}_{ghij,d} }\nonumber \\ 
&&{\sf+ (- \
d_{63} -  d_{82}  - 16 d_{99}) {F_4}_{abce}{}^{,f} {F_4}^{abcd,e} \
{F_4}_{f}{}^{ghi,j} {F_4}_{ghij,d} }\nonumber \\ 
&&{\sf+ (-6 d_{100} + 2 \
d_6) {F_4}_{ab}{}^{fg,h} {F_4}^{abcd,e} \
{F_4}_{cdf}{}^{i,j} {F_4}_{ghij,e} }\nonumber \\ 
&&{\sf + (-2 d_{101} + 16 d_{102} + d_{30}) \
{F_4}_{abc}{}^{f,g} {F_4}^{abcd,e} {F_4}_{d}{}^{hij}{}_{,f} \
{F_4}_{ghij,e} }\nonumber \\ 
&&{\sf+ (-9 d_{103} -  d_{32}  - 3 d_{68}) {F_4}_{abe}{}^{f,g} {F_4}^{abcd,e} \
{F_4}_{cd}{}^{hi,j} {F_4}_{ghij,f} }\nonumber \\ 
&&{\sf+ (-12 d_{104} -  \
d_{33} + 2 d_{85}) {F_4}_{abc}{}^{f,g} \
{F_4}^{abcd,e} {F_4}_{de}{}^{hi,j} {F_4}_{ghij,f} }\nonumber \\ 
&&{\sf + (-16 d_{105} -  d_{35} - 2 d_{86}) \
{F_4}_{abce}{}^{,f} {F_4}^{abcd,e} {F_4}_{d}{}^{ghi,j} \
{F_4}_{ghij,f} }\nonumber \\ 
&&{\sf+ (16 d_{106} + d_{36} + 2 d_{87}) {F_4}_{abc}{}^{f,g} {F_4}^{abcd,e} \
{F_4}_{d}{}^{hij}{}_{,e} {F_4}_{ghij,f} }\nonumber \\ 
&&{\sf + (-20 d_{107} + \
d_{83} + d_{89}) {F_4}_{abcd}{}^{,f} \
{F_4}^{abcd,e} {F_4}_{e}{}^{ghi,j} {F_4}_{ghij,f} }\nonumber \\ 
&&{\sf + (16 d_{108} + d_{51} + d_{88}) \
{F_4}_{abc}{}^{f,g} {F_4}^{abcd,e} {F_4}_{e}{}^{hij}{}_{,d} \
{F_4}_{ghij,f} }\nonumber \\ 
&&{\sf + (25 d_{109} + d_{84} + d_{90}) {F_4}_{abcd}{}^{,f} {F_4}^{abcd,e} \
{F_4}_{ghij,f} {F_4}^{ghij}{}_{,e} \left. \rp},\label{F44fromF54}
\eea
in eleven dimensions. Finally, we consider the basis (\ref{G5G5F4F4}) for the $ ({\pa {G_5}})^2 ({\pa {F_4}})^2 $ part of higher-order terms in twelve dimensions. Under dimensional reduction on a circle, one can get the $ ({\pa {F_4}})^4 $ terms in eleven dimensions, that are
\bea
&&{\sf \frac{2}{9}\lp\right. -2 (e_{26} + e_{33}) {F_4}_{ae}{}^{fg,h} \
	{F_4}^{abcd,e} {F_4}_{bfh}{}^{i,j} {F_4}_{cgij,d} }\nonumber \\ 
&&{\sf+ (-2 e_{30} - 3 e_{323} + 2 e_{34} -  e_{43} \
	- 2 e_{77} - 2 e_{88}) {F_4}_{ab}{}^{fg,h} \
	{F_4}^{abcd,e} {F_4}_{ceh}{}^{i,j} {F_4}_{dfij,g} }\nonumber \\ 
&&{\sf+ (-2 e_{31} -  e_{37} - 2 e_{68}  + 3 e_{78}) {F_4}_{ab}{}^{fg,h} {F_4}^{abcd,e} \
	{F_4}_{ce}{}^{ij}{}_{,h} {F_4}_{dfij,g} }\nonumber \\ 
&&{\sf+ (-2 e_{128} + 3 \
	e_{137} - 2 e_{58}  + 2 e_{89}) {F_4}_{ab}{}^{fg,h} {F_4}^{abcd,e} \
	{F_4}_{cef}{}^{i,j} {F_4}_{dghi,j} }\nonumber \\ 
&&{\sf+ (-2 e_{118} -  \
	e_{120} - 2 e_{121} + 3 e_{152} - 2 \
	e_{199} + 3 e_{90}) {F_4}_{abe}{}^{f,g} {F_4}^{abcd,e} \
	{F_4}_{cf}{}^{hi,j} {F_4}_{dghi,j} }\nonumber \\ 
&&{\sf+ (2 e_{122} + 3 \
	e_{91}) {F_4}_{ab}{}^{fg}{}_{,e} {F_4}^{abcd,e} \
	{F_4}_{cf}{}^{hi,j} {F_4}_{dghi,j} }\nonumber \\ 
&&{\sf + \bigl(e_{104} + 3 e_{138} - 2 (e_{146} + e_{60})\bigr) {F_4}_{ab}{}^{fg,h} {F_4}^{abcd,e} \
	{F_4}_{cef}{}^{i,j} {F_4}_{dghj,i} }\nonumber \\ 
&&{\sf+ (2 e_{105} - 2 e_{125} -  e_{126} - 2 e_{127} + \
	3 e_{153} - 2 e_{203}) {F_4}_{abe}{}^{f,g} \
	{F_4}^{abcd,e} {F_4}_{cf}{}^{hi,j} {F_4}_{dghj,i} }\nonumber \\ 
&&{\sf+ (2 e_{106} + 2 e_{131} + 3 e_{92}) \
	{F_4}_{ab}{}^{fg}{}_{,e} {F_4}^{abcd,e} {F_4}_{cf}{}^{hi,j} \
	{F_4}_{dghj,i} }\nonumber \\ 
&&{\sf+ 2 (e_{107} + e_{135}) \
	{F_4}_{ae}{}^{fg}{}_{,b} {F_4}^{abcd,e} {F_4}_{cf}{}^{hi,j} \
	{F_4}_{dghj,i} }\nonumber \\ 
&&{\sf + \bigl(2 e_{129} - 3 e_{136} - 2 (e_{139} + e_{144} + e_{44} -  e_{85})\bigr) {F_4}_{ab}{}^{fg,h} {F_4}^{abcd,e} \
	{F_4}_{cef}{}^{i,j} {F_4}_{dgij,h} }\nonumber \\ 
&&{\sf + (- e_{108} + 3 e_{140} + 3 e_{206} \
	- 2 e_{24} + 2 e_{318} + 2 e_{67}) \
	{F_4}_{ab}{}^{fg,h} {F_4}^{abcd,e} {F_4}_{ce}{}^{ij}{}_{,f} \
	{F_4}_{dgij,h} }\nonumber \\ 
&&{\sf + (2 e_{109} + 3 e_{141} + 2 e_{86}) \
	{F_4}_{ab}{}^{fg,h} {F_4}^{abcd,e} {F_4}_{cf}{}^{ij}{}_{,e} \
	{F_4}_{dgij,h} }\nonumber \\ 
&&{\sf+ (e_{101} + 2 e_{130} - 3 e_{148} - 2 e_{149} - 2 e_{209} \
	- 2 e_{45}) {F_4}_{ab}{}^{fg,h} {F_4}^{abcd,e} \
	{F_4}_{cef}{}^{i,j} {F_4}_{dhij,g} }\nonumber \\ 
&&{\sf+ (3 e_{150} -  \
	e_{191} + 3 e_{214} - 2 e_{25} + 2 e_{319} + \
	e_{69}) {F_4}_{ab}{}^{fg,h} {F_4}^{abcd,e} \
	{F_4}_{ce}{}^{ij}{}_{,f} {F_4}_{dhij,g} }\nonumber \\ 
&&{\sf+ (2 e_{119} - 3 \
	e_{151} - 2 e_{74} + e_{84}) {F_4}_{abe}{}^{f,g} \
	{F_4}^{abcd,e} {F_4}_{cf}{}^{hi,j} {F_4}_{dhij,g} }\nonumber \\ 
&&{\sf+ \
	\bigl(e_{102} + 3 e_{154} + 2 (e_{192} \
	+ e_{87})\bigr) {F_4}_{ab}{}^{fg,h} {F_4}^{abcd,e} \
	{F_4}_{cf}{}^{ij}{}_{,e} {F_4}_{dhij,g} }\nonumber \\ 
&&{\sf+ \bigl(e_{15} + \
	2 e_{164} + 3 (e_{205} + e_{321})\bigr) {F_4}_{ab}{}^{fg,h} {F_4}^{abcd,e} \
	{F_4}_{cdh}{}^{i,j} {F_4}_{efgi,j} }\nonumber \\ 
&&{\sf + \bigl(e_{165} + \
	e_{20} + 3 (e_{213} + e_{322})\bigr) {F_4}_{ab}{}^{fg,h} {F_4}^{abcd,e} \
	{F_4}_{cdh}{}^{i,j} {F_4}_{efgj,i} }\nonumber \\ 
&&{\sf+ (- e_{112} + 3 \
	e_{169} - 2 e_{198} + 4 e_{328}) {F_4}_{abc}{}^{f,g} {F_4}^{abcd,e} \
	{F_4}_{dg}{}^{hi,j} {F_4}_{efhi,j} }\nonumber \\ 
&&{\sf+ (- e_{123} + 2 \
	e_{170} - 2 e_{202} + 4 e_{329}) {F_4}_{abc}{}^{f,g} {F_4}^{abcd,e} \
	{F_4}_{dg}{}^{hi,j} {F_4}_{efhj,i} }\nonumber \\ 
&&{\sf+ (-2 e_{171} - 2 \
	e_{179} + 3 e_{271} + e_{29} - 3 \
	e_{320} + e_4) {F_4}_{ab}{}^{fg,h} {F_4}^{abcd,e} \
	{F_4}_{cdh}{}^{i,j} {F_4}_{efij,g} }\nonumber \\ 
&&{\sf+ (-3 e_{132} + 3 \
	e_{133} + 2 e_{178} + 2 e_{18} - 2 \
	e_{204} + e_{57}) {F_4}_{ab}{}^{fg,h} {F_4}^{abcd,e} \
	{F_4}_{cdf}{}^{i,j} {F_4}_{eghi,j} }\nonumber \\ 
&&{\sf+ (2 e_{115} + 4 \
	e_{156} + 3 e_{180} + e_{197}) {F_4}_{abc}{}^{f,g} {F_4}^{abcd,e} \
	{F_4}_{df}{}^{hi,j} {F_4}_{eghi,j} }\nonumber \\ 
&&{\sf+ (e_{166} - 3 \
	e_{181} + 2 e_{312} - 4 e_{332} -  \
	e_{63} -  e_{82}) {F_4}_{abc}{}^{f,g} {F_4}^{abcd,e} \
	{F_4}_{d}{}^{hij}{}_{,f} {F_4}_{eghi,j} }\nonumber \\ 
&&{\sf+ (3 e_{134} - 3 \
	e_{147} + e_{189} + 2 e_{21} - 2 \
	e_{212} + e_{59}) {F_4}_{ab}{}^{fg,h} {F_4}^{abcd,e} \
	{F_4}_{cdf}{}^{i,j} {F_4}_{eghj,i} }\nonumber \\ 
&&{\sf+ (2 e_{124} + 4 \
	e_{157} + 2 e_{190} + e_{201}) {F_4}_{abc}{}^{f,g} {F_4}^{abcd,e} \
	{F_4}_{df}{}^{hi,j} {F_4}_{eghj,i} }\nonumber \\ 
&&{\sf+ (e_{103} + 2 \
	e_{193} + 3 e_{200}) {F_4}_{ab}{}^{fg,h} \
	{F_4}^{abcd,e} {F_4}_{ch}{}^{ij}{}_{,f} {F_4}_{egij,d} }\nonumber \\ 
&&{\sf + (e_{177} - 3 e_{210} - 2 e_{257} - \
	e_{310} - 4 e_{327} -  e_{70}) \
	{F_4}_{abc}{}^{f,g} {F_4}^{abcd,e} {F_4}_{dg}{}^{hi,j} \
	{F_4}_{ehij,f} }\nonumber \\ 
&&{\sf + (-2 e_{116} - 4 e_{155} -  e_{176} \
	- 3 e_{215} + e_{335} + 2 e_{71}) \
	{F_4}_{abc}{}^{f,g} {F_4}^{abcd,e} {F_4}_{df}{}^{hi,j} \
	{F_4}_{ehij,g} }\nonumber \\ 
&&{\sf + (- e_{159} + 4 e_{216} - 2 e_{313} \
	-  e_{62}) {F_4}_{abc}{}^{f,g} {F_4}^{abcd,e} \
	{F_4}_{d}{}^{hij}{}_{,f} {F_4}_{ehij,g} }\nonumber \\ 
&&{\sf+ (3 e_{227} - 3 e_{282} - 2 e_{285} + 3 e_{295} \
	+ e_{52} + e_9) {F_4}_{abe}{}^{f,g} \
	{F_4}^{abcd,e} {F_4}_{cd}{}^{hi,j} {F_4}_{fghi,j} }\nonumber \\ 
&&{\sf+ \bigl(e_{12} + 3 (e_{228} + e_{286})\bigr) \
	{F_4}_{ab}{}^{fg}{}_{,e} {F_4}^{abcd,e} {F_4}_{cd}{}^{hi,j} \
	{F_4}_{fghi,j} }\nonumber \\ 
&&{\sf+ (-2 e_{225} - 3 e_{230} + \
	e_{28} + e_{41}) {F_4}_{abe}{}^{f,g} {F_4}^{abcd,e} \
	{F_4}_{c}{}^{hij}{}_{,d} {F_4}_{fghi,j} }\nonumber \\ 
&&{\sf+ (3 e_{231} + 2 \
	e_{277} + 4 e_{300} + e_{51}) {F_4}_{abc}{}^{f,g} {F_4}^{abcd,e} \
	{F_4}_{de}{}^{hi,j} {F_4}_{fghi,j} }\nonumber \\ 
&&{\sf+ (e_{187} + 4 \
	e_{232} - 2 e_{265} -  e_{267} + 4 \
	e_{344} + e_{96}) {F_4}_{abce}{}^{,f} {F_4}^{abcd,e} \
	{F_4}_{d}{}^{ghi,j} {F_4}_{fghi,j} }\nonumber \\ 
&&{\sf+ (4 e_{233} + 2 \
	e_{268} + e_{99}) {F_4}_{abc}{}^{f}{}_{,e} \
	{F_4}^{abcd,e} {F_4}_{d}{}^{ghi,j} {F_4}_{fghi,j} }\nonumber \\ 
&&{\sf + (e_{186} + 4 e_{235} + e_{261} + 5 \
	e_{350}) {F_4}_{abcd}{}^{,f} {F_4}^{abcd,e} \
	{F_4}_{e}{}^{ghi,j} {F_4}_{fghi,j} }\nonumber \\ 
&&{\sf+ (e_{14} + 2 \
	e_{247} - 3 e_{292} - 2 e_{293} + 3 e_{296} \
	+ e_{56}) {F_4}_{abe}{}^{f,g} {F_4}^{abcd,e} \
	{F_4}_{cd}{}^{hi,j} {F_4}_{fghj,i} }\nonumber \\ 
&&{\sf+ (e_{17} + 3 \
	e_{229}  + 2 e_{248} + 3 e_{297}) \
	{F_4}_{ab}{}^{fg}{}_{,e} {F_4}^{abcd,e} {F_4}_{cd}{}^{hi,j} \
	{F_4}_{fghj,i} }\nonumber \\ 
&&{\sf+ (2 e_{250} + 2 e_{291} + 4 \
	e_{301} + e_{55}) {F_4}_{abc}{}^{f,g} {F_4}^{abcd,e} \
	{F_4}_{de}{}^{hi,j} {F_4}_{fghj,i} }\nonumber \\ 
&&{\sf+ (e_{111} + e_{196} + 3 e_{251} - 2 e_{275} -  e_{276} + 4 e_{345}) {F_4}_{abce}{}^{,f} {F_4}^{abcd,e} \
	{F_4}_{d}{}^{ghi,j} {F_4}_{fghj,i} }\nonumber \\ 
&&{\sf+ (e_{114} + 4 \
	e_{234} + 3 e_{252}  + 2 e_{280}) {F_4}_{abc}{}^{f}{}_{,e} {F_4}^{abcd,e} \
	{F_4}_{d}{}^{ghi,j} {F_4}_{fghj,i} }\nonumber \\ 
&&{\sf+ (e_{117} + 3 \
	e_{253} + 2 e_{281}) {F_4}_{abe}{}^{f}{}_{,c} \
	{F_4}^{abcd,e} {F_4}_{d}{}^{ghi,j} {F_4}_{fghj,i} }\nonumber \\ 
&&{\sf + (e_{195} + 3 e_{254} + e_{274} + 5 \
	e_{351}) {F_4}_{abcd}{}^{,f} {F_4}^{abcd,e} \
	{F_4}_{e}{}^{ghi,j} {F_4}_{fghj,i} }\nonumber \\ 
&&{\sf+ (e_{16} + 3 \
	e_{172} - 3 e_{270} - 2 e_{272} + 3 e_{338} \
	+ e_6) {F_4}_{ab}{}^{fg,h} {F_4}^{abcd,e} \
	{F_4}_{cdh}{}^{i,j} {F_4}_{fgij,e} }\nonumber \\ 
	&&{\sf+ (3 e_{273} + e_7) {F_4}_{ab}{}^{fg,h} {F_4}^{abcd,e} \
		{F_4}_{cd}{}^{ij}{}_{,h} {F_4}_{fgij,e} }\nonumber \\ 
	&&{\sf+ (-2 e_{168} - 3 \
		e_{287} - 2 e_{316} + e_{83}) {F_4}_{abe}{}^{f,g} {F_4}^{abcd,e} \
		{F_4}_{cg}{}^{hi,j} {F_4}_{fhij,d} }\nonumber \\ 
	&&{\sf+ (- e_{113} + 2 \
		e_{167} - 4 e_{288} - 3 e_{289} + 2 \
		e_{336} -  e_{80}) {F_4}_{abc}{}^{f,g} {F_4}^{abcd,e} \
		{F_4}_{dg}{}^{hi,j} {F_4}_{fhij,e} }\nonumber \\ 
	&&{\sf+ (2 e_{259} + 4 \
		e_{290} + e_{81}) {F_4}_{abc}{}^{f,g} \
		{F_4}^{abcd,e} {F_4}_{d}{}^{hij}{}_{,g} {F_4}_{fhij,e} }\nonumber \\ 
	&&{\sf + \bigl(e_2 - 2 e_{226} - 3 (e_{283} \
		+ e_{294})\bigr) {F_4}_{abe}{}^{f,g} {F_4}^{abcd,e} \
		{F_4}_{cd}{}^{hi,j} {F_4}_{fhij,g} }\nonumber \\ 
	&&{\sf+ (e_{22} + 2 \
		e_{249} + 3 e_{298}) {F_4}_{ab}{}^{fg,h} {F_4}^{abcd,e} \
		{F_4}_{ce}{}^{ij}{}_{,d} {F_4}_{fhij,g} }\nonumber \\ 
	&&{\sf+ (e_{143} + 2 \
		e_{222} + 2 e_{278} - 4 e_{299} - 3 e_{302} + e_{39}) {F_4}_{abc}{}^{f,g} \
		{F_4}^{abcd,e} {F_4}_{de}{}^{hi,j} {F_4}_{fhij,g} }\nonumber \\ 
	&&{\sf+ (2 e_{223} + 4 e_{303} + e_{94}) {F_4}_{abc}{}^{f,g} {F_4}^{abcd,e} \
		{F_4}_{d}{}^{hij}{}_{,e} {F_4}_{fhij,g} }\nonumber \\ 
	&&{\sf+ (e_{241} + 5 \
		e_{304}) {F_4}_{abcd,e} {F_4}^{abcd,e} {F_4}_{fghi,j} \
		{F_4}^{fghi,j} }\nonumber \\ 
	&&{\sf + (e_{263} + 5 e_{305} + 4 e_{306}) \
		{F_4}_{abcd,e} {F_4}^{abcd,e} {F_4}_{fghj,i} {F_4}^{fghi,j} }\nonumber \\ 
	&&{\sf+ \
		(e_{264} + 4 e_{307}) {F_4}_{abce,d} \
		{F_4}^{abcd,e} {F_4}_{fghj,i} {F_4}^{fghi,j} }\nonumber \\ 
	&&{\sf + (-2 e_{317} - 2 e_{32} - 2 e_{324} \
		- 3 e_{35} -  e_{49} - 2 e_{75}) \
		{F_4}_{ab}{}^{fg,h} {F_4}^{abcd,e} {F_4}_{cef}{}^{i,j} \
		{F_4}_{ghij,d} }\nonumber \\ 
	&&{\sf + (3 e_{173} -  e_{23} - 2 e_{269} + \
		3 e_{325}) {F_4}_{ab}{}^{fg,h} {F_4}^{abcd,e} \
		{F_4}_{ce}{}^{ij}{}_{,f} {F_4}_{ghij,d} }\nonumber \\ 
	&&{\sf+ (-3 e_{145} + 2 e_{314} + e_{315} - 3 e_{326} + \
		2 e_{72} + e_{73}) {F_4}_{abe}{}^{f,g} \
		{F_4}^{abcd,e} {F_4}_{cf}{}^{hi,j} {F_4}_{ghij,d} }\nonumber \\ 
	&&{\sf+ (e_{255} + 4 e_{330}) {F_4}_{abc}{}^{f,g} {F_4}^{abcd,e} \
		{F_4}_{e}{}^{hij}{}_{,f} {F_4}_{ghij,d} }\nonumber \\ 
	&&{\sf+ (e_{162} -  \
		e_{220} + e_{309} - 4 e_{331}) {F_4}_{abce}{}^{,f} {F_4}^{abcd,e} \
		{F_4}_{f}{}^{ghi,j} {F_4}_{ghij,d} }\nonumber \\ 
	&&{\sf+ (2 e_{19} - 2 \
		e_{337} + e_{48} - 3 e_{76}) {F_4}_{ab}{}^{fg,h} {F_4}^{abcd,e} \
		{F_4}_{cdf}{}^{i,j} {F_4}_{ghij,e} }\nonumber \\ 
	&&{\sf+ (4 e_{211} - 2 \
		e_{256} -  e_{258} + e_{311} + 4 \
		e_{339} + e_{61}) {F_4}_{abc}{}^{f,g} {F_4}^{abcd,e} \
		{F_4}_{d}{}^{hij}{}_{,f} {F_4}_{ghij,e} }\nonumber \\ 
	&&{\sf+ (- e_{246} - 3 \
		e_{284} + e_3 - 3 e_{340}) {F_4}_{abe}{}^{f,g} {F_4}^{abcd,e} \
		{F_4}_{cd}{}^{hi,j} {F_4}_{ghij,f} }\nonumber \\ 
	&&{\sf+ (e_{208} + e_{243} + 2 e_{279} - 4 e_{341} - 3 e_{342} + e_{40}) {F_4}_{abc}{}^{f,g} {F_4}^{abcd,e} \
		{F_4}_{de}{}^{hi,j} {F_4}_{ghij,f} }\nonumber \\ 
	&&{\sf+ (- e_{221} - 2 \
		e_{266} - 4 e_{343} + e_{65}) {F_4}_{abce}{}^{,f} {F_4}^{abcd,e} \
		{F_4}_{d}{}^{ghi,j} {F_4}_{ghij,f} }\nonumber \\ 
	&&{\sf+ (e_{183} + 2 \
		e_{224} + e_{244} + 4 e_{346}) {F_4}_{abc}{}^{f,g} {F_4}^{abcd,e} \
		{F_4}_{d}{}^{hij}{}_{,e} {F_4}_{ghij,f} }\nonumber \\ 
	&&{\sf+ (e_{161} + \
		e_{218} + e_{262} + e_{334} - 4 \
		e_{347} - 5 e_{349}) {F_4}_{abcd}{}^{,f} {F_4}^{abcd,e} \
		{F_4}_{e}{}^{ghi,j} {F_4}_{ghij,f} }\nonumber \\ 
	&&{\sf+ (e_{184} + e_{245} + 4 e_{348}) {F_4}_{abc}{}^{f,g} {F_4}^{abcd,e} \
		{F_4}_{e}{}^{hij}{}_{,d} {F_4}_{ghij,f} }\nonumber \\ 
	&&{\sf + (e_{219} + e_{237} + 5 e_{352}) \
		{F_4}_{abcd}{}^{,f} {F_4}^{abcd,e} {F_4}_{ghij,f} \
		{F_4}^{ghij}{}_{,e} \left.\rp}.\label{F44fromG52F42}
	\eea
	
In the next section we shall attempt the truncation of the $ 12 $-dimensional Lagrangians to give type IIB supergravity. 

\section{Reduction to type IIB supergravity}\label{10d}

Here also we exploit the approach introduced in Ref. \cite{Khviengia:1997rh} for reduction of fields on a further circle. One can make a consistent truncation by setting to zero those field strengths that are not present in the field content of type IIB supergravity, \ie
\bea
F_4 = G_4^{(1)} = G_4^{(2)} = G_3^{(12)} = 0.\label{10drules}
\eea
As the eleven-dimensional case, it can be seen from the equations of motion for these fields that just when the conditions (\ref{10drules}) are imposed, the truncation of these fields is a consistent one.

By taking into account the Chern-Simons modifications to the various field strengths, one finds that the fields $ F_3^{(1)} $ and $ F_3^{(2)} $ are given by
\bea
F_3^{(1)} = dA_2^{(1)}, \, F_3^{(2)} = dA_2^{(2)} - \chi dA_2^{(1)}. 
\eea
These are precisely the same structures of the NSNS and RR $ 3 $-form field strengths respectively, in type IIB supergravity. \ie
\bea
F_3^{(1)} ={\tt H_3},\, F_3^{(2)} ={\tt F_3}. \label{10drules2}
\eea
Note that before applying the condition (\ref{10drules}) to the Lagrangian obtained from toroidal reduction, there are in total three $ 3 $-form field strengths in dimensionally-reduced theory, that are $ G_3^{(12)} $, $ F_3^{(1)} $ and $ F_3^{(2)} $. Since $ G_3^{(12)} $ is a singlet under the $ SL(2,\mathbb{R}) $ symmetry, it is clear that it should be excluded from the $ 10 $-dimensional theory; the remaining two $ 3 $-form field strengths form the required doublet under $ SL(2,\mathbb{R}) $. Furthermore, a consistent truncation to ten dimensions requires that the $ 5 $-form field strength $ G_5 $ is the ordinary RR $ 5 $-form in type IIB supergravity. \ie
\bea
G_5={\tt F_5}. \label{10drules3}
\eea
This implies that the RR $ 5 $-form in type IIB theory is given at linear order, which is sufficient for the purposes we follow in this paper. Due to the reasons noted in the previous section, for the linearized Riemann curvature we can set $ R_4={\tt R_4} $ in dimensional reduction directly from twelve to ten dimensions. At the following, we would like to truncate $ 12 $-dimensional Lagrangians to ten dimensions according to the compactification rules (\ref{10drules}), (\ref{10drules2}) and (\ref{10drules3}) for toroidal reduction. 

Let us first consider the ansatz (\ref{G5G5RR}) consists of $ (\pa {G_5})^2 R^2 $ terms in twelve dimensions. Our calculations shows that compactification of the ansatz on a torus results in $ (\pa {F_5})^2 R^2 $ coupling in ten dimensions upon the rule (\ref{10drules3}). The obtained coupling has the same form as $ (\pa {G_5})^2 R^2 $ in which the $ 12 $-dimensional $ 5 $-form field strength $ G_5 $ is replaced by the RR $ 5 $-form $ F_5 $ in ten dimensions.  

Similarly, toroidal reduction of the basis (\ref{F4F4RR}) containing the $ (\pa F_4)^2 R^2 $ terms in the supergravity action in twelve dimensions gives rise to the $ (\pa H_3)^2 R^2 $ and $ (\pa F_3)^2 R^2 $ terms in ten dimension. The $ (\pa F_3)^2 R^2 $ part has the same structure as $ (\pa H_3)^2 R^2 $ in which the Kalb-Ramond field strength $ H_3 $ is replaced by RR $ 3 $-form field strength $ F_3 $. However the $ (\pa H_3)^2 R^2 $ terms are given by 
\bea
&&{\tt -2 b_6 {H_3}_{a}{}^{ef,g} {H_3}^{abc,d} R_{bdg}{}^{h} R_{cefh} -  b_1 \
	{H_3}^{abc,d} {H_3}^{efg,h} R_{abde} R_{cfgh} }\nonumber \\ 
&&{\tt -  b_{10} {H_3}^{abc,d} {H_3}^{efg}{}_{,d} R_{abe}{}^{h} R_{cfgh} - 2 b_7 \
	{H_3}_{a}{}^{ef,g} {H_3}^{abc,d} R_{bde}{}^{h} R_{cfgh} }\nonumber \\ 
&&{\tt + 2 b_{19} {H_3}_{a}{}^{ef}{}_{,d} {H_3}^{abc,d} R_{be}{}^{gh} R_{cfgh} + b_3 \
	{H_3}^{abc,d} {H_3}^{efg,h} R_{aebf} R_{cgdh} }\nonumber \\ 
&&{\tt + 2 b_9 {H_3}_{a}{}^{ef,g} {H_3}^{abc,d} R_{bge}{}^{h} R_{chdf} + b_2 {H_3}^{abc,d} \
	{H_3}^{efg,h} R_{aebf} R_{chdg} }\nonumber \\ 
&&{\tt + 2 b_{13} {H_3}_{a}{}^{ef}{}_{,d} {H_3}^{abc,d} R_{b}{}^{g}{}_{e}{}^{h} R_{chfg} + 3 b_{21} {H_3}_{ab}{}^{e,f} {H_3}^{abc,d} R_{cf}{}^{gh} R_{degh} }\nonumber \\ 
&&{\tt + 3 b_{20} {H_3}_{ab}{}^{e,f} {H_3}^{abc,d} R_{ce}{}^{gh} R_{dfgh} - 2 b_8 \
	{H_3}_{a}{}^{ef,g} {H_3}^{abc,d} R_{bec}{}^{h} R_{dgfh} }\nonumber \\ 
&&{\tt + 3 b_{17} {H_3}_{ab}{}^{e,f} {H_3}^{abc,d} R_{c}{}^{g}{}_{e}{}^{h} R_{dgfh} - 2 b_{12} {H_3}_{a}{}^{ef,g} {H_3}^{abc,d} R_{bcg}{}^{h} R_{dhef} }\nonumber \\ 
&&{\tt + 3 b_{15} {H_3}_{ab}{}^{e,f} {H_3}^{abc,d} R_{c}{}^{g}{}_{f}{}^{h} R_{dheg} - 3 b_{23} {H_3}_{ab}{}^{e}{}_{,d} {H_3}^{abc,d} R_{c}{}^{fgh} R_{efgh} }\nonumber \\ 
&&{\tt - 4 b_{22} {H_3}_{abc}{}^{,e} {H_3}^{abc,d} R_{d}{}^{fgh} R_{efgh} + 2 b_5 \
	{H_3}_{a}{}^{ef,g} {H_3}^{abc,d} R_{bdc}{}^{h} R_{egfh} }\nonumber \\ 
&&{\tt - 2 b_{18} {H_3}_{ad}{}^{e,f} {H_3}^{abc,d} R_{b}{}^{g}{}_{c}{}^{h} R_{egfh} + 2 b_{14} {H_3}_{a}{}^{ef}{}_{,d} {H_3}^{abc,d} R_{b}{}^{g}{}_{c}{}^{h} R_{egfh} }\nonumber \\ 
&&{\tt + 3 b_{16} {H_3}_{ab}{}^{e,f} {H_3}^{abc,d} R_{c}{}^{g}{}_{d}{}^{h} R_{egfh} + 3 b_{24} {H_3}_{abd,c} {H_3}^{abc,d} R_{efgh} R^{efgh} }.\label{H32R22}
\eea

At the following, we are going to examine KK reduction of the ansatz (\ref{F4F4F4F4}), including $ (\pa {F_4})^4 $ terms in $ 12 $-dimensional theory, on a torus to find the couplings in ten dimensions. A consistent truncation leads to the couplings $ (\pa {F_3})^2 (\pa {H_3})^2 $, $ (\pa {H_3})^4 $ and $ (\pa {F_3})^4 $ in ten dimensions. Among other terms, the $ (\pa {F_3})^2 (\pa {H_3})^2 $ takes the following form
\bea
&&{\tt -4 c_1 {F_3}^{abc,d} {F_3}_{d}{}^{ef,g} {H_3}_{aeg}{}^{,h} \
	{H_3}_{bfh,c} - 2 c_1 {F_3}^{abc,d} {F_3}^{efg,h} \
	{H_3}_{ade,f} {H_3}_{bgh,c} }\nonumber \\ 
&&{\tt - 2 c_4 {F_3}^{abc,d} {F_3}_{d}{}^{ef,g} \
	{H_3}_{ae}{}^{h}{}_{,f} {H_3}_{bgh,c} -  c_5 {F_3}^{abc,d} \
	{F_3}^{efg,h} {H_3}_{abe,h} {H_3}_{cdf,g} }\nonumber \\ 
&&{\tt- 2 c_8 \
	{F_3}^{abc,d} {F_3}^{efg,h} {H_3}_{abd,h} {H_3}_{cef,g}  -  c_4 {F_3}^{abc,d} {F_3}_{d}{}^{ef,g} \
	{H_3}_{abg}{}^{,h} {H_3}_{ceh,f} }\nonumber \\ 
&&{\tt+ 2 c_3 {F_3}^{abc,d} \
	{F_3}^{efg,h} {H_3}_{abe,h} {H_3}_{cfg,d} + c_5 \
	{F_3}^{abc,d} {F_3}_{d}{}^{ef,g} {H_3}_{abe}{}^{,h} {H_3}_{cfg,,h} }\nonumber \\ 
&&{\tt + 2 c_3 {F_3}^{abc,d} {F_3}^{efg}{}_{,d} \
	{H_3}_{abe}{}^{,h} {H_3}_{cfg,h} + 2 c_7 {F_3}^{abc,d} \
	{F_3}_{d}{}^{ef}{}_{,a} {H_3}_{be}{}^{g,h} {H_3}_{cfg,h} }\nonumber \\ 
&&{\tt + 4 c_7 {F_3}^{abc,d} {F_3}^{efg,h} {H_3}_{abe,d} \
	{H_3}_{cfh,g} + c_5 {F_3}^{abc,d} {F_3}^{efg}{}_{,d} \
	{H_3}_{abe}{}^{,h} {H_3}_{cfh,g} }\nonumber \\ 
&&{\tt - 2 c_{15} {F_3}^{abc,d} {F_3}_{d}{}^{ef,g} \
	{H_3}_{ab}{}^{h}{}_{,e} {H_3}_{cfh,g} + 2 c_{15} \
	{F_3}^{abc,d} {F_3}^{efg}{}_{,d} {H_3}_{ab}{}^{h}{}_{,e} \
	{H_3}_{cfh,g} }\nonumber \\ 
&&{\tt + 4 c_6 {F_3}^{abc,d} {F_3}^{efg,h} {H_3}_{ade,b} \
	{H_3}_{cfh,g} - 2 c_5 {F_3}^{abc,d} {F_3}_{d}{}^{ef,g} \
	{H_3}_{ae}{}^{h}{}_{,b} {H_3}_{cfh,g} }\nonumber \\ 
&&{\tt - 2 c_5 {F_3}_{a}{}^{ef,g} {F_3}^{abc,d} \
	{H_3}_{bde}{}^{,h} {H_3}_{cfh,g} + 2 c_7 \
	{F_3}_{a}{}^{ef}{}_{,d} {F_3}^{abc,d} {H_3}_{be}{}^{g,h} \
	{H_3}_{cfh,g} }\nonumber \\ 
&&{\tt + 2 c_6 {F_3}^{abc,d} {F_3}_{d}{}^{ef}{}_{,a} \
	{H_3}_{be}{}^{g,h} {H_3}_{cfh,g} + 8 c_3 {F_3}_{a}{}^{ef,g} \
	{F_3}^{abc,d} {H_3}_{be}{}^{h}{}_{,d} {H_3}_{cfh,g} }\nonumber \\ 
&&{\tt -  c_5 {F_3}^{abc,d} {F_3}^{efg,h} {H_3}_{abe,f} \
	{H_3}_{cgh,d} + 4 c_{11} {F_3}_{ad}{}^{e,f} {F_3}^{abc,d} \
	{H_3}_{bf}{}^{g,h} {H_3}_{cgh,e} }\nonumber \\ 
&&{\tt - 8 c_8 {F_3}_{ad}{}^{e,f} {F_3}^{abc,d} \
	{H_3}_{be}{}^{g,h} {H_3}_{cgh,f} + 6 c_{18} \
	{F_3}_{ab}{}^{e,f} {F_3}^{abc,d} {H_3}_{cf}{}^{g,h} {H_3}_{deg,h} }\nonumber \\ 
&&{\tt + 2 c_{16} {F_3}_{a}{}^{ef,g} {F_3}^{abc,d} \
	{H_3}_{bcg}{}^{,h} {H_3}_{deh,f} + 4 c_{10} \
	{F_3}_{a}{}^{ef,g} {F_3}^{abc,d} {H_3}_{bg}{}^{h}{}_{,c} \
	{H_3}_{deh,f} }\nonumber \\ 
&&{\tt + 2 c_2 {F_3}^{abc,d} {F_3}^{efg,h} {H_3}_{abh,e} \
	{H_3}_{dfg,c} - 2 c_{18} {F_3}^{abc,d} {F_3}^{efg,h} \
	{H_3}_{abc,e} {H_3}_{dfg,h} }\nonumber \\ 
&&{\tt+ 2 c_{15} {F_3}_{a}{}^{ef,g} \
	{F_3}^{abc,d} {H_3}_{bce}{}^{,h} {H_3}_{dfg,h}  + 2 c_{16} {F_3}_{a}{}^{ef,g} {F_3}^{abc,d} \
	{H_3}_{bc}{}^{h}{}_{,e} {H_3}_{dfg,h} }\nonumber \\ 
&&{\tt+ 8 c_2 \
	{F_3}_{a}{}^{ef,g} {F_3}^{abc,d} {H_3}_{bg}{}^{h}{}_{,e} \
	{H_3}_{dfh,c}  - 2 c_{15} {F_3}_{a}{}^{ef,g} {F_3}^{abc,d} \
	{H_3}_{bce}{}^{,h} {H_3}_{dfh,g} }\nonumber \\ 
&&{\tt- 18 c_{13} \
	{F_3}_{ab}{}^{e,f} {F_3}^{abc,d} {H_3}_{c}{}^{gh}{}_{,e} \
	{H_3}_{dgh,f}  + 2 c_{19} {F_3}^{abc,d} {F_3}^{efg,h} {H_3}_{abc,d} \
	{H_3}_{efg,h} }\nonumber \\ 
&&{\tt+ c_{21} {F_3}^{abc,d} {F_3}_{d}{}^{ef,g} \
	{H_3}_{abc}{}^{,h} {H_3}_{efg,h}  + 6 c_{22} {F_3}_{ab}{}^{e,f} {F_3}^{abc,d} \
	{H_3}_{cd}{}^{g,h} {H_3}_{efg,h} }\nonumber \\ 
&&{\tt+ 18 c_{19} \
	{F_3}_{ab}{}^{e}{}_{,d} {F_3}^{abc,d} {H_3}_{c}{}^{fg,h} \
	{H_3}_{efg,h}  + 12 c_{23} {F_3}_{abc}{}^{,e} {F_3}^{abc,d} \
	{H_3}_{d}{}^{fg,h} {H_3}_{efg,h} }\nonumber \\ 
&&{\tt- 2 c_{10} {F_3}^{abc,d} \
	{F_3}_{d}{}^{ef,g} {H_3}_{abg}{}^{,h} {H_3}_{efh,c}  - 2 c_{16} {F_3}^{abc,d} {F_3}_{d}{}^{ef,g} \
	{H_3}_{ag}{}^{h}{}_{,b} {H_3}_{efh,c} }\nonumber \\ 
&&{\tt+ 2 c_{22} \
	{F_3}^{abc,d} {F_3}_{d}{}^{ef,g} {H_3}_{abc}{}^{,h} {H_3}_{efh,g}  + c_{21} {F_3}^{abc,d} {F_3}^{efg}{}_{,d} \
	{H_3}_{abc}{}^{,h} {H_3}_{efh,g} }\nonumber \\ 
&&{\tt+ 2 c_{17} \
	{F_3}_{ad}{}^{e,f} {F_3}^{abc,d} {H_3}_{bc}{}^{g,h} {H_3}_{efh,g}  + 2 c_{14} {F_3}^{abc,d} {F_3}_{d}{}^{ef}{}_{,a} \
	{H_3}_{bc}{}^{g,h} {H_3}_{efh,g} }\nonumber \\ 
&&{\tt- 2 c_{17} {F_3}^{abc,d} \
	{F_3}_{d}{}^{ef,g} {H_3}_{ab}{}^{h}{}_{,c} {H_3}_{egh,f}  - 2 c_{17} {F_3}_{a}{}^{ef,g} {F_3}^{abc,d} \
	{H_3}_{bcd}{}^{,h} {H_3}_{egh,f} }\nonumber \\ 
&&{\tt+ 2 c_{17} {F_3}^{abc,d} \
	{F_3}_{d}{}^{ef}{}_{,a} {H_3}_{bc}{}^{g,h} {H_3}_{egh,f}  + 8 c_{14} {F_3}_{a}{}^{ef,g} {F_3}^{abc,d} \
	{H_3}_{bd}{}^{h}{}_{,c} {H_3}_{egh,f} }\nonumber \\ 
&&{\tt- 6 c_{21} \
	{F_3}_{ad}{}^{e,f} {F_3}^{abc,d} {H_3}_{b}{}^{gh}{}_{,c} \
	{H_3}_{egh,f}  + 6 c_{22} {F_3}_{a}{}^{ef}{}_{,d} {F_3}^{abc,d} \
	{H_3}_{b}{}^{gh}{}_{,c} {H_3}_{egh,f} }\nonumber \\ 
&&{\tt- 6 c_{21} \
	{F_3}_{ab}{}^{e,f} {F_3}^{abc,d} {H_3}_{cd}{}^{g,h} {H_3}_{egh,f}  + 32 c_{24} {F_3}_{abc,d} {F_3}^{abc,d} {H_3}_{efg,h} \
	{H_3}^{efg,h} }\nonumber \\ 
&&{\tt-  c_4 {F_3}^{abc,d} {F_3}_{d}{}^{ef,g} \
	{H_3}_{abe}{}^{,h} {H_3}_{fgh,c}  + 2 c_{16} {F_3}^{abc,d} {F_3}_{d}{}^{ef,g} \
	{H_3}_{ab}{}^{h}{}_{,e} {H_3}_{fgh,c} }\nonumber \\ 
&&{\tt- 2 c_4 \
	{F_3}_{a}{}^{ef,g} {F_3}^{abc,d} {H_3}_{bde}{}^{,h} {H_3}_{fgh,c}  + 4 c_{10} {F_3}_{a}{}^{ef,g} {F_3}^{abc,d} \
	{H_3}_{bd}{}^{h}{}_{,e} {H_3}_{fgh,c} }\nonumber \\ 
&&{\tt+ 4 c_{11} \
	{F_3}_{ad}{}^{e,f} {F_3}^{abc,d} {H_3}_{be}{}^{g,h} {H_3}_{fgh,c}  + 6 c_{18} {F_3}_{a}{}^{ef}{}_{,d} {F_3}^{abc,d} \
	{H_3}_{b}{}^{gh}{}_{,e} {H_3}_{fgh,c} }\nonumber \\ 
&&{\tt- 18 c_9 \
	{F_3}_{ab}{}^{e,f} {F_3}^{abc,d} {H_3}_{d}{}^{gh}{}_{,e} \
	{H_3}_{fgh,c}  - 18 c_{20} {F_3}_{abd}{}^{,e} {F_3}^{abc,d} \
	{H_3}_{e}{}^{fg,h} {H_3}_{fgh,c} }\nonumber \\ 
&&{\tt+ 18 c_{12} \
	{F_3}_{ab}{}^{e,f} {F_3}^{abc,d} {H_3}_{c}{}^{gh}{}_{,d} \
	{H_3}_{fgh,e}  + 12 c_{23} {F_3}_{ab}{}^{e}{}_{,d} {F_3}^{abc,d} \
	{H_3}_{fgh,e} {H_3}^{fgh}{}_{,c} }\, ,\label{F32H32}
\eea
in 10-dimensional theory. We also observe that the $ (\pa {F_3})^4 $ terms has the same structure as $ (\pa {H_3})^4 $ in which the B-field strength $ H_3 $ is replaced by RR $ 3 $-form $ F_3 $. This result was expected because as remarked above, these fields transforms as a doublet under $ SL(2,{\mathbb R}) $ symmetry of type IIB theory. Anyway, $ (\pa {H_3})^4 $ terms are given as
\bea
&&{\tt -2 c_5 {H_3}_{ad}{}^{e,f} {H_3}^{abc,d} {H_3}_{be}{}^{g,h} \
	{H_3}_{cfg,h} + 4 c_3 {H_3}_{a}{}^{ef}{}_{,d} {H_3}^{abc,d} \
	{H_3}_{be}{}^{g,h} {H_3}_{cfg,h} }\nonumber \\ 
&&{\tt + 2 c_7 {H_3}_{a}{}^{ef}{}_{,d} {H_3}^{abc,d} \
	{H_3}_{be}{}^{g,h} {H_3}_{cfh,g} - 4 c_8 {H_3}_{ad}{}^{e,f} \
	{H_3}^{abc,d} {H_3}_{be}{}^{g,h} {H_3}_{cgh,f} }\nonumber \\ 
&&{\tt + 6 c_{18} {H_3}_{ab}{}^{e,f} {H_3}^{abc,d} \
	{H_3}_{cf}{}^{g,h} {H_3}_{deg,h} + 2 c_{15} \
	{H_3}_{ab}{}^{e,f} {H_3}^{abc,d} {H_3}_{ce}{}^{g,h} {H_3}_{dfg,h} }\nonumber \\ 
&&{\tt - 2 c_{16} {H_3}_{ab}{}^{e,f} {H_3}^{abc,d} \
	{H_3}_{c}{}^{gh}{}_{,e} {H_3}_{dfg,h} + (4 c_2 + c_6) {H_3}_{a}{}^{ef,g} {H_3}^{abc,d} {H_3}_{bg}{}^{h}{}_{,e} \
	{H_3}_{dfh,c} }\nonumber \\ 
&&{\tt + (2 c_{16} + c_4) {H_3}_{ab}{}^{e,f} \
	{H_3}^{abc,d} {H_3}_{cf}{}^{g,h} {H_3}_{dgh,e} + 2 c_{15} \
	{H_3}_{ab}{}^{e,f} {H_3}^{abc,d} {H_3}_{ce}{}^{g,h} {H_3}_{dgh,f} }\nonumber \\ 
&&{\tt - 9 c_{13} {H_3}_{ab}{}^{e,f} {H_3}^{abc,d} \
	{H_3}_{c}{}^{gh}{}_{,e} {H_3}_{dgh,f} -  c_1 \
	{H_3}_{ad}{}^{e,f} {H_3}^{abc,d} {H_3}_{b}{}^{gh}{}_{,c} \
	{H_3}_{efg,h} }\nonumber \\ 
&&{\tt + 6 c_{22} {H_3}_{ab}{}^{e,f} {H_3}^{abc,d} \
	{H_3}_{cd}{}^{g,h} {H_3}_{efg,h} + 9 c_{19} \
	{H_3}_{ab}{}^{e}{}_{,d} {H_3}^{abc,d} {H_3}_{c}{}^{fg,h} \
	{H_3}_{efg,h} }\nonumber \\ 
&&{\tt + (- c_{18}  + c_{22} + 12 c_{23}) {H_3}_{abc}{}^{,e} \
	{H_3}^{abc,d} {H_3}_{d}{}^{fg,h} {H_3}_{efg,h} - 2 c_{17} \
	{H_3}_{abd}{}^{,e} {H_3}^{abc,d} {H_3}_{c}{}^{fg,h} {H_3}_{efh,g} }\nonumber \\ 
&&{\tt + 2 c_7 {H_3}_{ab}{}^{e}{}_{,d} {H_3}^{abc,d} \
	{H_3}_{c}{}^{fg,h} {H_3}_{efh,g} + 2 (2 c_{14} + c_6) {H_3}_{ad}{}^{e}{}_{,b} {H_3}^{abc,d} {H_3}_{c}{}^{fg,h} \
	{H_3}_{efh,g} }\nonumber \\ 
&&{\tt - 2 (c_1 + 2 c_{10}) {H_3}_{ad}{}^{e,f} \
	{H_3}^{abc,d} {H_3}_{bf}{}^{g,h} {H_3}_{egh,c} -  c_5 \
	{H_3}_{ab}{}^{e,f} {H_3}^{abc,d} {H_3}_{cf}{}^{g,h} {H_3}_{egh,d} }\nonumber \\ 
&&{\tt + c_3 {H_3}_{ab}{}^{e,f} {H_3}^{abc,d} \
	{H_3}_{c}{}^{gh}{}_{,f} {H_3}_{egh,d} + (-6 c_{21} -  c_5) {H_3}_{ab}{}^{e,f} {H_3}^{abc,d} {H_3}_{cd}{}^{g,h} \
	{H_3}_{egh,f} }\nonumber \\ 
&&{\tt + c_3 {H_3}_{ab}{}^{e,f} {H_3}^{abc,d} \
	{H_3}_{c}{}^{gh}{}_{,d} {H_3}_{egh,f} + (c_{19} + 16 \
	c_{24}) {H_3}_{abc,d} {H_3}^{abc,d} {H_3}_{efg,h} \
	{H_3}^{efg,h} }\nonumber \\ 
&&{\tt+ 2 (2 c_{11}  + c_4) {H_3}_{ad}{}^{e,f} {H_3}^{abc,d} \
	{H_3}_{be}{}^{g,h} {H_3}_{fgh,c} + (- c_{10} - 9 c_9) {H_3}_{ab}{}^{e,f} {H_3}^{abc,d} {H_3}_{d}{}^{gh}{}_{,e} \
	{H_3}_{fgh,c} }\nonumber \\ 
&&{\tt - 9 c_{20} {H_3}_{abd}{}^{,e} {H_3}^{abc,d} \
	{H_3}_{e}{}^{fg,h} {H_3}_{fgh,c} + 2 c_{17} \
	{H_3}_{ab}{}^{e,f} {H_3}^{abc,d} {H_3}_{cd}{}^{g,h} {H_3}_{fgh,e} }\nonumber \\ 
&&{\tt -  c_8 {H_3}_{abd}{}^{,e} {H_3}^{abc,d} \
	{H_3}_{c}{}^{fg,h} {H_3}_{fgh,e} + 9 c_{12} \
	{H_3}_{ab}{}^{e,f} {H_3}^{abc,d} {H_3}_{c}{}^{gh}{}_{,d} \
	{H_3}_{fgh,e} }\nonumber \\ 
&&{\tt + c_{21} {H_3}_{abc}{}^{,e} {H_3}^{abc,d} \
	{H_3}_{d}{}^{fg,h} {H_3}_{fgh,e} + (c_{14} + c_2) \
	{H_3}_{ab}{}^{e,f} {H_3}^{abc,d} {H_3}_{d}{}^{gh}{}_{,c} {H_3}_{fgh,e} }.\label{H34}
\eea

Now, we consider a toroidal compactification of the basis (\ref{G5G5G5G5}). We find out that a consistent truncation to ten dimensions only gives rise to the coupling $ (\pa {F_5})^4 $ which has the same form as the ansatz (\ref{G5G5G5G5}) in which the $ 5 $-form field strength $ G_5 $ is replaced by the RR $ 5 $-form $ F_5 $. 

Finally, by a consistent reduction of the 12-dimensional basis (\ref{G5G5F4F4}) to $ D=10 $, we arrive at the couplings $ (\pa {F_5})^2  (\pa {H_3})^2 $ and $ (\pa {F_5})^2 (\pa {F_3})^2 $. The former has the same form as the latter in which the Kalb-Ramond field strength $ H_3 $ is replaced by RR $ 3 $-form field strength $ F_3 $. However, the $ (\pa {F_5})^2  (\pa {H_3})^2 $ part will be
\bea
&&{\tt e_{26} {H_3}^{abc,d} {H_3}_{d}{}^{ef,g} \
	{F_5}_{aeg}{}^{hi,j} {F_5}_{bfhij,c} + e_{33} \
	{H_3}^{abc,d} {H_3}^{efg,h} {F_5}_{ade}{}^{ij}{}_{,f} \
	{F_5}_{bghij,c} }\nonumber \\ 
&&{\tt + e_{34} {H_3}^{abc,d} {H_3}_{d}{}^{ef,g} \
	{F_5}_{aef}{}^{hi,j} {F_5}_{bghij,c} + e_{35} \
	{H_3}^{abc,d} {H_3}_{d}{}^{ef,g} {F_5}_{ae}{}^{hij}{}_{,f} \
	{F_5}_{bghij,c} }\nonumber \\ 
&&{\tt + e_{37} {H_3}^{abc,d} {H_3}^{efg,h} \
	{F_5}_{abeh}{}^{i,j} {F_5}_{cdfgi,j} + e_{42} \
	{H_3}^{abc,d} {H_3}^{efg,h} {F_5}_{abeh}{}^{i,j} {F_5}_{cdfgj,i} \
}\nonumber \\ 
&&{\tt + e_{43} {H_3}^{abc,d} {H_3}^{efg,h} \
	{F_5}_{abeh}{}^{i,j} {F_5}_{cdfij,g} + e_{44} \
	{H_3}^{abc,d} {H_3}^{efg,h} {F_5}_{abe}{}^{ij}{}_{,h} \
	{F_5}_{cdfij,g} }\nonumber \\ 
&&{\tt + e_{45} {H_3}^{abc,d} {H_3}^{efg,h} \
	{F_5}_{abh}{}^{ij}{}_{,e} {F_5}_{cdfij,g} + e_{48} \
	{H_3}^{abc,d} {H_3}^{efg,h} {F_5}_{abef}{}^{i,j} {F_5}_{cdghi,j} \
}\nonumber \\ 
&&{\tt + e_{49} {H_3}^{abc,d} {H_3}^{efg,h} \
	{F_5}_{abe}{}^{ij}{}_{,f} {F_5}_{cdghi,j} + e_{53} \
	{H_3}^{abc,d} {H_3}^{efg,h} {F_5}_{abef}{}^{i,j} {F_5}_{cdghj,i} \
}\nonumber \\ 
&&{\tt + e_{57} {H_3}^{abc,d} {H_3}^{efg,h} \
	{F_5}_{abef}{}^{i,j} {F_5}_{cdgij,h} + e_{58} \
	{H_3}^{abc,d} {H_3}^{efg,h} {F_5}_{abe}{}^{ij}{}_{,f} \
	{F_5}_{cdgij,h} }\nonumber \\ 
&&{\tt + e_{59} {H_3}^{abc,d} {H_3}^{efg,h} \
	{F_5}_{abef}{}^{i,j} {F_5}_{cdhij,g} + e_{60} \
	{H_3}^{abc,d} {H_3}^{efg,h} {F_5}_{abe}{}^{ij}{}_{,f} \
	{F_5}_{cdhij,g} }\nonumber \\ 
&&{\tt + e_{67} {H_3}^{abc,d} {H_3}_{d}{}^{ef,g} \
	{F_5}_{abg}{}^{hi,j} {F_5}_{cefhi,j} + e_{68} \
	{H_3}^{abc,d} {H_3}_{d}{}^{ef,g} {F_5}_{ab}{}^{hij}{}_{,g} \
	{F_5}_{cefhi,j} }\nonumber \\ 
&&{\tt + e_{69} {H_3}^{abc,d} {H_3}_{d}{}^{ef,g} \
	{F_5}_{abg}{}^{hi,j} {F_5}_{cefhj,i} + e_{73} \
	{H_3}^{abc,d} {H_3}^{efg,h} {F_5}_{abdh}{}^{i,j} {F_5}_{cefij,g} \
}\nonumber \\ 
&&{\tt + e_{74} {H_3}^{abc,d} {H_3}^{efg,h} \
	{F_5}_{abd}{}^{ij}{}_{,h} {F_5}_{cefij,g} + e_{75} \
	{H_3}^{abc,d} {H_3}_{d}{}^{ef,g} {F_5}_{abg}{}^{hi,j} \
	{F_5}_{cehij,f} }\nonumber \\ 
&&{\tt + e_{76} {H_3}^{abc,d} {H_3}_{d}{}^{ef,g} \
	{F_5}_{ab}{}^{hij}{}_{,g} {F_5}_{cehij,f} + 2 e_{77} \
	{H_3}_{a}{}^{ef,g} {H_3}^{abc,d} {F_5}_{bdg}{}^{hi,j} \
	{F_5}_{cehij,f} }\nonumber \\ 
&&{\tt + 2 e_{78} {H_3}_{a}{}^{ef,g} {H_3}^{abc,d} \
	{F_5}_{bd}{}^{hij}{}_{,g} {F_5}_{cehij,f} + e_{84} \
	{H_3}^{abc,d} {H_3}^{efg,h} {F_5}_{abde}{}^{i,j} {F_5}_{cfghi,j} \
}\nonumber \\ 
&&{\tt + e_{85} {H_3}^{abc,d} {H_3}_{d}{}^{ef,g} \
	{F_5}_{abe}{}^{hi,j} {F_5}_{cfghi,j} + e_{86} \
	{H_3}^{abc,d} {H_3}^{efg}{}_{,d} {F_5}_{abe}{}^{hi,j} \
	{F_5}_{cfghi,j} }\nonumber \\ 
&&{\tt + e_{88} {H_3}^{abc,d} {H_3}_{d}{}^{ef,g} \
	{F_5}_{ab}{}^{hij}{}_{,e} {F_5}_{cfghi,j} + 2 e_{89} \
	{H_3}_{a}{}^{ef,g} {H_3}^{abc,d} {F_5}_{bde}{}^{hi,j} \
	{F_5}_{cfghi,j} }\nonumber \\ 
&&{\tt + 2 e_{90} {H_3}_{ad}{}^{e,f} {H_3}^{abc,d} \
	{F_5}_{be}{}^{ghi,j} {F_5}_{cfghi,j} + 2 e_{91} \
	{H_3}_{a}{}^{ef}{}_{,d} {H_3}^{abc,d} {F_5}_{be}{}^{ghi,j} \
	{F_5}_{cfghi,j} }\nonumber \\ 
&&{\tt + e_{92} {H_3}^{abc,d} {H_3}_{d}{}^{ef}{}_{,a} \
	{F_5}_{be}{}^{ghi,j} {F_5}_{cfghi,j} + e_{100} \
	{H_3}^{abc,d} {H_3}^{efg,h} {F_5}_{abde}{}^{i,j} {F_5}_{cfghj,i} \
}\nonumber \\ 
&&{\tt + e_{101} {H_3}^{abc,d} {H_3}_{d}{}^{ef,g} \
	{F_5}_{abe}{}^{hi,j} {F_5}_{cfghj,i} + e_{102} \
	{H_3}^{abc,d} {H_3}^{efg}{}_{,d} {F_5}_{abe}{}^{hi,j} \
	{F_5}_{cfghj,i} }\nonumber \\ 
&&{\tt + 2 e_{104} {H_3}_{a}{}^{ef,g} {H_3}^{abc,d} \
	{F_5}_{bde}{}^{hi,j} {F_5}_{cfghj,i} + 2 e_{105} \
	{H_3}_{ad}{}^{e,f} {H_3}^{abc,d} {F_5}_{be}{}^{ghi,j} \
	{F_5}_{cfghj,i} }\nonumber \\ 
&&{\tt + 2 e_{106} {H_3}_{a}{}^{ef}{}_{,d} {H_3}^{abc,d} \
	{F_5}_{be}{}^{ghi,j} {F_5}_{cfghj,i} + e_{107} \
	{H_3}^{abc,d} {H_3}_{d}{}^{ef}{}_{,a} {F_5}_{be}{}^{ghi,j} \
	{F_5}_{cfghj,i} }\nonumber \\ 
&&{\tt + e_{108} {H_3}^{abc,d} {H_3}^{efg,h} \
	{F_5}_{abeh}{}^{i,j} {F_5}_{cfgij,d} + e_{109} \
	{H_3}^{abc,d} {H_3}^{efg,h} {F_5}_{abe}{}^{ij}{}_{,h} \
	{F_5}_{cfgij,d} }\nonumber \\ 
&&{\tt + e_{120} {H_3}^{abc,d} {H_3}^{efg,h} \
	{F_5}_{abde}{}^{i,j} {F_5}_{cfgij,h} + e_{121} \
	{H_3}^{abc,d} {H_3}^{efg,h} {F_5}_{abd}{}^{ij}{}_{,e} \
	{F_5}_{cfgij,h} }\nonumber \\ 
&&{\tt + e_{122} {H_3}^{abc,d} {H_3}^{efg,h} \
	{F_5}_{abe}{}^{ij}{}_{,d} {F_5}_{cfgij,h} + e_{126} \
	{H_3}^{abc,d} {H_3}^{efg,h} {F_5}_{abde}{}^{i,j} {F_5}_{cfhij,g} \
}\nonumber \\ 
&&{\tt + e_{127} {H_3}^{abc,d} {H_3}^{efg,h} \
	{F_5}_{abd}{}^{ij}{}_{,e} {F_5}_{cfhij,g} + e_{128} \
	{H_3}^{abc,d} {H_3}_{d}{}^{ef,g} {F_5}_{abe}{}^{hi,j} \
	{F_5}_{cfhij,g} }\nonumber \\ 
&&{\tt + e_{129} {H_3}^{abc,d} {H_3}^{efg}{}_{,d} \
	{F_5}_{abe}{}^{hi,j} {F_5}_{cfhij,g} + e_{131} \
	{H_3}^{abc,d} {H_3}^{efg,h} {F_5}_{abe}{}^{ij}{}_{,d} \
	{F_5}_{cfhij,g} }\nonumber \\ 
&&{\tt + e_{132} {H_3}^{abc,d} {H_3}_{d}{}^{ef,g} \
	{F_5}_{ab}{}^{hij}{}_{,e} {F_5}_{cfhij,g} + e_{133} \
	{H_3}^{abc,d} {H_3}^{efg}{}_{,d} {F_5}_{ab}{}^{hij}{}_{,e} \
	{F_5}_{cfhij,g} }\nonumber \\ 
&&{\tt + e_{135} {H_3}^{abc,d} {H_3}^{efg,h} \
	{F_5}_{ade}{}^{ij}{}_{,b} {F_5}_{cfhij,g} + e_{136} \
	{H_3}^{abc,d} {H_3}_{d}{}^{ef,g} {F_5}_{ae}{}^{hij}{}_{,b} \
	{F_5}_{cfhij,g} }\nonumber \\ 
&&{\tt + e_{137} {H_3}^{abc,d} {H_3}^{efg}{}_{,d} \
	{F_5}_{ae}{}^{hij}{}_{,b} {F_5}_{cfhij,g} + 2 e_{139} \
	{H_3}_{a}{}^{ef,g} {H_3}^{abc,d} {F_5}_{bde}{}^{hi,j} \
	{F_5}_{cfhij,g} }\nonumber \\ 
&&{\tt + 2 e_{140} {H_3}_{a}{}^{ef,g} {H_3}^{abc,d} \
	{F_5}_{bd}{}^{hij}{}_{,e} {F_5}_{cfhij,g} + 2 e_{141} \
	{H_3}_{a}{}^{ef,g} {H_3}^{abc,d} {F_5}_{be}{}^{hij}{}_{,d} \
	{F_5}_{cfhij,g} }\nonumber \\ 
&&{\tt + e_{144} {H_3}^{abc,d} {H_3}^{efg,h} \
	{F_5}_{abe}{}^{ij}{}_{,f} {F_5}_{cghij,d} + 2 e_{145} \
	{H_3}_{ad}{}^{e,f} {H_3}^{abc,d} {F_5}_{bf}{}^{ghi,j} \
	{F_5}_{cghij,e} }\nonumber \\ 
&&{\tt + e_{146} {H_3}^{abc,d} {H_3}_{d}{}^{ef,g} \
	{F_5}_{abe}{}^{hi,j} {F_5}_{cghij,f} + e_{147} \
	{H_3}^{abc,d} {H_3}_{d}{}^{ef,g} {F_5}_{ab}{}^{hij}{}_{,e} \
	{F_5}_{cghij,f} }\nonumber \\ 
&&{\tt + e_{148} {H_3}^{abc,d} {H_3}_{d}{}^{ef,g} \
	{F_5}_{ae}{}^{hij}{}_{,b} {F_5}_{cghij,f} + 2 e_{149} \
	{H_3}_{a}{}^{ef,g} {H_3}^{abc,d} {F_5}_{bde}{}^{hi,j} \
	{F_5}_{cghij,f} }\nonumber \\ 
&&{\tt + 2 e_{150} {H_3}_{a}{}^{ef,g} {H_3}^{abc,d} \
	{F_5}_{bd}{}^{hij}{}_{,e} {F_5}_{cghij,f} + 2 e_{151} \
	{H_3}_{ad}{}^{e,f} {H_3}^{abc,d} {F_5}_{be}{}^{ghi,j} \
	{F_5}_{cghij,f} }\nonumber \\ 
&&{\tt + 2 e_{152} {H_3}_{a}{}^{ef}{}_{,d} {H_3}^{abc,d} \
	{F_5}_{be}{}^{ghi,j} {F_5}_{cghij,f} + e_{153} \
	{H_3}^{abc,d} {H_3}_{d}{}^{ef}{}_{,a} {F_5}_{be}{}^{ghi,j} \
	{F_5}_{cghij,f} }\nonumber \\ 
&&{\tt + 2 e_{154} {H_3}_{a}{}^{ef,g} {H_3}^{abc,d} \
	{F_5}_{be}{}^{hij}{}_{,d} {F_5}_{cghij,f} + 2 e_{155} \
	{H_3}_{ad}{}^{e,f} {H_3}^{abc,d} {F_5}_{b}{}^{ghij}{}_{,e} \
	{F_5}_{cghij,f} }\nonumber \\ 
&&{\tt + 2 e_{156} {H_3}_{a}{}^{ef}{}_{,d} {H_3}^{abc,d} \
	{F_5}_{b}{}^{ghij}{}_{,e} {F_5}_{cghij,f} + e_{157} \
	{H_3}^{abc,d} {H_3}_{d}{}^{ef}{}_{,a} {F_5}_{b}{}^{ghi,j}{}_{,e} \
	{F_5}_{cghij,f} }\nonumber \\ 
&&{\tt + e_{159} {H_3}^{abc,d} {H_3}^{efg,h} \
	{F_5}_{abch}{}^{i,j} {F_5}_{defgi,j} + e_{163} \
	{H_3}^{abc,d} {H_3}^{efg,h} {F_5}_{abch}{}^{i,j} {F_5}_{defgj,i} \
}\nonumber \\ 
&&{\tt + 2 e_{164} {H_3}_{a}{}^{ef,g} {H_3}^{abc,d} \
	{F_5}_{bcg}{}^{hi,j} {F_5}_{defhi,j} + 2 e_{165} \
	{H_3}_{a}{}^{ef,g} {H_3}^{abc,d} {F_5}_{bcg}{}^{hi,j} \
	{F_5}_{defhj,i} }\nonumber \\ 
&&{\tt + e_{166} {H_3}^{abc,d} {H_3}^{efg,h} \
	{F_5}_{abch}{}^{i,j} {F_5}_{defij,g} + e_{167} \
	{H_3}^{abc,d} {H_3}^{efg,h} {F_5}_{abc}{}^{ij}{}_{,h} \
	{F_5}_{defij,g} }\nonumber \\ 
&&{\tt + e_{168} {H_3}^{abc,d} {H_3}^{efg,h} \
	{F_5}_{abh}{}^{ij}{}_{,c} {F_5}_{defij,g} + 3 e_{169} \
	{H_3}_{ab}{}^{e,f} {H_3}^{abc,d} {F_5}_{cf}{}^{ghi,j} \
	{F_5}_{deghi,j} }\nonumber \\ 
&&{\tt + 3 e_{170} {H_3}_{ab}{}^{e,f} {H_3}^{abc,d} \
	{F_5}_{cf}{}^{ghi,j} {F_5}_{deghj,i} + 2 e_{171} \
	{H_3}_{a}{}^{ef,g} {H_3}^{abc,d} {F_5}_{bcg}{}^{hi,j} \
	{F_5}_{dehij,f} }\nonumber \\ 
&&{\tt + 2 e_{172} {H_3}_{a}{}^{ef,g} {H_3}^{abc,d} \
	{F_5}_{bc}{}^{hij}{}_{,g} {F_5}_{dehij,f} + 2 e_{173} \
	{H_3}_{a}{}^{ef,g} {H_3}^{abc,d} {F_5}_{bg}{}^{hij}{}_{,c} \
	{F_5}_{dehij,f} }\nonumber \\ 
&&{\tt + e_{176} {H_3}^{abc,d} {H_3}^{efg,h} \
	{F_5}_{abce}{}^{i,j} {F_5}_{dfghi,j} + e_{177} \
	{H_3}^{abc,d} {H_3}^{efg,h} {F_5}_{abc}{}^{ij}{}_{,e} \
	{F_5}_{dfghi,j} }\nonumber \\ 
&&{\tt + 2 e_{178} {H_3}_{a}{}^{ef,g} {H_3}^{abc,d} \
	{F_5}_{bce}{}^{hi,j} {F_5}_{dfghi,j} + 2 e_{179} \
	{H_3}_{a}{}^{ef,g} {H_3}^{abc,d} {F_5}_{bc}{}^{hij}{}_{,e} \
	{F_5}_{dfghi,j} }\nonumber \\ 
&&{\tt + 3 e_{180} {H_3}_{ab}{}^{e,f} {H_3}^{abc,d} \
	{F_5}_{ce}{}^{ghi,j} {F_5}_{dfghi,j} + 3 e_{181} \
	{H_3}_{ab}{}^{e,f} {H_3}^{abc,d} {F_5}_{c}{}^{ghij}{}_{,e} \
	{F_5}_{dfghi,j} }\nonumber \\ 
&&{\tt + e_{188} {H_3}^{abc,d} {H_3}^{efg,h} \
	{F_5}_{abce}{}^{i,j} {F_5}_{dfghj,i} + 2 e_{189} \
	{H_3}_{a}{}^{ef,g} {H_3}^{abc,d} {F_5}_{bce}{}^{hi,j} \
	{F_5}_{dfghj,i} }\nonumber \\ 
&&{\tt + 3 e_{190} {H_3}_{ab}{}^{e,f} {H_3}^{abc,d} \
	{F_5}_{ce}{}^{ghi,j} {F_5}_{dfghj,i} + e_{191} \
	{H_3}^{abc,d} {H_3}^{efg,h} {F_5}_{abeh}{}^{i,j} {F_5}_{dfgij,c} \
}\nonumber \\ 
&&{\tt + e_{192} {H_3}^{abc,d} {H_3}^{efg,h} \
	{F_5}_{abe}{}^{ij}{}_{,h} {F_5}_{dfgij,c} + e_{193} \
	{H_3}^{abc,d} {H_3}^{efg,h} {F_5}_{abh}{}^{ij}{}_{,e} \
	{F_5}_{dfgij,c} }\nonumber \\ 
&&{\tt + e_{197} {H_3}^{abc,d} {H_3}^{efg,h} \
	{F_5}_{abce}{}^{i,j} {F_5}_{dfgij,h} + e_{198} \
	{H_3}^{abc,d} {H_3}^{efg,h} {F_5}_{abc}{}^{ij}{}_{,e} \
	{F_5}_{dfgij,h} }\nonumber \\ 
&&{\tt + e_{199} {H_3}^{abc,d} {H_3}^{efg,h} \
	{F_5}_{abe}{}^{ij}{}_{,c} {F_5}_{dfgij,h} + 2 e_{200} \
	{H_3}_{a}{}^{ef,g} {H_3}^{abc,d} {F_5}_{bg}{}^{hij}{}_{,e} \
	{F_5}_{dfhij,c} }\nonumber \\ 
&&{\tt + e_{201} {H_3}^{abc,d} {H_3}^{efg,h} \
	{F_5}_{abce}{}^{i,j} {F_5}_{dfhij,g} + e_{202} \
	{H_3}^{abc,d} {H_3}^{efg,h} {F_5}_{abc}{}^{ij}{}_{,e} \
	{F_5}_{dfhij,g} }\nonumber \\ 
&&{\tt + e_{203} {H_3}^{abc,d} {H_3}^{efg,h} \
	{F_5}_{abe}{}^{ij}{}_{,c} {F_5}_{dfhij,g} + 2 e_{204} \
	{H_3}_{a}{}^{ef,g} {H_3}^{abc,d} {F_5}_{bce}{}^{hi,j} \
	{F_5}_{dfhij,g} }\nonumber \\ 
&&{\tt + 2 e_{205} {H_3}_{a}{}^{ef,g} {H_3}^{abc,d} \
	{F_5}_{bc}{}^{hij}{}_{,e} {F_5}_{dfhij,g} + 2 e_{206} \
	{H_3}_{a}{}^{ef,g} {H_3}^{abc,d} {F_5}_{be}{}^{hij}{}_{,c} \
	{F_5}_{dfhij,g} }\nonumber \\ 
&&{\tt + e_{209} {H_3}^{abc,d} {H_3}^{efg,h} \
	{F_5}_{abe}{}^{ij}{}_{,f} {F_5}_{dghij,c} + 3 e_{210} \
	{H_3}_{ab}{}^{e,f} {H_3}^{abc,d} {F_5}_{cf}{}^{ghi,j} \
	{F_5}_{dghij,e} }\nonumber \\ 
&&{\tt + 3 e_{211} {H_3}_{ab}{}^{e,f} {H_3}^{abc,d} \
	{F_5}_{c}{}^{ghij}{}_{,f} {F_5}_{dghij,e} + 2 e_{212} \
	{H_3}_{a}{}^{ef,g} {H_3}^{abc,d} {F_5}_{bce}{}^{hi,j} \
	{F_5}_{dghij,f} }\nonumber \\ 
&&{\tt + 2 e_{213} {H_3}_{a}{}^{ef,g} {H_3}^{abc,d} \
	{F_5}_{bc}{}^{hij}{}_{,e} {F_5}_{dghij,f} + 2 e_{214} \
	{H_3}_{a}{}^{ef,g} {H_3}^{abc,d} {F_5}_{be}{}^{hij}{}_{,c} \
	{F_5}_{dghij,f} }\nonumber \\ 
&&{\tt + 3 e_{215} {H_3}_{ab}{}^{e,f} {H_3}^{abc,d} \
	{F_5}_{ce}{}^{ghi,j} {F_5}_{dghij,f} + 3 e_{216} \
	{H_3}_{ab}{}^{e,f} {H_3}^{abc,d} {F_5}_{c}{}^{ghij}{}_{,e} \
	{F_5}_{dghij,f} }\nonumber \\ 
&&{\tt + e_{221} {H_3}^{abc,d} {H_3}^{efg,h} \
	{F_5}_{abcd}{}^{i,j} {F_5}_{efghi,j} + e_{222} \
	{H_3}^{abc,d} {H_3}_{d}{}^{ef,g} {F_5}_{abc}{}^{hi,j} \
	{F_5}_{efghi,j} }\nonumber \\ 
&&{\tt + e_{223} {H_3}^{abc,d} {H_3}^{efg}{}_{,d} \
	{F_5}_{abc}{}^{hi,j} {F_5}_{efghi,j} + e_{225} \
	{H_3}^{abc,d} {H_3}_{d}{}^{ef,g} {F_5}_{ab}{}^{hij}{}_{,c} \
	{F_5}_{efghi,j} }\nonumber \\ 
&&{\tt + 2 e_{226} {H_3}_{a}{}^{ef,g} {H_3}^{abc,d} \
	{F_5}_{bcd}{}^{hi,j} {F_5}_{efghi,j} + 2 e_{227} \
	{H_3}_{ad}{}^{e,f} {H_3}^{abc,d} {F_5}_{bc}{}^{ghi,j} \
	{F_5}_{efghi,j} }\nonumber \\ 
&&{\tt + 2 e_{228} {H_3}_{a}{}^{ef}{}_{,d} {H_3}^{abc,d} \
	{F_5}_{bc}{}^{ghi,j} {F_5}_{efghi,j} + e_{229} \
	{H_3}^{abc,d} {H_3}_{d}{}^{ef}{}_{,a} {F_5}_{bc}{}^{ghi,j} \
	{F_5}_{efghi,j} }\nonumber \\ 
&&{\tt + 2 e_{230} {H_3}_{ad}{}^{e,f} {H_3}^{abc,d} \
	{F_5}_{b}{}^{ghij}{}_{,c} {F_5}_{efghi,j} + 3 e_{231} \
	{H_3}_{ab}{}^{e,f} {H_3}^{abc,d} {F_5}_{cd}{}^{ghi,j} \
	{F_5}_{efghi,j} }\nonumber \\ 
&&{\tt + 3 e_{232} {H_3}_{abd}{}^{,e} {H_3}^{abc,d} \
	{F_5}_{c}{}^{fghi,j} {F_5}_{efghi,j} + 3 e_{233} \
	{H_3}_{ab}{}^{e}{}_{,d} {H_3}^{abc,d} {F_5}_{c}{}^{fghi,j} \
	{F_5}_{efghi,j} }\nonumber \\ 
&&{\tt + 2 e_{234} {H_3}_{ad}{}^{e}{}_{,b} {H_3}^{abc,d} \
	{F_5}_{c}{}^{fghi,j} {F_5}_{efghi,j} + 4 e_{235} \
	{H_3}_{abc}{}^{,e} {H_3}^{abc,d} {F_5}_{d}{}^{fghi,j} \
	{F_5}_{efghi,j} }\nonumber \\ 
&&{\tt + e_{242} {H_3}^{abc,d} {H_3}^{efg,h} \
	{F_5}_{abcd}{}^{i,j} {F_5}_{efghj,i} + e_{243} \
	{H_3}^{abc,d} {H_3}_{d}{}^{ef,g} {F_5}_{abc}{}^{hi,j} \
	{F_5}_{efghj,i} }\nonumber \\ 
&&{\tt + e_{244} {H_3}^{abc,d} {H_3}^{efg}{}_{,d} \
	{F_5}_{abc}{}^{hi,j} {F_5}_{efghj,i} + 2 e_{246} \
	{H_3}_{a}{}^{ef,g} {H_3}^{abc,d} {F_5}_{bcd}{}^{hi,j} \
	{F_5}_{efghj,i} }\nonumber \\ 
&&{\tt + 2 e_{247} {H_3}_{ad}{}^{e,f} {H_3}^{abc,d} \
	{F_5}_{bc}{}^{ghi,j} {F_5}_{efghj,i} + 2 e_{248} \
	{H_3}_{a}{}^{ef}{}_{,d} {H_3}^{abc,d} {F_5}_{bc}{}^{ghi,j} \
	{F_5}_{efghj,i} }\nonumber \\ 
&&{\tt + e_{249} {H_3}^{abc,d} {H_3}_{d}{}^{ef}{}_{,a} \
	{F_5}_{bc}{}^{ghi,j} {F_5}_{efghj,i} + 3 e_{250} \
	{H_3}_{ab}{}^{e,f} {H_3}^{abc,d} {F_5}_{cd}{}^{ghi,j} \
	{F_5}_{efghj,i} }\nonumber \\ 
&&{\tt + 3 e_{251} {H_3}_{abd}{}^{,e} {H_3}^{abc,d} \
	{F_5}_{c}{}^{fghi,j} {F_5}_{efghj,i} + 3 e_{252} \
	{H_3}_{ab}{}^{e}{}_{,d} {H_3}^{abc,d} {F_5}_{c}{}^{fghi,j} \
	{F_5}_{efghj,i} }\nonumber \\ 
&&{\tt + 2 e_{253} {H_3}_{ad}{}^{e}{}_{,b} {H_3}^{abc,d} \
	{F_5}_{c}{}^{fghi,j} {F_5}_{efghj,i} + 4 e_{254} \
	{H_3}_{abc}{}^{,e} {H_3}^{abc,d} {F_5}_{d}{}^{fghi,j} \
	{F_5}_{efghj,i} }\nonumber \\ 
&&{\tt + e_{258} {H_3}^{abc,d} {H_3}^{efg,h} \
	{F_5}_{abch}{}^{i,j} {F_5}_{efgij,d} + e_{259} \
	{H_3}^{abc,d} {H_3}^{efg,h} {F_5}_{abc}{}^{ij}{}_{,h} \
	{F_5}_{efgij,d} }\nonumber \\ 
&&{\tt + e_{267} {H_3}^{abc,d} {H_3}^{efg,h} \
	{F_5}_{abcd}{}^{i,j} {F_5}_{efgij,h} + e_{268} \
	{H_3}^{abc,d} {H_3}^{efg,h} {F_5}_{abc}{}^{ij}{}_{,d} \
	{F_5}_{efgij,h} }\nonumber \\ 
&&{\tt + e_{269} {H_3}^{abc,d} {H_3}_{d}{}^{ef,g} \
	{F_5}_{abg}{}^{hi,j} {F_5}_{efhij,c} + e_{270} \
	{H_3}^{abc,d} {H_3}_{d}{}^{ef,g} {F_5}_{ab}{}^{hij}{}_{,g} \
	{F_5}_{efhij,c} }\nonumber \\ 
&&{\tt + e_{271} {H_3}^{abc,d} {H_3}_{d}{}^{ef,g} \
	{F_5}_{ag}{}^{hij}{}_{,b} {F_5}_{efhij,c} + 2 e_{272} \
	{H_3}_{a}{}^{ef,g} {H_3}^{abc,d} {F_5}_{bcg}{}^{hi,j} \
	{F_5}_{efhij,d} }\nonumber \\ 
&&{\tt + 2 e_{273} {H_3}_{a}{}^{ef,g} {H_3}^{abc,d} \
	{F_5}_{bc}{}^{hij}{}_{,g} {F_5}_{efhij,d} + e_{276} \
	{H_3}^{abc,d} {H_3}^{efg,h} {F_5}_{abcd}{}^{i,j} {F_5}_{efhij,g} \
}\nonumber \\ 
&&{\tt + e_{277} {H_3}^{abc,d} {H_3}_{d}{}^{ef,g} \
	{F_5}_{abc}{}^{hi,j} {F_5}_{efhij,g} + e_{278} \
	{H_3}^{abc,d} {H_3}^{efg}{}_{,d} {F_5}_{abc}{}^{hi,j} \
	{F_5}_{efhij,g} }\nonumber \\ 
&&{\tt + e_{280} {H_3}^{abc,d} {H_3}^{efg,h} \
	{F_5}_{abc}{}^{ij}{}_{,d} {F_5}_{efhij,g} + e_{281} \
	{H_3}^{abc,d} {H_3}^{efg,h} {F_5}_{abd}{}^{ij}{}_{,c} \
	{F_5}_{efhij,g} }\nonumber \\ 
&&{\tt + e_{282} {H_3}^{abc,d} {H_3}_{d}{}^{ef,g} \
	{F_5}_{ab}{}^{hij}{}_{,c} {F_5}_{efhij,g} + e_{283} \
	{H_3}^{abc,d} {H_3}^{efg}{}_{,d} {F_5}_{ab}{}^{hij}{}_{,c} \
	{F_5}_{efhij,g} }\nonumber \\ 
&&{\tt + 2 e_{285} {H_3}_{a}{}^{ef,g} {H_3}^{abc,d} \
	{F_5}_{bcd}{}^{hi,j} {F_5}_{efhij,g} + 2 e_{286} \
	{H_3}_{a}{}^{ef,g} {H_3}^{abc,d} {F_5}_{bc}{}^{hij}{}_{,d} \
	{F_5}_{efhij,g} }\nonumber \\ 
&&{\tt + 2 e_{287} {H_3}_{ad}{}^{e,f} {H_3}^{abc,d} \
	{F_5}_{bf}{}^{ghi,j} {F_5}_{eghij,c} + 2 e_{288} \
	{H_3}_{ad}{}^{e,f} {H_3}^{abc,d} {F_5}_{b}{}^{ghij}{}_{,f} \
	{F_5}_{eghij,c} }\nonumber \\ 
&&{\tt + 3 e_{289} {H_3}_{ab}{}^{e,f} {H_3}^{abc,d} \
	{F_5}_{cf}{}^{ghi,j} {F_5}_{eghij,d} + 3 e_{290} \
	{H_3}_{ab}{}^{e,f} {H_3}^{abc,d} {F_5}_{c}{}^{ghij}{}_{,f} \
	{F_5}_{eghij,d} }\nonumber \\ 
&&{\tt + e_{291} {H_3}^{abc,d} {H_3}_{d}{}^{ef,g} \
	{F_5}_{abc}{}^{hi,j} {F_5}_{eghij,f} + e_{292} \
	{H_3}^{abc,d} {H_3}_{d}{}^{ef,g} {F_5}_{ab}{}^{hij}{}_{,c} \
	{F_5}_{eghij,f} }\nonumber \\ 
&&{\tt + 2 e_{293} {H_3}_{a}{}^{ef,g} {H_3}^{abc,d} \
	{F_5}_{bcd}{}^{hi,j} {F_5}_{eghij,f} + 2 e_{294} \
	{H_3}_{ad}{}^{e,f} {H_3}^{abc,d} {F_5}_{bc}{}^{ghi,j} \
	{F_5}_{eghij,f} }\nonumber \\ 
&&{\tt + 2 e_{295} {H_3}_{a}{}^{ef}{}_{,d} {H_3}^{abc,d} \
	{F_5}_{bc}{}^{ghi,j} {F_5}_{eghij,f} + e_{296} \
	{H_3}^{abc,d} {H_3}_{d}{}^{ef}{}_{,a} {F_5}_{bc}{}^{ghi,j} \
	{F_5}_{eghij,f} }\nonumber \\ 
&&{\tt + 2 e_{297} {H_3}_{a}{}^{ef,g} {H_3}^{abc,d} \
	{F_5}_{bc}{}^{hij}{}_{,d} {F_5}_{eghij,f} + 2 e_{298} \
	{H_3}_{a}{}^{ef,g} {H_3}^{abc,d} {F_5}_{bd}{}^{hij}{}_{,c} \
	{F_5}_{eghij,f} }\nonumber \\ 
&&{\tt + 2 e_{299} {H_3}_{ad}{}^{e,f} {H_3}^{abc,d} \
	{F_5}_{b}{}^{ghij}{}_{,c} {F_5}_{eghij,f} + 2 e_{300} \
	{H_3}_{a}{}^{ef}{}_{,d} {H_3}^{abc,d} {F_5}_{b}{}^{ghij}{}_{,c} \
	{F_5}_{eghij,f} }\nonumber \\ 
&&{\tt + e_{301} {H_3}^{abc,d} {H_3}_{d}{}^{ef}{}_{,a} \
	{F_5}_{b}{}^{ghij}{}_{,c} {F_5}_{eghij,f} + 3 e_{302} \
	{H_3}_{ab}{}^{e,f} {H_3}^{abc,d} {F_5}_{cd}{}^{ghi,j} \
	{F_5}_{eghij,f} }\nonumber \\ 
&&{\tt + 3 e_{303} {H_3}_{ab}{}^{e,f} {H_3}^{abc,d} \
	{F_5}_{c}{}^{ghij}{}_{,d} {F_5}_{eghij,f} + 4 e_{304} \
	{H_3}_{abc,d} {H_3}^{abc,d} {F_5}_{efghi,j} {F_5}^{efghi,j} }\nonumber \\ 
&&{\tt + 3 e_{305} {H_3}_{abd,c} {H_3}^{abc,d} \
	{F_5}_{efghi,j} {F_5}^{efghi,j} + 4 e_{306} \
	{H_3}_{abc,d} {H_3}^{abc,d} {F_5}_{efghj,i} {F_5}^{efghi,j} }\nonumber \\ 
&&{\tt + 3 e_{307} {H_3}_{abd,c} {H_3}^{abc,d} \
	{F_5}_{efghj,i} {F_5}^{efghi,j} + e_{315} {H_3}^{abc,d} \
	{H_3}^{efg,h} {F_5}_{abde}{}^{i,j} {F_5}_{fghij,c} }\nonumber \\ 
&&{\tt + e_{316} {H_3}^{abc,d} {H_3}^{efg,h} \
	{F_5}_{abd}{}^{ij}{}_{,e} {F_5}_{fghij,c} + e_{317} \
	{H_3}^{abc,d} {H_3}_{d}{}^{ef,g} {F_5}_{abe}{}^{hi,j} \
	{F_5}_{fghij,c} }\nonumber \\ 
&&{\tt + e_{318} {H_3}^{abc,d} {H_3}^{efg}{}_{,d} \
	{F_5}_{abe}{}^{hi,j} {F_5}_{fghij,c} + e_{320} \
	{H_3}^{abcd} {H_3}_{d}{}^{ef,g} {F_5}_{ab}{}^{hij}{}_{,e} \
	{F_5}_{fghij,c} }\nonumber \\ 
&&{\tt + e_{321} {H_3}^{abc,d} {H_3}^{efg}{}_{,d} \
	{F_5}_{ab}{}^{hij}{}_{,e} {F_5}_{fghij,c} + e_{323} \
	{H_3}^{abc,d} {H_3}_{d}{}^{ef,g} {F_5}_{ae}{}^{hij}{}_{,b} \
	{F_5}_{fghij,c} }\nonumber \\ 
&&{\tt + 2 e_{324} {H_3}_{a}{}^{ef,g} {H_3}^{abc,d} \
	{F_5}_{bde}{}^{hi,j} {F_5}_{fghij,c} + 2 e_{325} \
	{H_3}_{a}{}^{ef,g} {H_3}^{abc,d} {F_5}_{bd}{}^{hij}{}_{,e} \
	{F_5}_{fghij,c} }\nonumber \\ 
&&{\tt + 2 e_{326} {H_3}_{ad}{}^{e,f} {H_3}^{abc,d} \
	{F_5}_{be}{}^{ghi,j} {F_5}_{fghij,c} + 2 e_{327} \
	{H_3}_{ad}{}^{e,f} {H_3}^{abc,d} {F_5}_{b}{}^{ghij}{}_{,e} \
	{F_5}_{fghij,c} }\nonumber \\ 
&&{\tt + 2 e_{328} {H_3}_{a}{}^{ef}{}_{,d} {H_3}^{abc,d} \
	{F_5}_{b}{}^{ghij}{}_{,e} {F_5}_{fghij,c} + e_{329} \
	{H_3}^{abc,d} {H_3}_{d}{}^{ef}{}_{,a} {F_5}_{b}{}^{ghij}{}_{,e} \
	{F_5}_{fghij,c} }\nonumber \\ 
&&{\tt + 3 e_{330} {H_3}_{ab}{}^{e,f} {H_3}^{abc,d} \
	{F_5}_{d}{}^{ghij}{}_{,e} {F_5}_{fghij,c} + 3 e_{331} \
	{H_3}_{abd}{}^{,e} {H_3}^{abc,d} {F_5}_{e}{}^{fghi,j} \
	{F_5}_{fghij,c} }\nonumber \\ 
&&{\tt + 2 e_{332} {H_3}_{ad}{}^{e,f} {H_3}^{abc,d} \
	{F_5}_{e}{}^{ghij}{}_{,b} {F_5}_{fghij,c} + e_{335} \
	{H_3}^{abc,d} {H_3}^{efg,h} {F_5}_{abce}{}^{i,j} {F_5}_{fghij,d} \
}\nonumber \\ 
&&{\tt + e_{336} {H_3}^{abc,d} {H_3}^{efg,h} \
	{F_5}_{abc}{}^{ij}{}_{,e} {F_5}_{fghij,d} + 2 e_{337} \
	{H_3}_{a}{}^{ef,g} {H_3}^{abc,d} {F_5}_{bce}{}^{hi,j} \
	{F_5}_{fghij,d} }\nonumber \\ 
&&{\tt + 2 e_{338} {H_3}_{a}{}^{ef,g} {H_3}^{abc,d} \
	{F_5}_{bc}{}^{hij}{}_{,e} {F_5}_{fghij,d} + 3 e_{339} \
	{H_3}_{ab}{}^{e,f} {H_3}^{abc,d} {F_5}_{c}{}^{ghij}{}_{,e} \
	{F_5}_{fghij,d} }\nonumber \\ 
&&{\tt + 2 e_{340} {H_3}_{ad}{}^{e,f} {H_3}^{abc,d} \
	{F_5}_{bc}{}^{ghi,j} {F_5}_{fghij,e} + 2 e_{341} \
	{H_3}_{ad}{}^{e,f} {H_3}^{abc,d} {F_5}_{b}{}^{ghij}{}_{,c} \
	{F_5}_{fghij,e} }\nonumber \\ 
&&{\tt + 3 e_{342} {H_3}_{ab}{}^{e,f} {H_3}^{abc,d} \
	{F_5}_{cd}{}^{ghi,j} {F_5}_{fghij,e} + 3 e_{343} \
	{H_3}_{abd}{}^{,e} {H_3}^{abc,d} {F_5}_{c}{}^{fghi,j} \
	{F_5}_{fghij,e} }\nonumber \\ 
&&{\tt + 3 e_{344} {H_3}_{ab}{}^{e}{}_{,d} {H_3}^{abc,d} \
	{F_5}_{c}{}^{fghi,j} {F_5}_{fghij,e} + 2 e_{345} \
	{H_3}_{ad}{}^{e}{}_{,b} {H_3}^{abc,d} {F_5}_{c}{}^{fghi,j} \
	{F_5}_{fghij,e} }\nonumber \\ 
&&{\tt + 3 e_{346} {H_3}_{ab}{}^{e,f} {H_3}^{abc,d} \
	{F_5}_{c}{}^{ghij}{}_{,d} {F_5}_{fghij,e} + 4 e_{347} \
	{H_3}_{abc}{}^{,e} {H_3}^{abc,d} {F_5}_{d}{}^{fghi,j} \
	{F_5}_{fghij,e} }\nonumber \\ 
&&{\tt + 3 e_{348} {H_3}_{ab}{}^{e,f} {H_3}^{abc,d} \
	{F_5}_{d}{}^{ghij}{}_{,c} {F_5}_{fghij,e} + 3 e_{349} \
	{H_3}_{abd}{}^{,e} {H_3}^{abc,d} {F_5}_{fghij,e} \
	{F_5}^{fghij}{}_{,c} }\nonumber \\ 
&&{\tt + 3 e_{350} {H_3}_{ab}{}^{e}{}_{,d} {H_3}^{abc,d} \
	{F_5}_{fghij,e} {F_5}^{fghij}{}_{,c} + 2 e_{351} \
	{H_3}_{ad}{}^{e}{}_{,b} {H_3}^{abc,d} {F_5}_{fghij,e} \
	{F_5}^{fghij}{}_{,c} }\nonumber \\ 
&&{\tt + 4 e_{352} {H_3}_{abc}{}^{,e} {H_3}^{abc,d} \
	{F_5}_{fghij,e} {F_5}^{fghij}{}_{,d} }.\label{F52H32}
\eea

Having obtained the complete dimensionally-reduced Lagrangians at order $ \alpha'^3 $ in $ D = 11 $ and $ D = 10 $ dimensions, we shall now find the unknown coefficients in $ 12 $-dimensional bases by comparing the dimensionally-reduced Lagrangians with the corresponding known ones in type IIB and $ 11 $-dimensional supergravity theories, respectively.

\section{Higher-derivative corrections} 

\subsection{Two 5-form-two Riemann curvature corrections}\label{G52R210}

To find the terms containing two $ 5 $-form field strengths and two Riemann curvatures in effective action of F-theory, we use the fact that $ (\pa {F_5})^2 R^2 $ terms in $ D=10 $ obtained from toroidal reduction of the basis (\ref{G5G5RR}) should be match with the corresponding ones in type IIB supergravity where have already been found in \cite{Policastro:2006vt,Garousi:2013nfw,Bakhtiarizadeh:2013zia,Bakhtiarizadeh:2017bpl}. Demanding this hold fixes the unknown coefficients to the values
\bea
&& \left\{a_{37}\to -128-a_{110}+2 a_{114}-3 a_{34}-a_{35}-3 a_{36}-\frac{a_{111}}{2},a_{41}\to +\frac{64}{3}+2 a_{18}+\frac{a_{19}}{2}\right.\nn\\&&+4 a_{3}+\frac{a_{38}}{4}-a_{39}+a_{4}+\frac{a_{40}}{4}-\frac{a_{34}}{2}-\frac{a_{36}}{2},a_{44}\to 40 a_{1}+2 a_{10}+2 a_{15}+20 a_{16}+4 a_{17}\nn\\&&+8 a_{2}+a_{25}+a_{32}+a_{33}-a_{39},a_{46}\to -2 a_{11}+a_{13}-2 a_{14}-2 a_{18}-a_{26}+\frac{a_{27}}{2}+\frac{a_{29}}{2}\nn\\&&-4 a_{3}-a_{30}-a_{4}+\frac{a_{45}}{4}-\frac{a_{19}}{2}-\frac{a_{42}}{2},a_{47}\to -80 a_{1}-2 a_{10}-2 a_{15}-40 a_{16}-8 a_{17}\nn\\&&-2 a_{18}-16 a_{2}-a_{25}-4 a_{3}-a_{32}-a_{33}+\frac{a_{34}}{2}+\frac{a_{36}}{2}+a_{39}-a_{4}-\frac{a_{19}}{2}-\frac{a_{38}}{4}-\frac{a_{40}}{4},\nn\\&&a_{48}\to 40 a_{1}+2 a_{11}-a_{13}+2 a_{14}+20 a_{16}+4 a_{17}+2 a_{18}+\frac{a_{19}}{2}+8 a_{2}+a_{26}+4 a_{3}\nn\\&&+a_{30}+a_{4}+\frac{a_{42}}{2}-\frac{a_{27}}{2}-\frac{a_{29}}{2}-\frac{a_{45}}{4}, a_{6}\to -8 a_{1}-4 a_{16}-\frac{a_{21}}{2}-\frac{4 a_{17}}{5}-\frac{8 a_{2}}{5}-\frac{a_{3}}{5}\nn\\&&-\frac{a_{5}}{5}-\frac{a_{18}}{10}-\frac{a_{20}}{10},a_{69}\to -\frac{256}{3}-\frac{4 a_{110}}{3}+\frac{8 a_{114}}{3}+\frac{2 a_{55}}{3}+a_{56}+\frac{a_{58}}{3}+\frac{a_{60}}{3}-a_{62}\nn\\&&+\frac{2 a_{63}}{3}+\frac{a_{67}}{3}-\frac{2 a_{111}}{3}-\frac{4 a_{49}}{3}-\frac{2 a_{65}}{3}, a_{73}\to +\frac{256}{3} +4 a_{18}+a_{19}+8 a_{3}+2 a_{4}-a_{62}\nn\\&&-a_{72},a_{8}\to -6 a_{11}-2 a_{12}-6 a_{18}-3 a_{26}-a_{28}-12 a_{3}+\frac{3 a_{34}}{2}+\frac{a_{35}}{2}-3 a_{4}-a_{49}+\frac{a_{55}}{2}\nn\\&&+\frac{3 a_{56}}{4}+\frac{3 a_{57}}{4}+\frac{a_{58}}{4}+\frac{3 a_{62}}{4}+\frac{a_{63}}{2}+\frac{a_{67}}{4}-3 a_{7}+\frac{3 a_{72}}{4}-\frac{3 a_{19}}{2}-\frac{3 a_{22}}{2}-\frac{a_{23}}{2}-\frac{3 a_{42}}{2}\nn\\&&-\frac{a_{43}}{2}-\frac{a_{51}}{2}-\frac{a_{70}}{4}-64,a_{81}\to -512-24 a_{18}-6 a_{19}-48 a_{3}-12 a_{4}-a_{51}+2 a_{52}\nn\\&&+3 a_{53}+3 a_{54}+a_{55}+3 a_{56}+3 a_{57}+3 a_{62}-a_{64}+\frac{a_{67}}{2}+\frac{3 a_{71}}{2}+3 a_{72}-\frac{3 a_{68}}{2}-\frac{a_{70}}{2}\nn\\&&-\frac{a_{78}}{2},a_{82}\to +\frac{256}{3}+6 a_{18}+\frac{3 a_{19}}{2}+12 a_{3}+3 a_{4}-a_{53}-a_{54}+\frac{a_{68}}{2}+2 a_{80}-\frac{a_{56}}{2}-\frac{a_{57}}{2}\nn\\&&-\frac{a_{62}}{2}-\frac{a_{71}}{2}-\frac{a_{72}}{2}-\frac{a_{79}}{2},a_{83}\to -128-\frac{2 a_{110}}{3}+\frac{4 a_{114}}{3}-8 a_{18}-2 a_{19}-16 a_{3}-4 a_{4}\nn\\&&+a_{54}+\frac{a_{55}}{3}+\frac{a_{56}}{2}+\frac{a_{57}}{2}+\frac{a_{58}}{6}+\frac{a_{60}}{6}+\frac{a_{61}}{2}+\frac{a_{62}}{2}+\frac{a_{63}}{3}+\frac{a_{67}}{6}+\frac{a_{71}}{2}+\frac{a_{72}}{2}-\frac{a_{111}}{3}\nn\\&&-\frac{2 a_{49}}{3}-\frac{a_{65}}{3}, a_{86}\to a_{18}+\frac{a_{19}}{4}+2 a_{3}+\frac{a_{4}}{2}+\frac{a_{85}}{2}-\frac{a_{77}}{4},a_{9}\to \frac{128}{3}+80 a_{1}+4 a_{10}\nn\\&&+40 a_{16}+8 a_{17}+4 a_{18}+a_{19}+16 a_{2}+a_{22}+2 a_{25}+8 a_{3}+\frac{a_{31}}{2}+2 a_{33}-a_{34}+\frac{a_{38}}{2}\nn\\&&-2 a_{39}+2 a_{4}+2 a_{7}-\frac{a_{24}}{2}, a_{93}\to 4 a_{101}-4 a_{102}-2 a_{104}-4 a_{105}-2 a_{110}-a_{111}+4 a_{114}\nn\\&&+2 a_{91}-2 a_{92},a_{97}\to 192-a_{102}+\frac{a_{110}}{2}+\frac{a_{111}}{4}-a_{114}+6 a_{18}+\frac{3 a_{19}}{2}+12 a_{3}+3 a_{4}\nn\\&&+a_{49}+\frac{a_{51}}{2}+\frac{a_{70}}{4}-\frac{a_{55}}{2}-\frac{a_{63}}{2}-\frac{a_{90}}{2}-\frac{a_{92}}{2}-\frac{3 a_{56}}{4}-\frac{3 a_{57}}{4}-\frac{a_{58}}{4}-\frac{3 a_{62}}{4}-\frac{a_{67}}{4}\nn\\&&-\frac{3 a_{72}}{4}, a_{98}\to -32+\frac{a_{100}}{4}+\frac{a_{102}}{4}+\frac{a_{103}}{8}+\frac{a_{58}}{8}+\frac{a_{59}}{16}+\frac{a_{63}}{4}+\frac{a_{66}}{8}+\frac{a_{89}}{16}+\frac{a_{92}}{8}+\frac{a_{96}}{8}\nn\\&&\left.-\frac{a_{49}}{2}-\frac{a_{94}}{2}-\frac{a_{101}}{4}-\frac{a_{50}}{4}-\frac{a_{87}}{4}-\frac{a_{91}}{8}\right\}.\label{F52R2cond}
\eea
These conditions should also be able to ensure a consistent truncation to the couplings (\ref{F4F4RR11}) in eleven dimensions. To justify this, we put them into the couplings (\ref{F4F4RR11}) and rewrite the result in terms of independent variables. It can be seen that no unknown coefficient remains unfixed. This indicates that the above conditions properly fix the coefficients. Inserting the conditions (\ref{F52R2cond}) into the basis (\ref{G5G5RR}) yields  
\bea
&& e^{-1}{\cal L}_{(\pa {G_5})^2 R^2}=\nn\\&&- \frac{32}{3} (6 {G_5}_{ce}{}^{ghi,j} {G_5}_{dfghj,i} R_{ab}{}^{ef} R^{abcd} - 4 {G_5}_{ce}{}^{ghi,j} \
{G_5}_{dghij,f} R_{ab}{}^{ef} R^{abcd} \nonumber \\ 
&& + 12 {G_5}_{be}{}^{ghi,j} {G_5}_{dfghj,i} R_{a}{}^{e}{}_{c}{}^{f} R^{abcd} - 2 \
{G_5}_{b}{}^{ghij}{}_{,e} {G_5}_{dghij,f} R_{a}{}^{e}{}_{c}{}^{f} R^{abcd} \nonumber \\ 
&& + 8 {G_5}_{be}{}^{hij}{}_{,c} {G_5}_{dfhij,g} R_{a}{}^{efg} R^{abcd} - 8 \
{G_5}_{cf}{}^{hij}{}_{,b} {G_5}_{dghij,e} R_{a}{}^{efg} R^{abcd} \nonumber \\ 
&& + 48 {G_5}_{bce}{}^{hi,j} {G_5}_{fghij,d} R_{a}{}^{efg} R^{abcd} - 8 \
{G_5}_{bc}{}^{hij}{}_{,e} {G_5}_{fghij,d} R_{a}{}^{efg} R^{abcd} \nonumber \\ 
&& + 12 {G_5}_{be}{}^{hij}{}_{,c} {G_5}_{fghij,d} R_{a}{}^{efg} R^{abcd} - 18 \
{G_5}_{abe}{}^{ij}{}_{,g} {G_5}_{cdfij,h} R^{abcd} \
R^{efgh} \nonumber \\ 
&& + 3 {G_5}_{abef}{}^{i,j} {G_5}_{cdghi,j} R^{abcd} \
R^{efgh}),
\eea
for the couplings between two $ 5 $-form field strengths and two Riemann curvatures in twelve dimensions. Note that in writing the above result, the self-duality condition is also imposed, automatically. Because the dimensionally-reduced coupling $ (\pa {F_5})^2 R^2 $ has been compared with the corresponding self-dual one in type IIB supergravity.
 
\subsection{Two 4-form-two Riemann curvature corrections}\label{F42R210}

Now we are in a position to determine the corrections consist of two $ 4 $-form field strengths and two Riemann tensors to the $ 12 $-dimensional supergravity action. As previously noted in Sec. \ref{10d}, dimensional reduction of the ansatz (\ref{F4F4RR}) on a torus leads to the $ (\pa H_3)^2 R^2 $ and $ (\pa F_3)^2 R^2 $ couplings in $ 10 $-dimensional theory. It now should be possible to match the result (\ref{H32R22}) for the former with known expressions for the two-B-field-two-graviton amplitudes in type IIB superstring theory, where were computed along time ago by Gross and Sloan \cite{Gross:1986mw}. This fixes the unknown coefficients as
\bea
&&\left\{b_{13}\to \frac{b_{1}}{2}-128,b_{16}\to -b_{1}+\frac{2 b_{10}}{3}-b_{15}+256,b_{17}\to b_{1}-\frac{2 b_{10}}{3}-\frac{512}{3},\right. \nn\\&& b_{18}\to -b_{1}-2 b_{14}+3 b_{15}+256,b_{19}\to -\frac{b_{1}}{4}+b_{12}-\frac{b_{14}}{2}+128,b_{2}\to b_{1},\nn\\&& b_{20}\to \frac{b_{1}}{6}-\frac{2 b_{12}}{3}+\frac{b_{14}}{3}+\frac{b_{15}}{2}-128,b_{21}\to \frac{b_{1}}{3}-\frac{b_{10}}{3}+\frac{2 b_{12}}{3}-\frac{b_{14}}{3}+\frac{b_{15}}{2},\nn\\&& b_{22}\to -\frac{b_{1}}{6}+\frac{b_{10}}{8}-\frac{b_{15}}{4}+\frac{128}{3},b_{23}\to -\frac{b_{1}}{3}+\frac{b_{10}}{6}+\frac{256}{3},b_{24}\to -\frac{b_{1}}{8}+\frac{b_{10}}{12}-\frac{b_{15}}{8}+32,\nn\\&& b_{3}\to -4 b_{12}-256,b_{5}\to 2 b_{1}-b_{10}-512,b_{6}\to -4 b_{1}+2 b_{10}+8 b_{12}+1024,\nn\\&& \left. b_{7}\to 4 b_{12}+256,b_{8}\to 4 b_{1}-2 b_{10}-4 b_{12}-1280,b_{9}\to 2 b_{1}-b_{10}-256\right\}.\label{H32R2cond}
\eea
Since $ (\pa F_3)^2 R^2 $ terms have the same structure as $ (\pa H_3)^2 R^2 $ in which the B-field strength is replaced by RR $ 3 $-form field strength, it is obvious that comparing them with their $ 10 $-dimensional counterparts results in the same constraints on the unknown coefficients.    

By inserting the above conditions into the basis (\ref{F4F4RR}) and writing the result in terms of independent variables, we are left with some unknown coefficients that have not yet been fixed. This means there should be at least one extra constraint on the coefficients. The remaining coefficients can be found with this requirement that the $ (\pa {F_4})^2 R^2 $ couplings, arising from circular reduction of $ 12 $-dimensional theory, should be consistent with the corresponding ones in $ 11 $-dimensional supergravity. A precise investigation shows that this dimensionally-reduced coupling results from two contributions. The first one comes from reduction of the basis (\ref{F4F4RR}), which has the same structure as $ (\pa {F_4})^2 R^2 $ ansatz in twelve dimensions but with an overall factor $ 2/3 $, and the other one arises from reduction of $ (\pa {F_5})^2 R^2 $ terms to eleven dimensions which is given by (\ref{F4F4RR11}). By adding these two contributions together and comparing the result with the corresponding expression in type IIB supergravity  \cite{Peeters:2005tb,Bakhtiarizadeh:2017ojz}, one ends up with the following extra constraint on unknown coefficients 
\bea
b_{4}\to -\frac{b_{1}}{4}+\frac{b_{10}}{2}+\frac{b_{11}}{2}+128
\eea
Applying the above two sets of conditions to the basis (\ref{F4F4RR}) properly gives the $ (\pa {F_4})^2 R^2 $ couplings in twelve dimensions, that are
\bea
&& e^{-1} {\cal L}_{(\pa {F_4})^2 R^2}=\nn\\&& \frac{32}{3} (3 {F_4}^{eghi,f} {F_4}_{fghi,e} R_{abcd} R^{abcd} + 8 \
{F_4}^{b}{}_{fgh,i} {F_4}^{efgh,i} R_{abcd} R_{e}{}^{acd} \nonumber \\ 
&&+ \
4 {F_4}_{fghi}{}^{,b} {F_4}^{fghi,e} R_{abcd} R_{e}{}^{acd} \
- 12 {F_4}^{bf}{}_{gh,i} {F_4}^{degh,i} R_{abcd} R_{e}{}^{a}{}_{f}{}^{c} \nonumber \\ 
&&+ 24 {F_4}^{bghi,d} {F_4}^{ef}{}_{gh,i} R_{abcd} R_{e}{}^{a}{}_{f}{}^{c}  - 16 {F_4}^{e}{}_{ghi}{}^{,b} {F_4}^{fghi,d} R_{abcd} R_{e}{}^{a}{}_{f}{}^{c} \nonumber \\ 
&&+ 24 {F_4}^{d}{}_{ghi}{}^{,b} {F_4}^{fghi,e} R_{abcd} R_{e}{}^{a}{}_{f}{}^{c}  + 12 {F_4}^{ae}{}_{gh,i} {F_4}^{bfgh,i} R_{abcd} R_{ef}{}^{cd} \nonumber \\ 
&&- 12 {F_4}^{e}{}_{ghi}{}^{,a} {F_4}^{fghi,b} R_{abcd} \
R_{ef}{}^{cd} - 24 {F_4}^{bghi,f} {F_4}^{ce}{}_{hi}{}^{,a} R_{abcd} R_{efg}{}^{d}  \nonumber \\ 
&&+ 24 {F_4}^{be}{}_{hi}{}^{,a} {F_4}^{cfhi,g} R_{abcd} R_{efg}{}^{d} - 120 {F_4}^{ce}{}_{hi}{}^{,a} {F_4}^{fghi,b} R_{abcd} \
R_{efg}{}^{d} \nonumber \\ 
&&+ 96 {F_4}^{be}{}_{hi}{}^{,a} {F_4}^{fghi,c} R_{abcd} R_{efg}{}^{d} - 48 {F_4}^{bc}{}_{hi}{}^{,a} {F_4}^{fghi,e} R_{abcd} R_{efg}{}^{d} \nonumber \\ 
&&- 24 {F_4}^{acg}{}_{i}{}^{,e} {F_4}^{bdhi,f} R_{abcd} \
R_{efgh} + 12 {F_4}^{cdgh}{}_{,i} {F_4}^{iabe,f} R_{abcd} R_{efgh}).
\eea
 
\subsection{Four 4-form corrections}\label{F4410}

To find the corrections including four $ 4 $-form field strengths to effective action of F-theory, we first match the coupling $ (\pa {F_3})^2 (\pa {H_3})^2 $ presented in (\ref{F32H32}), which has been obtained from toroidal reduction of the basis (\ref{F4F4F4F4}), with the corresponding one in type IIB supergravity founded in \cite{Policastro:2006vt,Garousi:2013nfw,Bakhtiarizadeh:2013zia,Bakhtiarizadeh:2017bpl}. This fixes the unknown coefficients to be
\bea
&&\left\{c_{11}\to 0,c_{16}\to -128+c_{10}-2 c_{14},c_{19}\to \frac{64}{9}+\frac{c_{14}}{9},c_{2}\to 128-c_{10}+3 c_{14},\right. \nn\\&& c_{20}\to -\frac{128}{9}+2 c_{12}-c_{13}-\frac{c_{14}}{9}+\frac{c_{18}}{3},c_{21}\to 128-\frac{2 c_{10}}{3}-3 c_{12}+3 c_{13}+\frac{4 c_{14}}{3}-c_{18},\nn\\&& c_{22}\to -\frac{448}{3}+\frac{2 c_{10}}{3}+3 c_{12}-3 c_{13}-2 c_{14}+c_{18},c_{23}\to -\frac{32}{9}-\frac{c_{13}}{4}-\frac{c_{14}}{18}+\frac{c_{18}}{12},\nn\\&& c_{24}\to -\frac{2}{9}+\frac{c_{12}}{32},c_{3}\to -64+c_{10}-2 c_{14},c_{5}\to 512-8 c_{10}+16 c_{14}-2 c_{15}+c_{4},\nn\\&& c_{6}\to -768+\frac{c_{1}}{2}+6 c_{10}+18 c_{12}-18 c_{13}-16 c_{14}-2 c_{17},c_{7}\to 192-9 c_{12}+9 c_{13}\nn\\&&\left.+2 c_{14} +c_{15}+c_{17}-\frac{c_{4}}{2},c_{8}\to 128+2 c_{14}-3 c_{18},c_{9}\to -\frac{128}{9}+c_{12}-c_{13}\right\}.\label{F44cond}
\eea
Comparison of the dimensionally-reduced coupling $ (\pa {F_3})^4 $ with its $ 10 $-dimensional counterpart also yields the similar constraints. Before putting the above conditions into the basis (\ref{F4F4F4F4}), we have to make sure that they correctly gives the other couplings in eleven and ten dimensions. In doing so, we first substitute them into the $ (\pa {F_4})^4 $ coupling in eleven dimensions, which has the same structure as the $ 12 $-dimensional one, but with an overall factor $ 4/9 $. After writing the result in terms of independent variables, we observe that they exactly produces the known $ (\pa {F_4})^4 $ couplings in eleven dimensions obtained in \cite{Peeters:2005tb,Bakhtiarizadeh:2017ojz}. As a further consistency check, applying the constraints (\ref{F44cond}) to the couplings (\ref{H34}) also yields the correct result for $ (\pa {H_3})^4 $ couplings in ten dimensions calculated in \cite{Gross:1986mw}. By inserting the above conditions into the basis (\ref{F4F4F4F4}), one ends up with the following coupling for four $ 4 $-form field strengths in twelve dimensions:
\bea
&& e^{-1} {\cal L}_{(\pa {F_4})^4}=\nn\\&&\frac{2}{9} (576 {F_4}^{a}{}_{fgh}{}^{,b} {F_4}_{bcde,a} {F_4}^{cdf}{}_{j,i} \
{F_4}^{eghi,j} - 288 {F_4}_{bcde,a} {F_4}^{b}{}_{fgh}{}^{,a} {F_4}^{cdf}{}_{j,i} \
{F_4}^{eghj,i} \nonumber \\ 
&&+ 2304 {F_4}_{bcde,a} {F_4}^{b}{}_{fgh}{}^{,a} \
{F_4}^{df}{}_{ij}{}^{,c} {F_4}^{eghj,i}  - 3456 {F_4}^{ac}{}_{fg}{}^{,b} {F_4}_{bcde,a} {F_4}^{df}{}_{ij,h} {F_4}^{eghj,i} \
\nonumber \\ 
&&+ 864 {F_4}_{bcde,a} {F_4}^{bc}{}_{fg}{}^{,a} {F_4}^{df}{}_{ij,h} {F_4}^{eghj,i} + \
576 {F_4}_{bcde,a} {F_4}^{cdf}{}_{j}{}^{,b} {F_4}^{ehij,g} {F_4}_{fghi}{}^{,a} \nonumber \\ 
&& - 64 {F_4}^{aghi,j} {F_4}_{bcde,a} {F_4}^{cde}{}_{j}{}^{,f} {F_4}_{fghi}{}^{,b} - \
576 {F_4}^{aeij,h} {F_4}_{bcde,a} {F_4}^{c}{}_{fgh}{}^{,b} {F_4}^{fg}{}_{ij}{}^{,d} \nonumber \\ 
&&+ \
32 {F_4}_{bcde,a} {F_4}^{bcd}{}_{f}{}^{,a} {F_4}^{e}{}_{hij,g} {F_4}^{fhij,g} - 64 {F_4}^{acde,j} {F_4}_{bcde,a} {F_4}_{fghi}{}^{,b} {F_4}^{ghi}{}_{j}{}^{,f} \nonumber \\ 
&&+ \
576 {F_4}_{bcde,a} {F_4}^{b}{}_{fgh}{}^{,a} {F_4}^{cde}{}_{j,i} {F_4}^{ghij,f} - 672 \
{F_4}_{bcde,a} {F_4}^{bc}{}_{fg}{}^{,a} {F_4}^{e}{}_{hij}{}^{,d} {F_4}^{ghij,f} \nonumber \\ 
&& - 16 {F_4}_{bcde,a} {F_4}^{bcd}{}_{f}{}^{,a} {F_4}_{ghij}{}^{,e} {F_4}^{ghij,f} -  \
{F_4}_{bcde,a} {F_4}^{bcde,a} {F_4}_{ghij,f} {F_4}^{ghij,f}).
\eea

\subsection{Four 5-form corrections}\label{G5410}

As already mentioned in Sec. \ref{11d}, dimensional reduction of the 12-dimensional basis (\ref{G5G5G5G5}) on a 2-torus only gives rise the coupling $ (\pa {F_5})^4 $ in ten dimensions. It has the same form as $ (\pa {G_5})^4 $ in which the $ 5 $-form field strength $ G_5 $ is replaced by ordinary RR $ 5 $-form field strength $ F_5 $ in ten dimensions. The corresponding self-dual coupling in type IIB supergravity have been obtained previously in \cite{Policastro:2006vt,Bakhtiarizadeh:2015exa}. Matching the results yields the following values for unknown coefficients:
\bea
&&\left\{d_{21}\to \frac{4 d_{13}}{3}+\frac{16 d_{14}}{3}-2 d_{16}+\frac{2 d_{19}}{3}+\frac{4 d_{20}}{3}-\frac{2048}{3},d_{28}\to -4 d_{14}+18 d_{18}-2 d_{20}\right.\nn\\&&+6 d_{23}+2 d_{24}+1536,d_{3}\to -9 d_{18}-3 d_{23}-d_{24}+d_{25}-4 d_{26}-\frac{d_{2}}{2}-768,d_{31}\to -\frac{d_{1}}{9}\nn\\&&+\frac{d_{16}}{3}-4 d_{18}+\frac{d_{29}}{3}-\frac{4 d_{23}}{3}-\frac{512}{3}-\frac{2 d_{13}}{9}-\frac{4 d_{14}}{9}-\frac{4 d_{24}}{9}-\frac{2 d_{27}}{9},d_{71}\to 2 d_{26}+2 d_{49}\nn\\&&+2 d_{50}-9 d_{62}-3 d_{70},d_{75}\to -\frac{d_{1}}{9}+\frac{d_{16}}{3}-4 d_{18}+\frac{d_{29}}{3}+d_{36}+d_{51}-16 d_{66}-4 d_{74}\nn\\&&-\frac{4 d_{23}}{3}-\frac{2 d_{13}}{9}-\frac{4 d_{14}}{9}-\frac{4 d_{24}}{9}-\frac{2 d_{27}}{9},d_{79}\to -\frac{d_{1}}{36}+\frac{d_{101}}{2}+\frac{d_{11}}{12}+\frac{d_{12}}{4}+\frac{d_{37}}{3}-d_{39}\nn\\&&+\frac{d_{41}}{6}+2 d_{62}+d_{64}+\frac{d_{70}}{3}+\frac{d_{72}}{3}-\frac{d_{53}}{2}-\frac{d_{61}}{2}-\frac{d_{40}}{3}-\frac{d_{17}}{4}-\frac{d_{23}}{6}-\frac{d_{42}}{6}-\frac{d_{52}}{6}-\frac{d_{69}}{6}\nn\\&&-\frac{d_{14}}{9}-\frac{d_{24}}{9}+\frac{512}{9}-\frac{d_{22}}{12}-\frac{d_{13}}{18}-\frac{d_{27}}{18},d_{8}\to -\frac{d_{1}}{2}-4 d_{14}+\frac{3 d_{16}}{2}-6 d_{23}-2 d_{26}-d_{27}\nn\\&&+\frac{3 d_{29}}{4}+\frac{d_{32}}{2}+\frac{9 d_{36}}{4}-9 d_{39}+\frac{d_{4}}{2}-3 d_{40}-2 d_{49}-4 d_{50}+18 d_{62}+9 d_{64}+3 d_{70}+3 d_{72}\nn\\&&-d_{77}-\frac{3 d_{13}}{2}-\frac{d_{15}}{2}-\frac{27 d_{18}}{2}-\frac{5 d_{24}}{2}-\frac{3 d_{42}}{2}-\frac{3 d_{43}}{2}-\frac{d_{48}}{2}-\frac{3 d_{52}}{2}-\frac{9 d_{53}}{2}-\frac{9 d_{61}}{2}\nn\\&&-\frac{3 d_{69}}{2}-\frac{9 d_{17}}{4}-\frac{d_{2}}{4}-\frac{3 d_{22}}{4}-\frac{3 d_{33}}{4}-128,d_{81}\to -\frac{2 d_{26}}{9}-\frac{d_{60}}{3}-\frac{d_{80}}{3}-\frac{2 d_{49}}{9}-\frac{2 d_{50}}{9}\nn\\&&-\frac{d_{59}}{9}-\frac{256}{9},d_{83}\to -4 d_{102}-2 d_{39}-2 d_{44}+d_{45}+d_{54}+d_{57}-8 d_{58}+8 d_{65}+16 d_{66}\nn\\&&+40 d_{67}+2 d_{73}+2 d_{74}+10 d_{76}-\frac{512}{3},d_{84}\to \frac{d_{1}}{72}-d_{102}+\frac{d_{13}}{36}+\frac{d_{14}}{18}+\frac{d_{17}}{16}+\frac{3 d_{18}}{8}\nn\\&&+\frac{d_{23}}{6}+\frac{d_{24}}{18}+\frac{d_{27}}{36}+\frac{d_{45}}{4}+\frac{d_{54}}{4}+\frac{d_{57}}{4}-2 d_{58}+2 d_{65}+4 d_{66}+10 d_{67}+\frac{d_{73}}{2}+\frac{d_{74}}{2}\nn\\&&+\frac{5 d_{76}}{2}-\frac{d_{44}}{2}-\frac{d_{39}}{4}-\frac{d_{36}}{8}-\frac{d_{51}}{8}-\frac{d_{35}}{16}-\frac{d_{82}}{16}-\frac{d_{16}}{24}-\frac{d_{29}}{24}-32,d_{87}\to \frac{2 d_{1}}{9}+\frac{5 d_{13}}{9}\nn\\&&+\frac{8 d_{14}}{9}+\frac{d_{15}}{9}+d_{17}+6 d_{18}+\frac{d_{2}}{9}+\frac{d_{22}}{3}+\frac{8 d_{23}}{3}+\frac{10 d_{24}}{9}+\frac{8 d_{26}}{9}+\frac{d_{27}}{3}+\frac{d_{33}}{3}-d_{36}\nn\\&&+4 d_{39}+\frac{4 d_{40}}{3}+\frac{d_{42}}{3}+\frac{d_{43}}{3}+\frac{2 d_{48}}{9}+\frac{8 d_{49}}{9}+\frac{16 d_{50}}{9}+\frac{d_{52}}{3}+d_{53}+\frac{4 d_{59}}{9}+\frac{2 d_{60}}{3}+2 d_{61}\nn\\&&-8 d_{62}-4 d_{64}+\frac{2 d_{69}}{3}+\frac{2 d_{77}}{9}+\frac{2 d_{80}}{3}-\frac{2 d_{16}}{3}-\frac{d_{29}}{3}-\frac{2 d_{37}}{3}-\frac{d_{38}}{3}-\frac{d_{68}}{3}-\frac{4 d_{70}}{3}\nn\\&&-\frac{4 d_{72}}{3}-\frac{d_{85}}{3}+\frac{512}{3}-\frac{d_{10}}{9}-\frac{d_{20}}{9}-\frac{2 d_{32}}{9}-\frac{2 d_{4}}{9},d_{88}\to -\frac{5 d_{1}}{9}+\frac{2 d_{10}}{9}+\frac{5 d_{16}}{3}-2 d_{17}\nn\\&&-16 d_{18}+\frac{2 d_{20}}{9}+d_{29}+\frac{4 d_{32}}{9}+3 d_{36}+\frac{4 d_{37}}{3}+\frac{2 d_{38}}{3}-8 d_{39}+\frac{4 d_{4}}{9}+d_{51}-2 d_{53}-4 d_{61}\nn\\&&+16 d_{62}+8 d_{64}+\frac{2 d_{68}}{3}+\frac{8 d_{70}}{3}+\frac{8 d_{72}}{3}+\frac{2 d_{85}}{3}-\frac{4 d_{13}}{3}-\frac{2 d_{22}}{3}-\frac{20 d_{23}}{3}-\frac{8 d_{24}}{3}-\frac{2 d_{33}}{3}\nn\\&&-\frac{8 d_{40}}{3}-\frac{2 d_{42}}{3}-\frac{2 d_{43}}{3}-\frac{8 d_{50}}{3}-\frac{2 d_{52}}{3}-\frac{4 d_{69}}{3}-\frac{20 d_{14}}{9}-\frac{2 d_{15}}{9}-\frac{2 d_{2}}{9}-\frac{8 d_{26}}{9}-\frac{8 d_{27}}{9}\nn\\&&-\frac{4 d_{48}}{9}-\frac{8 d_{49}}{9}-\frac{4 d_{59}}{9}-\frac{4 d_{77}}{9}-\frac{4096}{9},d_{89}\to -\frac{d_{1}}{18}+d_{104}-4 d_{106}+\frac{d_{16}}{6}-2 d_{18}+\frac{d_{29}}{6}\nn\\&&+\frac{3 d_{36}}{4}+d_{51}-d_{61}+2 d_{62}+\frac{d_{63}}{2}+2 d_{64}-4 d_{65}-8 d_{66}-20 d_{67}-d_{73}-d_{74}-5 d_{76}\nn\\&&+\frac{d_{86}}{2}-\frac{d_{103}}{2}-\frac{d_{53}}{2}-\frac{d_{56}}{2}-\frac{2 d_{23}}{3}-\frac{d_{30}}{4}-\frac{d_{34}}{4}-\frac{d_{13}}{9}-\frac{2 d_{14}}{9}-\frac{2 d_{24}}{9}-\frac{d_{27}}{9}+\frac{256}{9}\nn\\&&,d_{9}\to \frac{3 d_{1}}{4}+2 d_{13}+4 d_{14}+\frac{d_{15}}{2}+\frac{9 d_{17}}{2}+\frac{27 d_{18}}{2}+\frac{d_{2}}{4}+\frac{3 d_{22}}{2}+\frac{15 d_{23}}{2}+\frac{7 d_{24}}{2}-2 d_{26}\nn\\&&+\frac{3 d_{27}}{2}+\frac{3 d_{33}}{4}-6 d_{37}+18 d_{39}+6 d_{40}+3 d_{42}+\frac{3 d_{43}}{2}+\frac{d_{48}}{2}-2 d_{49}+3 d_{52}+9 d_{53}\nn\\&&+9 d_{61}-36 d_{62}-18 d_{64}+3 d_{69}-6 d_{70}-6 d_{72}-3 d_{78}-\frac{9 d_{101}}{2}-\frac{3 d_{16}}{2}-\frac{d_{32}}{2}-\frac{d_{4}}{2}\nn\\&&-\frac{3 d_{41}}{2}-\frac{3 d_{11}}{4}-\frac{9 d_{12}}{4}-\frac{3 d_{29}}{4}-\frac{9 d_{36}}{4}-384,d_{90}\to -\frac{d_{103}}{8}+\frac{d_{104}}{4}-d_{106}+\frac{d_{36}}{16}+\frac{d_{51}}{8}\nn\\&&+\frac{d_{62}}{4}+\frac{d_{63}}{16}+\frac{d_{64}}{4}-d_{65}-2 d_{66}-5 d_{67}-\frac{d_{73}}{4}-\frac{d_{74}}{4}-\frac{5 d_{76}}{4}-\frac{d_{53}}{8}-\frac{d_{56}}{8}-\frac{d_{61}}{8}+\frac{320}{9},\nn\\&& d_{93}\to-\frac{d_{30}}{16}-\frac{d_{34}}{16}, -\frac{d_{103}}{16}+\frac{d_{104}}{8}+\frac{d_{108}}{2}+\frac{d_{36}}{32}+\frac{d_{51}}{16}+\frac{d_{62}}{8}+\frac{d_{63}}{32}+\frac{d_{64}}{8}-d_{66}\nn\\&&-25 d_{91}-5 d_{92}-\frac{d_{65}}{2}-\frac{5 d_{67}}{2}-\frac{d_{73}}{8}-\frac{d_{74}}{8}-\frac{5 d_{76}}{8}+\frac{32}{9}-\frac{d_{53}}{16}-\frac{d_{56}}{16}-\frac{d_{61}}{16}-\frac{d_{30}}{32}\nn\\&&-\frac{d_{34}}{32}, d_{94}\to -3 d_{100}+9 d_{101}+3 d_{11}+9 d_{12}-3 d_{16}+27 d_{18}+\frac{d_{2}}{2}+6 d_{23}+d_{24}+4 d_{26}\nn\\&&-d_{32}+\frac{3 d_{33}}{2}+12 d_{37}-18 d_{39}-d_{4}-6 d_{40}+3 d_{41}-6 d_{42}+d_{48}+4 d_{49}+8 d_{50}-3 d_{52}\nn\\&&-9 d_{53}-9 d_{61}+36 d_{62}+18 d_{64}-3 d_{69}+6 d_{70}+6 d_{72}+2 d_{77}+6 d_{78}-\frac{9 d_{17}}{2}-\frac{3 d_{22}}{2}\nn\\&&-\frac{3 d_{29}}{2}-\frac{9 d_{36}}{2}+2304,d_{95}\to -2 d_{100}+6 d_{101}+d_{11}+6 d_{12}-2 d_{16}-6 d_{17}+24 d_{18}\nn\\&&-2 d_{22}+4 d_{23}-d_{29}-3 d_{30}+8 d_{37}-24 d_{39}-8 d_{40}+2 d_{41}-4 d_{42}-2 d_{43}-2 d_{52}-12 d_{53}\nn\\&&-12 d_{61}+48 d_{62}+24 d_{64}-4 d_{69}+8 d_{70}+8 d_{72}+4 d_{78}+2048,d_{96}\to -\frac{4 d_{1}}{9}+\frac{d_{11}}{3}+d_{16}\nn\\&&-d_{17}-12 d_{18}+d_{29}+d_{30}+2 d_{36}+\frac{4 d_{37}}{3}-4 d_{39}+\frac{2 d_{41}}{3}+2 d_{51}-2 d_{61}+8 d_{62}+4 d_{64}\nn\\&&+\frac{4 d_{70}}{3}+\frac{4 d_{72}}{3}-\frac{d_{22}}{3}-\frac{14 d_{23}}{3}-\frac{4 d_{40}}{3}-\frac{2 d_{42}}{3}-\frac{2 d_{52}}{3}-\frac{2 d_{69}}{3}-\frac{1024}{3}-\frac{8 d_{13}}{9}-\frac{16 d_{14}}{9}\nn\\&&-\frac{16 d_{24}}{9}-\frac{8 d_{27}}{9},d_{97}\to -\frac{2 d_{1}}{9}+\frac{2 d_{16}}{3}-2 d_{17}-4 d_{18}+\frac{2 d_{29}}{3}+d_{35}+2 d_{36}+8 d_{44}-4 d_{45}\nn\\&&+2 d_{51}-4 d_{57}+16 d_{58}-16 d_{65}-32 d_{66}-80 d_{67}-4 d_{73}-4 d_{74}-20 d_{76}+d_{82}-\frac{8 d_{23}}{3}\nn\\&&-\frac{4 d_{13}}{9}-\frac{8 d_{14}}{9}-\frac{8 d_{24}}{9}-\frac{4 d_{27}}{9}+\frac{2048}{9},d_{98}\to -\frac{d_{1}}{72}+\frac{d_{103}}{8}-d_{108}+\frac{d_{16}}{24}+\frac{d_{29}}{24}+\frac{d_{30}}{16}\nn\\&&+\frac{d_{34}}{16}+\frac{d_{35}}{16}+\frac{d_{36}}{16}+\frac{d_{39}}{4}+\frac{d_{44}}{2}+\frac{d_{53}}{8}+\frac{d_{56}}{8}+d_{58}+\frac{d_{61}}{8}-d_{65}-2 d_{66}-5 d_{67}+\frac{d_{82}}{16}\nn\\&&-\frac{d_{104}}{4}-\frac{d_{45}}{4}-\frac{d_{54}}{4}-\frac{d_{57}}{4}-\frac{d_{62}}{4}-\frac{d_{64}}{4}-\frac{d_{73}}{4}-\frac{d_{74}}{4}-\frac{5 d_{76}}{4}-\frac{d_{23}}{6}-\frac{3 d_{18}}{8}+\frac{32}{9}-\frac{d_{17}}{16}\nn\\&&-\frac{d_{63}}{16}-\frac{d_{14}}{18}-\frac{d_{24}}{18}-\frac{d_{13}}{36}-\frac{d_{27}}{36},d_{99}\to \frac{d_{103}}{4}-d_{105}-5 d_{107}-2 d_{108}-25 d_{109}+\frac{d_{30}}{8}\nn\\&&+\frac{d_{34}}{8}+\frac{d_{53}}{4}+\frac{d_{56}}{4}+\frac{d_{61}}{4}+d_{65}+3 d_{66}+5 d_{67}+\frac{d_{73}}{2}+\frac{d_{74}}{2}+\frac{5 d_{76}}{2}-\frac{d_{104}}{2}-\frac{d_{62}}{2}-\frac{d_{64}}{2}\nn\\&&\left.-\frac{d_{51}}{4}-\frac{d_{36}}{8}-\frac{d_{63}}{8}-\frac{512}{9}\right\}.\label{G54cond}
\eea
Before applying these conditions to the ansatz (\ref{G5G5G5G5}), we have to make sure that these conditions provide a consistent truncation to $ (\pa {F_4})^4 $ couplings in $ 11 $-dimensional supergravity. To this end, we first substitute them into the dimensionally-reduced coupling (\ref{F44fromF54}) and then write the result in terms of independent variables. We find out that no unfixed coefficient remains and as a result the above conditions precisely produce the $ (\pa {F_4})^4 $ couplings in $ 11 $-dimensional supergravity. This indicates that we are safe to apply the conditions (\ref{G54cond}) to obtain the $ (\pa {G_5})^4 $ couplings in twelve dimensions. In doing so, we put them into the basis (\ref{G5G5G5G5}). This leads to the following Lagrangian for $ (\pa {G_5})^4 $ in twelve dimensions:
\bea
&& e^{-1} {\cal L}_{(\pa {G_5})^4}=\nn\\&&- \frac{32}{9} (216 {G_5}_{ab}{}^{ghi,j} {G_5}^{abcde,f} {G_5}_{cdgj}{}^{k,l} \
{G_5}_{efhik,l} + 36 {G_5}_{abf}{}^{gh,i} {G_5}^{abcde,f} {G_5}_{cdi}{}^{jk,l} \
{G_5}_{eghjk,l} \nonumber \\ 
&&+ 108 {G_5}_{abf}{}^{gh,i} {G_5}^{abcde,f} {G_5}_{cd}{}^{jkl}{}_{,i} \
{G_5}_{eghjk,l}  + 192 {G_5}_{abc}{}^{gh,i} {G_5}^{abcde,f} {G_5}_{dfg}{}^{jk,l} {G_5}_{ehijl,k} \nonumber \\ 
&&- \
432 {G_5}_{ab}{}^{ghi,j} {G_5}^{abcde,f} {G_5}_{cfg}{}^{kl}{}_{,d} {G_5}_{ehjkl,i} + \
48 {G_5}_{abc}{}^{gh,i} {G_5}^{abcde,f} {G_5}_{dg}{}^{jkl}{}_{,f} {G_5}_{ehjkl,i} \nonumber \\ 
&& - 16 {G_5}_{abc}{}^{gh,i} {G_5}^{abcde,f} {G_5}_{de}{}^{jkl}{}_{,i} \
{G_5}_{ghjkl,f} + 8 {G_5}_{abc}{}^{gh,i} {G_5}^{abcde,f} {G_5}_{de}{}^{jkl}{}_{,f} \
{G_5}_{ghjkl,i} \nonumber \\ 
&&+ 48 {G_5}_{abcd}{}^{g,h} {G_5}^{abcde,f} {G_5}_{eh}{}^{ijk,l} \
{G_5}_{gijkl,f}  + 9 {G_5}_{abcd}{}^{g,h} {G_5}^{abcde,f} {G_5}_{e}{}^{ijkl}{}_{,h} {G_5}_{gijkl,f} \
\nonumber \\ 
&&- 48 {G_5}_{abc}{}^{gh,i} {G_5}^{abcde,f} {G_5}_{de}{}^{jkl}{}_{,f} {G_5}_{gijkl,h} + \
128 {G_5}_{abc}{}^{gh,i} {G_5}^{abcde,f} {G_5}_{df}{}^{jkl}{}_{,e} {G_5}_{gijkl,h} \nonumber \\ 
&& - 8 {G_5}_{abcd}{}^{g,h} {G_5}^{abcde,f} {G_5}_{ef}{}^{ijk,l} {G_5}_{gijkl,h} - 10 \
{G_5}_{abcd}{}^{g,h} {G_5}^{abcde,f} {G_5}_{e}{}^{ijkl}{}_{,f} {G_5}_{gijkl,h} \nonumber \\ 
&&-  \
{G_5}_{abcdf,e} {G_5}^{abcde,f} {G_5}_{ghijl,k} {G_5}^{ghijk,l} - 648 {G_5}_{abf}{}^{gh,i} {G_5}^{abcde,f} {G_5}_{cdg}{}^{jk,l} {G_5}_{hijkl,e} \nonumber \\ 
&&- \
576 {G_5}_{abc}{}^{gh,i} {G_5}^{abcde,f} {G_5}_{dfg}{}^{jk,l} {G_5}_{hijkl,e} + 96 \
{G_5}_{abc}{}^{gh,i} {G_5}^{abcde,f} {G_5}_{df}{}^{jkl}{}_{,g} {G_5}_{hijkl,e} \nonumber \\ 
&& - 64 {G_5}_{abcf}{}^{g,h} {G_5}^{abcde,f} {G_5}_{dg}{}^{ijk,l} {G_5}_{hijkl,e} -  \
{G_5}_{abcd}{}^{g,h} {G_5}^{abcde,f} {G_5}_{f}{}^{ijkl}{}_{,g} {G_5}_{hijkl,e} \nonumber \\ 
&& + 16 \
{G_5}_{abcdf}{}^{,g} {G_5}^{abcde,f} {G_5}_{g}{}^{hijk,l} {G_5}_{hijkl,e}).\label{G54coup}
\eea 

\subsection{Two 5-form-two 4-form corrections}\label{G52F4210}

Finally, we are going to find the couplings including two $ 5 $-form field strengths and two $ 4 $-form field strengths in the low-energy effective action of F-theory at eight-derivative level. As discussed in Sec. \ref{10d}, a consistent truncation of the $ 12 $-dimensional ansatz (\ref{G5G5F4F4}) to ten dimensions leads to the couplings $ (\pa {F_5})^2 (\pa {H_3})^2 $ given by Eq. (\ref{F52H32}) and $ (\pa {F_5})^2 (\pa {F_3})^2 $. The latter has the same form as the former in which the B-field strength $ H_3 $ is replaced by the RR $ 3 $-form $ F_3 $. By comparing the couplings (\ref{F52H32}) or $ (\pa {F_5})^2 (\pa {F_3})^2 $ with the corresponding self-dual ones in type IIB supergravity, which have already been obtained in Refs. \cite{Policastro:2006vt,Garousi:2013nfw,Bakhtiarizadeh:2013zia,Bakhtiarizadeh:2017bpl,Bakhtiarizadeh:2015exa}, we easily find the following relations between unknown coefficients
\bea
&&\left\{e_{249}\to e_{109}+e_{192}+e_{193}-6 e_{228}-3 e_{229}-2 e_{248}-64, e_{253}\to e_{141}+e_{154}+e_{200}\right.\nn\\&&-12 e_{233}-4 e_{234}-3 e_{252},e_{281}\to e_{109}+e_{192}+e_{193}-9 e_{268}-3 e_{280},e_{286}\to -\frac{e_{109}}{3}\nn\\&&-\frac{e_{222}}{2}-\frac{3 e_{223}}{2}-\frac{e_{277}}{2}-\frac{e_{278}}{2}-\frac{e_{282}}{2}-\frac{e_{283}}{2}-\frac{e_{192}}{3}-\frac{e_{193}}{3}-\frac{e_{226}}{3}-\frac{e_{285}}{3}+\frac{32}{3}\nn\\&&-\frac{e_{225}}{6},e_{289}\to -2 e_{169}-2 e_{180}+e_{181}+e_{210}-4 e_{211}+e_{215}-4 e_{216}+8 e_{232}+8 e_{233}\nn\\&&+32 e_{235}+2 e_{251}+2 e_{252}+8 e_{254},e_{298}\to \frac{4 e_{109}}{3}+e_{141}+e_{154}+\frac{4 e_{192}}{3}+\frac{4 e_{193}}{3}+e_{200}\nn\\&&+2 e_{222}+6 e_{223}+\frac{2 e_{225}}{3}+\frac{4 e_{226}}{3}+2 e_{277}+2 e_{278}+2 e_{282}+2 e_{283}+\frac{4 e_{285}}{3}-2 e_{297}\nn\\&&-\frac{256}{3},e_{303}\to -\frac{e_{141}}{6}+\frac{e_{227}}{6}+\frac{e_{295}}{6}+\frac{e_{299}}{3}+\frac{e_{302}}{4}-\frac{e_{228}}{3}-\frac{e_{300}}{3}-\frac{e_{231}}{4}-\frac{e_{154}}{6}\nn\\&&-\frac{e_{200}}{6}-\frac{e_{294}}{6}+\frac{32}{9}-\frac{e_{230}}{12},e_{329}\to -\frac{e_{106}}{3}+\frac{e_{122}}{3}+\frac{e_{131}}{6}+\frac{e_{135}}{12}+\frac{e_{140}}{2}+e_{141}\nn\\&&+\frac{e_{154}}{2}+\frac{e_{200}}{2}+2 e_{228}+e_{229}-e_{295}+2 e_{300}+e_{301}+\frac{e_{320}}{2}+e_{321}+\frac{e_{323}}{4}-2 e_{328}\nn\\&&-\frac{e_{132}}{2}-\frac{e_{133}}{2}-\frac{e_{152}}{2}-\frac{e_{296}}{2}-\frac{e_{325}}{2}+\frac{64}{3}-\frac{e_{136}}{4}-\frac{e_{137}}{4}-\frac{e_{153}}{4}-\frac{e_{107}}{6},e_{332}\to -e_{141}\nn\\&&-e_{154}-e_{155}+2 e_{156}-3 e_{169}-3 e_{180}+\frac{3 e_{181}}{2}-e_{200}+3 e_{210}-9 e_{211}+\frac{3 e_{215}}{2}-6 e_{216}\nn\\&&+18 e_{232}+18 e_{233}+72 e_{235}+\frac{9 e_{251}}{2}+\frac{9 e_{252}}{2}+18 e_{254}+3 e_{288}-9 e_{290}+120 e_{304}\nn\\&&+30 e_{305}+24 e_{306}+6 e_{307}+e_{327}+2 e_{328}+3 e_{330}-\frac{e_{145}}{2}-\frac{e_{151}}{2}-\frac{e_{287}}{2}-\frac{e_{326}}{2}-\frac{32}{3},\nn\\&& e_{339}\to -e_{211}-e_{216}-e_{290}-40 e_{304}-10 e_{305}-8 e_{306}-2 e_{307}-e_{330}+\frac{32}{9},\nn\\&& e_{346}\to -\frac{e_{140}}{12}+\frac{e_{154}}{12}+\frac{e_{181}}{4}+\frac{e_{200}}{6}+\frac{e_{210}}{2}-e_{211}+\frac{e_{215}}{4}-e_{216}+\frac{2 e_{228}}{3}+\frac{e_{230}}{6}\nn\\&&+\frac{e_{231}}{2}+2 e_{232}+2 e_{233}+8 e_{235}+\frac{e_{251}}{2}+\frac{e_{252}}{2}+2 e_{254}+\frac{e_{288}}{3}-e_{290}+\frac{e_{294}}{6}+\frac{2 e_{300}}{3}\nn\\&&+40 e_{304}+10 e_{305}+8 e_{306}+2 e_{307}+\frac{e_{327}}{3}+e_{330}+\frac{e_{341}}{3}+\frac{e_{342}}{4}-\frac{e_{169}}{2}-\frac{e_{180}}{2}-\frac{e_{227}}{3}\nn\\&&-\frac{e_{295}}{3}-\frac{e_{299}}{3}-\frac{e_{302}}{4}-\frac{e_{205}}{6}-\frac{e_{213}}{6}-\frac{e_{340}}{6}+\frac{32}{9}-\frac{e_{145}}{12}-\frac{e_{150}}{12}-\frac{e_{151}}{12}-\frac{e_{206}}{12}-\frac{e_{214}}{12}\nn\\&&-\frac{e_{287}}{12}-\frac{e_{326}}{12},e_{348}\to \frac{e_{140}}{12}+\frac{e_{141}}{6}+\frac{e_{145}}{12}+\frac{e_{150}}{12}+\frac{e_{151}}{12}+\frac{e_{154}}{12}+\frac{e_{169}}{2}+\frac{e_{180}}{2}+\frac{e_{205}}{6}\nn\\&&+\frac{e_{206}}{12}+e_{211}+\frac{e_{213}}{6}+\frac{e_{214}}{12}+e_{216}+\frac{e_{227}}{6}-2 e_{232}-2 e_{233}-8 e_{235}-2 e_{254}+\frac{e_{287}}{12}\nn\\&&+e_{290}+\frac{e_{295}}{6}+\frac{e_{326}}{12}-e_{330}+\frac{e_{340}}{6}-\frac{e_{210}}{2}-\frac{e_{251}}{2}-\frac{e_{252}}{2}-\frac{e_{228}}{3}-\frac{e_{288}}{3}-\frac{e_{300}}{3}-\frac{e_{327}}{3}\nn\\&&-\frac{e_{341}}{3}-\frac{e_{181}}{4}-\frac{e_{215}}{4}-\frac{e_{231}}{4}-\frac{e_{342}}{4}-\frac{32}{9}-\frac{e_{230}}{12},e_{351}\to \frac{e_{141}}{10}+\frac{e_{145}}{20}+\frac{e_{151}}{20}+\frac{e_{154}}{10}\nn\\&&+\frac{3 e_{169}}{10}+\frac{3 e_{180}}{10}+\frac{e_{200}}{10}+\frac{3 e_{211}}{5}+\frac{3 e_{216}}{5}+\frac{e_{287}}{20}+\frac{3 e_{290}}{5}-24 e_{304}-6 e_{305}+\frac{e_{326}}{20}\nn\\&&-3 e_{350}-\frac{9 e_{232}}{5}-\frac{12 e_{233}}{5}-\frac{e_{234}}{5}-\frac{36 e_{235}}{5}-\frac{9 e_{254}}{5}-\frac{e_{288}}{5}-\frac{24 e_{306}}{5}-\frac{6 e_{307}}{5}-\frac{e_{327}}{5}\nn\\&&-\frac{3 e_{330}}{5}-\frac{3 e_{344}}{5}-\frac{e_{345}}{5}-\frac{3 e_{210}}{10}+\frac{32}{15}-\frac{3 e_{181}}{20}-\frac{3 e_{215}}{20}-\frac{9 e_{251}}{20}-\frac{9 e_{252}}{20},e_{352}\to -\frac{e_{232}}{20}\nn\\&&-4 e_{304}-e_{305}-\frac{e_{349}}{4}-\frac{e_{350}}{4}-\frac{e_{235}}{5}-\frac{4 e_{306}}{5}-\frac{e_{307}}{5}-\frac{e_{347}}{5}-\frac{e_{233}}{20}-\frac{e_{331}}{20}-\frac{e_{343}}{20}\nn\\&&-\frac{e_{344}}{20}-\frac{8}{45},e_{57}\to -e_{101}-2 e_{102}+e_{104}+e_{108}-2 e_{109}+6 e_{122}+2 e_{128}+2 e_{129}+3 e_{131}\nn\\&&+\frac{3 e_{135}}{2}-3 e_{136}-3 e_{137}-2 e_{139}+6 e_{141}+e_{144}+e_{26}-2 e_{37}+\frac{e_{43}}{2}+2 e_{48}-\frac{e_{33}}{2}-\frac{e_{49}}{2},\nn\\&& e_{58}\to -2 e_{109}-2 e_{122}-e_{131}+e_{144}+e_{44}-\frac{e_{135}}{2}-\frac{e_{33}}{2},e_{60}\to e_{105}-2 e_{106}+6 e_{109}\nn\\&&-2 e_{121}-2 e_{122}-e_{127}-e_{128}-3 e_{132}-3 e_{133}+\frac{e_{135}}{2}+e_{139}+3 e_{140}+6 e_{141}-e_{146}\nn\\&&+\frac{3 e_{148}}{2}+e_{149}+3 e_{151}-3 e_{152}+6 e_{154}+18 e_{169}+3 e_{170}+18 e_{180}-9 e_{181}+6 e_{192}\nn\\&&+8 e_{193}-6 e_{198}-2 e_{199}+9 e_{200}-3 e_{202}-e_{203}+e_{209}-9 e_{210}+36 e_{211}-9 e_{215}+36 e_{216}\nn\\&&+12 e_{222}+36 e_{223}+4 e_{225}+8 e_{226}+6 e_{227}-12 e_{228}-9 e_{231}-72 e_{232}-72 e_{233}-288 e_{235}\nn\\&&+2 e_{247}-4 e_{248}-3 e_{250}-18 e_{251}-18 e_{252}-72 e_{254}+6 e_{277}+12 e_{278}+6 e_{282}+12 e_{283}\nn\\&&+4 e_{285}+3 e_{287}-12 e_{288}+36 e_{290}-3 e_{291}-3 e_{292}-2 e_{293}-6 e_{297}+3 e_{320}+6 e_{321}\nn\\&&+\frac{3 e_{323}}{2}-3 e_{325}-12 e_{328}+\frac{e_{33}}{2}+e_{45}-\frac{3 e_{137}}{2}-640,e_{73}\to -e_{120}-2 e_{122}-3 e_{166}\nn\\&&-2 e_{168}+3 e_{177}-3 e_{197}-2 e_{199}-9 e_{258}+18 e_{259}-e_{315}+2 e_{316}-3 e_{335}+6 e_{336},\nn\\&& e_{74}\to -2 e_{109}-e_{121}-e_{122}-3 e_{167}-e_{168}-2 e_{192}-2 e_{193}-3 e_{198}-e_{199}-9 e_{259}\nn\\&&-e_{316}-3 e_{336},e_{75}\to -4 e_{109}+2 e_{121}+4 e_{122}+e_{127}+e_{128}+e_{131}+3 e_{140}+3 e_{141}-2 e_{171}\nn\\&&+3 e_{173}-4 e_{192}-4 e_{193}+6 e_{198}+2 e_{199}+3 e_{202}+e_{203}+2 e_{204}+6 e_{205}-12 e_{222}-36 e_{223}\nn\\&&-4 e_{225}-8 e_{226}+2 e_{269}-3 e_{271}+4 e_{272}+12 e_{273}-6 e_{277}-12 e_{278}-6 e_{282}-12 e_{283}\nn\\&&-4 e_{285}+3 e_{291}+3 e_{292}+2 e_{293}+6 e_{297}-e_{317}+\frac{3 e_{323}}{2}-3 e_{325}-2 e_{337}-6 e_{338}+\frac{3 e_{35}}{2}\nn\\&&-\frac{3 e_{136}}{2}+384,e_{76}\to e_{132}+\frac{e_{136}}{2}-e_{140}-e_{141}-2 e_{172}-e_{173}+2 e_{270}+e_{271}+4 e_{273}\nn\\&&-e_{320}+e_{325}-\frac{e_{323}}{2}-\frac{e_{35}}{2}-\frac{128}{3},e_{77}\to e_{139}-3 e_{141}-2 e_{171}+3 e_{173}-2 e_{179}+2 e_{204}\nn\\&&-3 e_{206}+4 e_{272}+12 e_{273}-e_{324}-3 e_{325}-2 e_{337}-6 e_{338}-128,e_{78}\to e_{140}+3 e_{141}\nn\\&&+2 e_{154}-2 e_{172}-e_{173}+2 e_{200}+2 e_{205}+e_{206}+4 e_{273}-e_{325}-2 e_{338}-\frac{128}{3},\nn\\&& e_{84}\to e_{102}-2 e_{109}-9 e_{159}+2 e_{165}-3 e_{176}+3 e_{177}-2 e_{192}-2 e_{193}+6 e_{222}+18 e_{223}\nn\\&&+4 e_{226}+3 e_{243}+9 e_{244}+2 e_{246}-9 e_{258}+18 e_{259}-2 e_{269}-6 e_{270}-4 e_{272}-12 e_{273}\nn\\&&-e_{315}+2 e_{316}-2 e_{318}-6 e_{321}-3 e_{335}+6 e_{336}-2 e_{68}+e_{69}-128,e_{85}\to -e_{105}+2 e_{106}\nn\\&&+4 e_{109}-4 e_{122}-e_{128}-e_{129}-2 e_{131}-e_{135}+\frac{3 e_{136}}{2}+\frac{3 e_{137}}{2}+e_{139}-6 e_{141}-3 e_{151}\nn\\&&+3 e_{152}-3 e_{154}+8 e_{164}-18 e_{169}-3 e_{170}-2 e_{178}-2 e_{179}-18 e_{180}+9 e_{181}+4 e_{192}\nn\\&&+4 e_{193}-3 e_{200}+9 e_{210}-36 e_{211}+9 e_{215}-36 e_{216}-6 e_{227}+12 e_{228}+9 e_{231}+72 e_{232}\nn\\&&+72 e_{233}+288 e_{235}-2 e_{247}+4 e_{248}+3 e_{250}+18 e_{251}+18 e_{252}+72 e_{254}+2 e_{269}+6 e_{270}\nn\\&&+8 e_{272}+24 e_{273}-3 e_{287}+12 e_{288}-36 e_{290}+2 e_{318}+6 e_{321}-e_{324}-3 e_{325}+12 e_{328}\nn\\&&-2 e_{337}-6 e_{338}+2 e_{67}+2 e_{68}-\frac{e_{26}}{2}-128,e_{86}\to -2 e_{109}-2 e_{164}-2 e_{192}-2 e_{193}-e_{269}\nn\\&&-3 e_{270}-2 e_{272}-6 e_{273}-e_{318}-3 e_{321}-e_{67}-e_{68}+64,e_{88}\to e_{105}-2 e_{106}+8 e_{109}\nn\\&&-2 e_{121}-e_{127}+e_{131}-3 e_{132}-3 e_{133}+e_{135}-e_{139}+6 e_{141}+3 e_{151}-3 e_{152}+3 e_{154}\nn\\&&+18 e_{169}+3 e_{170}+18 e_{180}-9 e_{181}+8 e_{192}+8 e_{193}-6 e_{198}-2 e_{199}+3 e_{200}-3 e_{202}-e_{203}\nn\\&&-2 e_{204}-6 e_{205}-9 e_{210}+36 e_{211}-9 e_{215}+36 e_{216}+12 e_{222}+36 e_{223}+4 e_{225}+8 e_{226}\nn\\&&+6 e_{227}-12 e_{228}-9 e_{231}-72 e_{232}-72 e_{233}-288 e_{235}+2 e_{247}-4 e_{248}-3 e_{250}-18 e_{251}\nn\\&&-18 e_{252}-72 e_{254}+\frac{e_{26}}{2}+6 e_{270}+3 e_{271}+6 e_{277}+12 e_{278}+6 e_{282}+12 e_{283}+4 e_{285}\nn\\&&+3 e_{287}-12 e_{288}+36 e_{290}-3 e_{291}-3 e_{292}-2 e_{293}-6 e_{297}+6 e_{321}+e_{324}+3 e_{325}\nn\\&&-12 e_{328}+2 e_{337}+6 e_{338}+e_{34}+2 e_{68}-\frac{3 e_{136}}{2}-\frac{3 e_{137}}{2}-640,e_{89}\to -e_{105}+2 e_{106}\nn\\&&-3 e_{141}-3 e_{151}+3 e_{152}-3 e_{154}+4 e_{164}-18 e_{169}-3 e_{170}-2 e_{178}-2 e_{179}-18 e_{180}\nn\\&&+9 e_{181}-3 e_{200}+9 e_{210}-36 e_{211}+9 e_{215}-36 e_{216}-6 e_{227}+12 e_{228}+9 e_{231}+72 e_{232}\nn\\&&+72 e_{233}+288 e_{235}-2 e_{247}+4 e_{248}+3 e_{250}+18 e_{251}+18 e_{252}+72 e_{254}+4 e_{272}+12 e_{273}\nn\\&&-3 e_{287}+12 e_{288}-36 e_{290}-e_{324}-3 e_{325}+12 e_{328}-2 e_{337}-6 e_{338},e_{91}\to -e_{141}+\frac{e_{152}}{2}\nn\\&&-e_{154}-3 e_{180}+\frac{3 e_{181}}{2}-e_{200}+\frac{3 e_{210}}{2}-6 e_{211}+\frac{3 e_{215}}{2}-6 e_{216}+12 e_{232}+12 e_{233}\nn\\&&+48 e_{235}+3 e_{251}+3 e_{252}+12 e_{254}+2 e_{288}-6 e_{290}+2 e_{328}+\frac{e_{90}}{2}-\frac{e_{151}}{2}-\frac{3 e_{169}}{2}-\frac{e_{287}}{2},\nn\\&& e_{92}\to -\frac{2 e_{106}}{3}+\frac{2 e_{122}}{3}+\frac{e_{131}}{3}+\frac{e_{135}}{6}+2 e_{141}+e_{151}-e_{152}+2 e_{154}+3 e_{169}+6 e_{180}\nn\\&&-3 e_{181}+2 e_{200}-3 e_{210}+12 e_{211}-3 e_{215}+12 e_{216}-24 e_{232}-24 e_{233}-96 e_{235}-6 e_{251}\nn\\&&\left.-6 e_{252}-24 e_{254}+e_{287}-4 e_{288}+12 e_{290}-4 e_{328}-e_{90}-\frac{e_{107}}{3}\right\}.\label{cond1}
\eea

By substituting these conditions into the ansatz (\ref{G5G5F4F4}) and writing it in terms of independent variables, it can be seen that the conditions above do not fix all unknown coefficients. To find the other ones, we use the fact that the ansatz (\ref{G5G5F4F4}) should also be able to produce the couplings $ (\pa {F_4})^4 $ in eleven dimensions. A detailed calculation shows that the $ (\pa {F_4})^4 $ terms in eleven dimensions arise from three distinct contributions. The first one is obtained from circular reduction of $ 12 $-dimensional $ (\pa {F_4})^4 $ couplings to eleven dimensions. The second one comes from compactification of $ (\pa {G_5})^4 $ on a circle, which is given by (\ref{F44fromF54}) and the last contribution results from reduction of the $ 12 $-dimensional basis (\ref{G5G5F4F4}) to eleven dimensions, \ie the couplings (\ref{F44fromG52F42}). Adding these three contributions together, along with this constraint that the third part must satisfy the conditions (\ref{cond1}), truly gives the coupling $ (\pa {F_4})^4 $ in eleven dimensions. The counterpart of this coupling in $ D=11 $ has already been calculated in \cite{Peeters:2005tb,Bakhtiarizadeh:2017ojz}. Comparing them yields the following constraints on those coefficients which have not been fixed in conditions (\ref{cond1}), that are           
\bea
&&\left\{e_{43}\to e_{101}-2 e_{103}-e_{104}+2 e_{108}-4 e_{109}-8 e_{12}-4 e_{122}-2 e_{128}+2 e_{130}-2 e_{131}\right.\nn\\&&-e_{135}+3 e_{136}-3 e_{138}+2 e_{139}-6 e_{141}-4 e_{15}-4 e_{17}-2 e_{22}-2 e_{23}+4 e_{24}-e_{26}\nn\\&&+2 e_{29}-2 e_{30}+e_{33}+2 e_{4}+704,e_{44}\to -2 e_{102}-2 e_{103}+2 e_{109}+8 e_{12}+2 e_{122}\nn\\&&+2 e_{129}+2 e_{130}+e_{131}+\frac{e_{135}}{2}-3 e_{137}-3 e_{138}+4 e_{17}+2 e_{18}-4 e_{19}+2 e_{22}-2 e_{24}\nn\\&&-e_{30}+4 e_{31}+e_{32}+\frac{e_{33}}{2}-704, e_{56}\to \frac{e_{101}}{2}-e_{103}+3 e_{112}+2 e_{118}+e_{120}+\frac{3 e_{123}}{2}\nn\\&&+e_{125}+\frac{e_{126}}{2}-e_{128}+e_{130}-e_{131}-e_{135}+\frac{3 e_{136}}{2}+e_{139}-e_{14}-3 e_{141}-3 e_{143}\nn\\&&+9 e_{183}+9 e_{184}+2 e_{2}-e_{22}+18 e_{223}+18 e_{224}-288 e_{241}+9 e_{244}+9 e_{245}-72 e_{263}\nn\\&&-18 e_{264}+e_{28}-3 e_{39}+e_{41}+\frac{3 e_{55}}{2}-\frac{e_{104}}{2}-\frac{3 e_{138}}{2}-\frac{e_{26}}{2}-480,e_{69}\to e_{103}+e_{108}\nn\\&&-3 e_{111}+e_{117}-8 e_{12}-3 e_{143}-2 e_{15}-36 e_{161}-9 e_{162}-2 e_{165}-4 e_{17}+9 e_{183}+18 e_{184}\nn\\&&+e_{191}+2 e_{192}+4 e_{193}-12 e_{195}-3 e_{196}+2 e_{2}-2 e_{20}-3 e_{208}-36 e_{218}+144 e_{219}\nn\\&&-2 e_{22}+9 e_{220}+9 e_{221}-6 e_{222}+18 e_{224}-4 e_{226}+144 e_{237}+2 e_{24}-3 e_{243}+9 e_{244}\nn\\&&+18 e_{245}-2 e_{246}+2 e_{25}-36 e_{262}+18 e_{266}-18 e_{268}+2 e_{269}+6 e_{270}+4 e_{272}+12 e_{273}\nn\\&&-12 e_{274}+6 e_{275}+3 e_{276}-6 e_{278}-6 e_{279}-6 e_{280}-6 e_{283}-6 e_{284}+2 e_{3}-9 e_{309}-2 e_{319}\nn\\&&-6 e_{322}-36 e_{334}-3 e_{39}-3 e_{40}-9 e_{65}+2 e_{68}-96, e_{7}\to \frac{e_{108}}{8}-e_{12}+\frac{e_{16}}{4}+\frac{e_{24}}{4}+\frac{e_{6}}{4}\nn\\&&-\frac{e_{17}}{2}-\frac{e_{109}}{4}-\frac{e_{15}}{4}-\frac{e_{22}}{4}-\frac{e_{31}}{4}-\frac{e_{23}}{8}-\frac{e_{37}}{8}+44, e_{72}\to -\frac{3 e_{111}}{2}+\frac{e_{117}}{2}-e_{118}-2 e_{12}\nn\\&&-e_{121}-e_{122}-18 e_{161}+\frac{3 e_{166}}{2}+e_{168}-e_{17}-6 e_{195}+\frac{3 e_{197}}{2}-3 e_{198}-18 e_{218}+72 e_{219}\nn\\&&+\frac{9 e_{220}}{2}+\frac{9 e_{221}}{2}+72 e_{237}+144 e_{241}-3 e_{257}+\frac{9 e_{258}}{2}-9 e_{259}-18 e_{262}+36 e_{263}+9 e_{264}\nn\\&&+9 e_{266}-9 e_{268}-6 e_{274}+3 e_{275}+\frac{3 e_{276}}{2}-3 e_{280}-e_{314}-e_{316}-18 e_{334}+\frac{3 e_{335}}{2}-3 e_{336}\nn\\&&-\frac{3 e_{112}}{2}-\frac{9 e_{162}}{2}-\frac{3 e_{196}}{2}-\frac{e_{22}}{2}-\frac{9 e_{255}}{2}-\frac{9 e_{309}}{2}-\frac{3 e_{310}}{2}-\frac{9 e_{65}}{2}-\frac{3 e_{70}}{2}-48,\nn\\&& e_{81}\to e_{159}-32 e_{241}-e_{255}+2 e_{256}+e_{258}-2 e_{259}-8 e_{263}-2 e_{264}-e_{311}+2 e_{313}-e_{61}\nn\\&&+e_{62}-\frac{224}{9},e_{82}\to -2 e_{111}-2 e_{112}+e_{113}+4 e_{115}-2 e_{116}+\frac{2 e_{117}}{3}-6 e_{159}-24 e_{161}\nn\\&&-6 e_{162}+e_{166}-2 e_{167}-e_{176}+e_{177}-8 e_{195}-2 e_{196}+2 e_{197}-4 e_{198}-24 e_{218}+96 e_{219}\nn\\&&+6 e_{220}+6 e_{221}+96 e_{237}+384 e_{241}-6 e_{256}-2 e_{257}-3 e_{258}-24 e_{262}+96 e_{263}+24 e_{264}\nn\\&&+12 e_{266}-12 e_{268}-8 e_{274}+4 e_{275}+2 e_{276}-4 e_{280}-6 e_{309}-e_{310}+3 e_{311}+2 e_{312}\nn\\&&-12 e_{313}-24 e_{334}+e_{335}-2 e_{336}+3 e_{61}-6 e_{62}-e_{63}-6 e_{65}-e_{70}+2 e_{71}+e_{80}\nn\\&&-\frac{8 e_{12}}{3}-\frac{4 e_{17}}{3}-\frac{2 e_{22}}{3}+\frac{256}{3},e_{83}\to -e_{102}-e_{103}-e_{108}+2 e_{109}+3 e_{111}+3 e_{112}\nn\\&&-3 e_{113}-e_{117}+2 e_{118}-2 e_{119}+16 e_{12}+e_{120}+2 e_{122}+3 e_{143}+2 e_{15}+18 e_{159}+36 e_{161}\nn\\&&+9 e_{162}+8 e_{17}+3 e_{176}-3 e_{177}-9 e_{183}-18 e_{184}-e_{191}-2 e_{193}+12 e_{195}+3 e_{196}-2 e_{2}\nn\\&&+2 e_{20}+3 e_{208}+36 e_{218}-144 e_{219}+4 e_{22}-9 e_{220}-9 e_{221}-18 e_{223}-18 e_{224}-144 e_{237}\nn\\&&-2 e_{24}-288 e_{241}-18 e_{244}-18 e_{245}-2 e_{25}-9 e_{255}+18 e_{256}+18 e_{258}-36 e_{259}+36 e_{262}\nn\\&&-72 e_{263}-18 e_{264}-18 e_{266}+18 e_{268}+12 e_{274}-6 e_{275}-3 e_{276}+6 e_{278}+6 e_{279}+6 e_{280}\nn\\&&+6 e_{283}+6 e_{284}-2 e_{3}+9 e_{309}-9 e_{311}+18 e_{313}+e_{315}-2 e_{316}+2 e_{318}+2 e_{319}+6 e_{321}\nn\\&&+6 e_{322}+36 e_{334}+3 e_{335}-6 e_{336}+3 e_{39}+3 e_{40}-9 e_{61}+9 e_{62}+9 e_{65}-3 e_{80}-704,\nn\\&& e_{87}\to -\frac{e_{102}}{2}+2 e_{109}+2 e_{12}+2 e_{164}+e_{17}+2 e_{192}+2 e_{193}+\frac{e_{22}}{2}+e_{269}+3 e_{270}+2 e_{272}\nn\\&&+6 e_{273}+e_{318}+3 e_{321}+e_{67}+e_{68}-\frac{e_{103}}{2}-240,e_{9}\to 6 e_{12}+\frac{3 e_{143}}{2}+2 e_{17}-e_{2}+e_{22}\nn\\&&-9 e_{223}-9 e_{224}+144 e_{241}+36 e_{263}+9 e_{264}+\frac{3 e_{39}}{2}+\frac{3 e_{51}}{2}-e_{52}-\frac{9 e_{183}}{2}-\frac{9 e_{184}}{2}\nn\\&&-\frac{9 e_{244}}{2}-\frac{9 e_{245}}{2}-\frac{e_{28}}{2}-\frac{e_{41}}{2}-288,e_{94}\to -e_{183}-e_{184}-2 e_{223}-2 e_{224}+32 e_{241}-e_{244}\nn\\&&-e_{245}+8 e_{263}+2 e_{264}-\frac{224}{9},e_{96}\to \frac{e_{114}}{3}+\frac{e_{117}}{9}+4 e_{161}+e_{162}-4 e_{186}-e_{187}+4 e_{218}\nn\\&&-16 e_{219}-e_{220}-e_{221}-16 e_{237}-64 e_{241}-4 e_{261}+4 e_{262}-16 e_{263}-4 e_{264}+2 e_{265}\nn\\&&-2 e_{266}+e_{267}-2 e_{268}+e_{309}+4 e_{334}+e_{65}-\frac{4 e_{12}}{9}-\frac{2 e_{17}}{9}-\frac{e_{22}}{9}+64,e_{99}\to -\frac{e_{114}}{3}\nn\\&&\left.+\frac{4 e_{12}}{9}+\frac{2 e_{17}}{9}+\frac{e_{22}}{9}-\frac{e_{117}}{9}-\frac{352}{9}\right\}.\label{cond2}
\eea
The above two sets of conditions are direct consequence of consistent truncation of the $ 12 $-dimensional theory to ten and eleven dimensions, respectively. It is possible to substitute them into the basis (\ref{G5G5F4F4}) to find the couplings between two $ 5 $-form and two $ 4 $-form field strengths. By doing so, one arrives at:
\bea
&& e^{-1} {\cal L}_{(\pa {G_5})^2 (\pa {F_4})^2}=\nn\\&& \frac{4}{45} (495 {F_4}^{abcd,e} {F_4}^{fghi}{}_{,e} {G_5}_{abfg}{}^{j,k} \
{G_5}_{cdhij,k} - 3240 {F_4}^{abcd,e} {F_4}^{fghi,j} {G_5}_{abef}{}^{k}{}_{,g} \
{G_5}_{cdhik,j} \nonumber \\ 
&&+ 7920 {F_4}_{a}{}^{fgh,i} {F_4}^{abcd,e} {G_5}_{bcfi}{}^{j,k} \
{G_5}_{degjk,h}  - 7920 {F_4}_{a}{}^{fgh,i} {F_4}^{abcd,e} {G_5}_{bcf}{}^{jk}{}_{,i} \
{G_5}_{degjk,h} \nonumber \\ 
&&- 5400 {F_4}^{abcd,e} {F_4}^{fghi,j} {G_5}_{abfg}{}^{k}{}_{,c} \
{G_5}_{dehjk,i} + 3960 {F_4}_{a}{}^{fgh,i} {F_4}^{abcd,e} {G_5}_{bcfg}{}^{j,k} \
{G_5}_{dehjk,i} \nonumber \\ 
&& - 7920 {F_4}_{a}{}^{fgh,i} {F_4}^{abcd,e} {G_5}_{bcf}{}^{jk}{}_{,g} \
{G_5}_{dehjk,i} - 7200 {F_4}_{a}{}^{fgh,i} {F_4}^{abcd,e} {G_5}_{bcf}{}^{jk}{}_{,g} \
{G_5}_{deijk,h} \nonumber \\ 
&&- 1080 {F_4}_{ae}{}^{fg,h} {F_4}^{abcd,e} {G_5}_{bch}{}^{ij,k} \
{G_5}_{dfgik,j} - 540 {F_4}^{abcd,e} {F_4}_{e}{}^{fgh,i} {G_5}_{abi}{}^{jk}{}_{,c} {G_5}_{dfgjk,h} \
\nonumber \\ 
&&+ 4320 {F_4}_{ae}{}^{fg,h} {F_4}^{abcd,e} {G_5}_{bch}{}^{ij,k} {G_5}_{dfijk,g} - 480 \
{F_4}_{ae}{}^{fg,h} {F_4}^{abcd,e} {G_5}_{bc}{}^{ijk}{}_{,h} {G_5}_{dfijk,g} \nonumber \\ 
&& - 1440 {F_4}_{ab}{}^{fg,h} {F_4}^{abcd,e} {G_5}_{ceh}{}^{ij,k} {G_5}_{dfijk,g} - \
480 {F_4}_{ab}{}^{fg,h} {F_4}^{abcd,e} {G_5}_{ce}{}^{ijk}{}_{,h} {G_5}_{dfijk,g} \nonumber \\ 
&&- \
280 {F_4}^{abcd,e} {F_4}^{fghi}{}_{,e} {G_5}_{abcf}{}^{j,k} {G_5}_{dghij,k}  + 960 {F_4}^{abcd,e} {F_4}_{e}{}^{fgh,i} {G_5}_{abc}{}^{jk}{}_{,f} {G_5}_{dghij,k} \
\nonumber \\ 
&&- 7920 {F_4}^{abcd,e} {F_4}_{e}{}^{fgh,i} {G_5}_{abf}{}^{jk}{}_{,c} {G_5}_{dghij,k} - \
2520 {F_4}_{a}{}^{fgh,i} {F_4}^{abcd,e} {G_5}_{bcef}{}^{j,k} {G_5}_{dghij,k} \nonumber \\ 
&& - 1440 {F_4}_{ae}{}^{fg,h} {F_4}^{abcd,e} {G_5}_{bcf}{}^{ij,k} {G_5}_{dghij,k} + \
720 {F_4}_{a}{}^{fgh}{}_{,e} {F_4}^{abcd,e} {G_5}_{bcf}{}^{ij,k} {G_5}_{dghij,k} \nonumber \\ 
&& - \
2700 {F_4}^{abcd,e} {F_4}_{e}{}^{fgh}{}_{,a} {G_5}_{bcf}{}^{ij,k} {G_5}_{dghij,k} - 7200 {F_4}_{ae}{}^{fg,h} {F_4}^{abcd,e} {G_5}_{bc}{}^{ijk}{}_{,f} \
{G_5}_{dghij,k} \nonumber \\ 
&&- 280 {F_4}^{abcd,e} {F_4}^{fghi,j} {G_5}_{abcf}{}^{k}{}_{,j} \
{G_5}_{dghik,e} + 720 {F_4}^{abcd,e} {F_4}^{fghi,j} {G_5}_{abce}{}^{k}{}_{,f} \
{G_5}_{dghik,j} \nonumber \\ 
&& - 440 {F_4}^{abcd,e} {F_4}^{fghi,j} {G_5}_{abcf}{}^{k}{}_{,e} {G_5}_{dghik,j} - \
720 {F_4}_{ae}{}^{fg}{}_{,b} {F_4}^{abcd,e} {G_5}_{cd}{}^{hij,k} {G_5}_{fghik,j} \nonumber \\ 
&&+ \
120 {F_4}_{ab}{}^{fg,h} {F_4}^{abcd,e} {G_5}_{cd}{}^{ijk}{}_{,e} {G_5}_{fgijk,h} - 960 {F_4}_{ab}{}^{fg,h} {F_4}^{abcd,e} {G_5}_{ce}{}^{ijk}{}_{,d} {G_5}_{fhijk,g} \
\nonumber \\ 
&&+ 40 {F_4}_{abc}{}^{f,g} {F_4}^{abcd,e} {G_5}_{d}{}^{hijk}{}_{,e} {G_5}_{fhijk,g} + \
240 {F_4}_{ae}{}^{fg}{}_{,b} {F_4}^{abcd,e} {G_5}_{c}{}^{hijk}{}_{,f} {G_5}_{ghijk,d} \
\nonumber \\ 
&& - 120 {F_4}_{abe}{}^{f,g} {F_4}^{abcd,e} {G_5}_{f}{}^{hijk}{}_{,c} {G_5}_{ghijk,d} \
+ 40 {F_4}_{abc}{}^{f,g} {F_4}^{abcd,e} {G_5}_{d}{}^{hijk}{}_{,f} {G_5}_{ghijk,e} \nonumber \\ 
&&+ \
40 {F_4}_{abc}{}^{f,g} {F_4}^{abcd,e} {G_5}_{d}{}^{hijk}{}_{,e} {G_5}_{ghijk,f}  - 40 {F_4}_{abc}{}^{f,g} {F_4}^{abcd,e} {G_5}_{e}{}^{hijk}{}_{,d} {G_5}_{ghijk,f} \
\nonumber \\ 
&&+ 24 {F_4}_{abe}{}^{f}{}_{,c} {F_4}^{abcd,e} {G_5}_{ghijk,f} {G_5}^{ghijk}{}_{,d} - 2 \
{F_4}_{abcd}{}^{,f} {F_4}^{abcd,e} {G_5}_{ghijk,f} {G_5}^{ghijk}{}_{,e}).\label{G52F42coup}
\eea
This coupling obviously is self-dual, because the result is a consequence of comparison with the self-dual coupling in type IIB supergravity. 

\section{Discussion}\label{dis}

In this paper, we have shown that there exists a possible candidate field theory for F-theory at order $ \alpha'^3 $, whose dimensional reduction to $ D=11 $ and $ D=10 $ admits a consistent truncation to the bosonic sector of eleven-dimensional and type IIB supergravity theories. Let us conclude by discussing some issues and possible applications of our results.

As already discussed in \cite{Khviengia:1997rh} for $ 12 $-dimensional supergravity at leading order, we have obtained a twelve-dimensional theory at eight-derivative level that contains more bosonic degrees of freedom than are in its dimensionally-reduced M-theory or type IIB theory. Alternatively, one could choose an another method, and carry out precisely these field truncations already in twelve dimensions. But, they cannot be performed in a twelve-dimensionally covariant manner, and thus this formulation of field theory in twelve dimensions would have only eleven-dimensional or ten-dimensional covariance. In summary, we have two ways in formulation of the theory, a covariant twelve-dimensional theory with extra degrees of freedom, or non-covariant one with the correct degrees of freedom. But in this paper we have employed the former, since the truncation required to obtain eleven-dimensional supergravity is different from the one required to obtain type IIB theory, it would sound that the two would only be unified in twelve dimensions if the covariant twelve-dimensional theory is taken as the starting point \cite{Khviengia:1997rh}.

Finally, let us comment on the applications of our results and future directions. It also would be interesting to obtain the couplings containing the dilaton field with the approach introduced in this paper. Upon compactification on elliptically fibered Calabi-Yau fourfolds, one can obtain the non-trivial vacuum for the axio-dilaton which leads to a new, $ {\cal N} = 1 $, $ \alpha'^3 $ correction to the four-dimensional effective action, as noted in \cite{Minasian:2015bxa}, which makes it phenomenologically attractive. One also can explore the implications of flux compactifications for the moduli-space problem \cite{Grimm:2011sk}. Brane solutions \cite{Khviengia:1997rh,Lu:1995cs,Lu:1995yn,Gueven:1992hh} and applications in cosmology \cite{Ciupke:2016vkc} will become interesting future directions. 

Furthermore, it is possible to employ the algorithm introduced in \cite{Bakhtiarizadeh:2013zia} for reducing the number of terms of the couplings obtained in this paper and write them in its minimal-term form. This imposes all symmetries including mono-term (antisymmetry property of field strengths and symmetries of Riemann tensors) as well as multi-term symmetries (the Bianchi identities for field strengths and Riemann tensors).

\appendix

\section{The bases for higher-order terms}

Here we collect the bases for higher-order terms that have been used in this paper, but due to their size are unpleasant to include in the main body of the paper. Note that the bases in sections \ref{G52R2}, \ref{G54} and \ref{G52F42} have been calculated by the ``AllContractions" command in ``xTras" package of Mathematica that does not take the Bianchi identities into account and accordingly does not necessarily return an irreducible basis of contractions, but it does always gives a complete basis.

\subsection{The $ ({\partial {G_5}})^2 R^2 $ basis}\label{G52R2}

There are 120 possible $ ({\partial {G_5}})^2 R^2 $ terms
in the action, which satisfy the on-shell conditions, and those are:
\bea
&& a_1 {G_5}_{efghi,j} {G_5}^{efghi,j} R_{abcd} R^{abcd} + a_2 {G_5}_{efghj,i} {G_5}^{efghi,j} R_{abcd} R^{abcd} \nonumber \\ 
&&+ a_3 {G_5}_{d}{}^{fghi,j} {G_5}_{efghi,j} R_{abc}{}^{e} R^{abcd}  + a_4 {G_5}_{d}{}^{fghi,j} {G_5}_{efghj,i} R_{abc}{}^{e} R^{abcd} \nonumber \\ 
&&+ a_5 {G_5}_{d}{}^{fghi,j} \
{G_5}_{fghij,e} R_{abc}{}^{e} R^{abcd}  + a_6 {G_5}_{fghij,e} {G_5}^{fghij}{}_{,d} R_{abc}{}^{e} R^{abcd} \nonumber \\ 
&&+ a_7 {G_5}_{ce}{}^{ghi,j} \
{G_5}_{dfghi,j} R_{ab}{}^{ef} R^{abcd}  + a_8 {G_5}_{ce}{}^{ghi,j} {G_5}_{dfghj,i} R_{ab}{}^{ef} R^{abcd} \nonumber \\ 
&&+ a_9 {G_5}_{ce}{}^{ghi,j} \
{G_5}_{dghij,f} R_{ab}{}^{ef} R^{abcd}  + a_{10} {G_5}_{c}{}^{ghij}{}_{,e} {G_5}_{dghij,f} R_{ab}{}^{ef} R^{abcd} \nonumber \\ 
&&+ a_{11} {G_5}_{cd}{}^{ghi,j} \
{G_5}_{efghi,j} R_{ab}{}^{ef} R^{abcd}  + a_{12} {G_5}_{cd}{}^{ghi,j} {G_5}_{efghj,i} R_{ab}{}^{ef} R^{abcd} \nonumber \\ 
&&+ a_{13} {G_5}_{cd}{}^{ghi,j} \
{G_5}_{eghij,f} R_{ab}{}^{ef} R^{abcd}  + a_{14} {G_5}_{c}{}^{ghij}{}_{,d} {G_5}_{eghij,f} R_{ab}{}^{ef} R^{abcd} \nonumber \\ 
&&+ a_{15} \
{G_5}_{c}{}^{ghij}{}_{,e} {G_5}_{fghij,d} R_{ab}{}^{ef} R^{abcd}  + a_{16} {G_5}_{efghi,j} {G_5}^{efghi,j} R_{acbd} R^{abcd} \nonumber \\ 
&&+ a_{17} {G_5}_{efghj,i} {G_5}^{efghi,j} R_{acbd} R^{abcd}  + a_{18} {G_5}_{d}{}^{fghi,j} {G_5}_{efghi,j} R_{acb}{}^{e} R^{abcd} \nonumber \\ 
&&+ a_{19} {G_5}_{d}{}^{fghi,j} \
{G_5}_{efghj,i} R_{acb}{}^{e} R^{abcd}  + a_{20} {G_5}_{d}{}^{fghi,j} {G_5}_{fghij,e} R_{acb}{}^{e} R^{abcd} \nonumber \\ 
&&+ a_{21} {G_5}_{fghij,e} \
{G_5}^{fghij}{}_{,d} R_{acb}{}^{e} R^{abcd}  + a_{22} {G_5}_{be}{}^{ghi,j} {G_5}_{dfghi,j} R_{ac}{}^{ef} R^{abcd} \nonumber \\ 
&&+ a_{23} {G_5}_{be}{}^{ghi,j} \
{G_5}_{dfghj,i} R_{ac}{}^{ef} R^{abcd}  + a_{24} {G_5}_{be}{}^{ghi,j} {G_5}_{dghij,f} R_{ac}{}^{ef} R^{abcd} \nonumber \\ 
&&+ a_{25} \
{G_5}_{b}{}^{ghij}{}_{,e} {G_5}_{dghij,f} R_{ac}{}^{ef} R^{abcd} + a_{26} {G_5}_{bd}{}^{ghi,j} {G_5}_{efghi,j} R_{ac}{}^{ef} R^{abcd} \nonumber \\ 
&&+ a_{27} \
{G_5}_{b}{}^{ghij}{}_{,d} {G_5}_{efghi,j} R_{ac}{}^{ef} R^{abcd}  + a_{28} {G_5}_{bd}{}^{ghi,j} {G_5}_{efghj,i} R_{ac}{}^{ef} R^{abcd} \nonumber \\ 
&&+ a_{29} {G_5}_{bd}{}^{ghi,j} \
{G_5}_{eghij,f} R_{ac}{}^{ef} R^{abcd}  + a_{30} {G_5}_{b}{}^{ghij}{}_{,d} {G_5}_{eghij,f} R_{ac}{}^{ef} R^{abcd} \nonumber \\ 
&&+ a_{31} {G_5}_{be}{}^{ghi,j} \
{G_5}_{fghij,d} R_{ac}{}^{ef} R^{abcd}  + a_{32} {G_5}_{b}{}^{ghij}{}_{,e} {G_5}_{fghij,d} R_{ac}{}^{ef} R^{abcd} \nonumber \\ 
&&+ a_{33} \
{G_5}_{e}{}^{ghij}{}_{,b} {G_5}_{fghij,d} R_{ac}{}^{ef} R^{abcd}  + a_{34} {G_5}_{bf}{}^{ghi,j} {G_5}_{deghi,j} R_{a}{}^{e}{}_{c}{}^{f} R^{abcd} \nonumber \\ 
&&+ a_{35} \
{G_5}_{bf}{}^{ghi,j} {G_5}_{deghj,i} R_{a}{}^{e}{}_{c}{}^{f} R^{abcd}  + a_{36} {G_5}_{be}{}^{ghi,j} {G_5}_{dfghi,j} R_{a}{}^{e}{}_{c}{}^{f} R^{abcd} \nonumber \\ 
&&+ a_{37} \
{G_5}_{be}{}^{ghi,j} {G_5}_{dfghj,i} R_{a}{}^{e}{}_{c}{}^{f} R^{abcd}  + a_{38} {G_5}_{bf}{}^{ghi,j} {G_5}_{dghij,e} R_{a}{}^{e}{}_{c}{}^{f} R^{abcd} \nonumber \\ 
&&+ a_{39} \
{G_5}_{b}{}^{ghij}{}_{,f} {G_5}_{dghij,e} R_{a}{}^{e}{}_{c}{}^{f} R^{abcd} + a_{40} {G_5}_{be}{}^{ghi,j} {G_5}_{dghij,f} R_{a}{}^{e}{}_{c}{}^{f} R^{abcd}  \nonumber \\ 
&&+ a_{41} \
{G_5}_{b}{}^{ghij}{}_{,e} {G_5}_{dghij,f} R_{a}{}^{e}{}_{c}{}^{f} R^{abcd}  + a_{42} {G_5}_{bd}{}^{ghi,j} {G_5}_{efghi,j} R_{a}{}^{e}{}_{c}{}^{f} R^{abcd} \nonumber \\ 
&&+ a_{43} \
{G_5}_{bd}{}^{ghi,j} {G_5}_{efghj,i} R_{a}{}^{e}{}_{c}{}^{f} R^{abcd}  + a_{44} {G_5}_{b}{}^{ghij}{}_{,f} {G_5}_{eghij,d} R_{a}{}^{e}{}_{c}{}^{f} R^{abcd} \nonumber \\ 
&&+ a_{45} \
{G_5}_{bd}{}^{ghi,j} {G_5}_{eghij,f} R_{a}{}^{e}{}_{c}{}^{f} R^{abcd}  + a_{46} {G_5}_{b}{}^{ghij}{}_{,d} {G_5}_{eghij,f} R_{a}{}^{e}{}_{c}{}^{f} R^{abcd} \nonumber \\ 
&&+ a_{47} \
{G_5}_{b}{}^{ghij}{}_{,e} {G_5}_{fghij,d} R_{a}{}^{e}{}_{c}{}^{f} R^{abcd} + a_{48} {G_5}_{b}{}^{ghij}{}_{,d} {G_5}_{fghij,e} R_{a}{}^{e}{}_{c}{}^{f} R^{abcd} \nonumber \\ 
&& + a_{49} \
{G_5}_{bfg}{}^{hi,j} {G_5}_{cdehi,j} R_{a}{}^{efg} R^{abcd} \
+ a_{50} {G_5}_{bfg}{}^{hi,j} {G_5}_{cdehj,i} R_{a}{}^{efg} R^{abcd} \nonumber \\ 
&&+ a_{51} \
{G_5}_{bf}{}^{hij}{}_{,e} {G_5}_{cdghi,j} R_{a}{}^{efg} R^{abcd}  + a_{52} {G_5}_{bfg}{}^{hi,j} {G_5}_{cdhij,e} R_{a}{}^{efg} R^{abcd} \nonumber \\ 
&&+ a_{53} \
{G_5}_{bf}{}^{hij}{}_{,g} {G_5}_{cdhij,e} R_{a}{}^{efg} R^{abcd}  + a_{54} {G_5}_{bf}{}^{hij}{}_{,e} {G_5}_{cdhij,g} R_{a}{}^{efg} R^{abcd} \nonumber \\ 
&&+ a_{55} {G_5}_{bfg}{}^{hi,j} \
{G_5}_{cehij,d} R_{a}{}^{efg} R^{abcd}  + a_{56} {G_5}_{bf}{}^{hij}{}_{,g} {G_5}_{cehij,d} R_{a}{}^{efg} R^{abcd} \nonumber \\ 
&&+ a_{57} \
{G_5}_{bf}{}^{hij}{}_{,e} {G_5}_{cghij,d} R_{a}{}^{efg} R^{abcd}  + a_{58} {G_5}_{bcf}{}^{hi,j} {G_5}_{deghi,j} R_{a}{}^{efg} R^{abcd} \nonumber \\ 
&&+ a_{59} {G_5}_{bcf}{}^{hi,j} \
{G_5}_{deghj,i} R_{a}{}^{efg} R^{abcd}  + a_{60} {G_5}_{bcf}{}^{hi,j} {G_5}_{dehij,g} R_{a}{}^{efg} R^{abcd} \nonumber \\ 
&&+ a_{61} \
{G_5}_{bc}{}^{hij}{}_{,f} {G_5}_{dehij,g} R_{a}{}^{efg} R^{abcd}  + a_{62} {G_5}_{bf}{}^{hij}{}_{,c} {G_5}_{dehij,g} R_{a}{}^{efg} R^{abcd} \nonumber \\ 
&&+ a_{63} {G_5}_{bce}{}^{hi,j} \
{G_5}_{dfghi,j} R_{a}{}^{efg} R^{abcd}  + a_{64} {G_5}_{bc}{}^{hij}{}_{,e} {G_5}_{dfghi,j} R_{a}{}^{efg} R^{abcd} \nonumber \\ 
&&+ a_{65} \
{G_5}_{be}{}^{hij}{}_{,c} {G_5}_{dfghi,j} R_{a}{}^{efg} R^{abcd}  + a_{66} {G_5}_{bce}{}^{hi,j} {G_5}_{dfghj,i} R_{a}{}^{efg} R^{abcd} \nonumber \\ 
&&+ a_{67} {G_5}_{bce}{}^{hi,j} \
{G_5}_{dfhij,g} R_{a}{}^{efg} R^{abcd}  + a_{68} {G_5}_{bc}{}^{hij}{}_{,e} {G_5}_{dfhij,g} R_{a}{}^{efg} R^{abcd} \nonumber \\ 
&&+ a_{69} \
{G_5}_{be}{}^{hij}{}_{,c} {G_5}_{dfhij,g} R_{a}{}^{efg} R^{abcd}  + a_{70} {G_5}_{bcf}{}^{hi,j} {G_5}_{dghij,e} R_{a}{}^{efg} R^{abcd} \nonumber \\ 
&&+ a_{71} \
{G_5}_{bc}{}^{hij}{}_{,f} {G_5}_{dghij,e} R_{a}{}^{efg} R^{abcd}  + a_{72} {G_5}_{bf}{}^{hij}{}_{,c} {G_5}_{dghij,e} R_{a}{}^{efg} R^{abcd} \nonumber \\
&&+ a_{73} \
{G_5}_{cf}{}^{hij}{}_{,b} {G_5}_{dghij,e} R_{a}{}^{efg} R^{abcd}  + a_{74} {G_5}_{bcd}{}^{hi,j} {G_5}_{efghi,j} R_{a}{}^{efg} R^{abcd} \nonumber \\ 
&&+ a_{75} {G_5}_{bcd}{}^{hi,j} \
{G_5}_{efghj,i} R_{a}{}^{efg} R^{abcd}  + a_{76} {G_5}_{bcd}{}^{hi,j} {G_5}_{efhij,g} R_{a}{}^{efg} R^{abcd} \nonumber \\ 
&&+ a_{77} \
{G_5}_{bc}{}^{hij}{}_{,d} {G_5}_{efhij,g} R_{a}{}^{efg} R^{abcd}  + a_{78} {G_5}_{bcf}{}^{hi,j} {G_5}_{eghij,d} R_{a}{}^{efg} R^{abcd} \nonumber \\ 
&&+ a_{79} \
{G_5}_{bc}{}^{hij}{}_{,f} {G_5}_{eghij,d} R_{a}{}^{efg} R^{abcd}  + a_{80} {G_5}_{cd}{}^{hij}{}_{,e} {G_5}_{fghij,b} R_{a}{}^{efg} R^{abcd} \nonumber \\ 
&&+ a_{81} {G_5}_{bce}{}^{hi,j} \
{G_5}_{fghij,d} R_{a}{}^{efg} R^{abcd}  + a_{82} {G_5}_{bc}{}^{hij}{}_{,e} {G_5}_{fghij,d} R_{a}{}^{efg} R^{abcd} \nonumber \\ 
&&+ a_{83} \
{G_5}_{be}{}^{hij}{}_{,c} {G_5}_{fghij,d} R_{a}{}^{efg} R^{abcd}  + a_{84} {G_5}_{bcd}{}^{hi,j} {G_5}_{fghij,e} R_{a}{}^{efg} R^{abcd} \nonumber \\ 
&&+ a_{85} \
{G_5}_{bc}{}^{hij}{}_{,d} {G_5}_{fghij,e} R_{a}{}^{efg} R^{abcd}  + a_{86} {G_5}_{cd}{}^{hij}{}_{,b} {G_5}_{fghij,e} R_{a}{}^{efg} R^{abcd} \nonumber \\ 
&&+ a_{87} {G_5}_{aceg}{}^{i,j} \
{G_5}_{bdfhi,j} R^{abcd} R^{efgh}  + a_{88} {G_5}_{aceg}{}^{i,j} {G_5}_{bdfhj,i} R^{abcd} \
R^{efgh} \nonumber \\ 
&&+ a_{89} {G_5}_{aceg}{}^{i,j} {G_5}_{bdfij,h} R^{abcd} R^{efgh}  + a_{90} {G_5}_{ace}{}^{ij}{}_{,g} {G_5}_{bdfij,h} R^{abcd} R^{efgh} \nonumber \\ 
&&+ a_{91} {G_5}_{ace}{}^{ij}{}_{,f} \
{G_5}_{bdgij,h} R^{abcd} R^{efgh}  + a_{92} {G_5}_{ace}{}^{ij}{}_{,g} {G_5}_{bdhij,f} R^{abcd} R^{efgh} \nonumber \\ 
&&+ a_{93} {G_5}_{ace}{}^{ij}{}_{,g} \
{G_5}_{bfhij,d} R^{abcd} R^{efgh}  + a_{94} {G_5}_{abeg}{}^{i,j} {G_5}_{cdfhi,j} R^{abcd} \
R^{efgh} \nonumber \\ 
&&+ a_{95} {G_5}_{abeg}{}^{i,j} {G_5}_{cdfhj,i} R^{abcd} R^{efgh}  + a_{96} {G_5}_{abeg}{}^{i,j} {G_5}_{cdfij,h} R^{abcd} \
R^{efgh} \nonumber \\ 
&&+ a_{97} {G_5}_{abe}{}^{ij}{}_{,g} \
{G_5}_{cdfij,h} R^{abcd} R^{efgh}  + a_{98} {G_5}_{abef}{}^{i,j} {G_5}_{cdghi,j} R^{abcd} \
R^{efgh} \nonumber \\ 
&&+ a_{99} {G_5}_{abef}{}^{i,j} {G_5}_{cdghj,i} R^{abcd} R^{efgh}  + a_{100} {G_5}_{abef}{}^{i,j} {G_5}_{cdgij,h} R^{abcd} \
R^{efgh} \nonumber \\ 
&&+ a_{101} {G_5}_{abe}{}^{ij}{}_{,f} \
{G_5}_{cdgij,h} R^{abcd} R^{efgh}  + a_{102} {G_5}_{abe}{}^{ij}{}_{,g} {G_5}_{cdhij,f} R^{abcd} R^{efgh} \nonumber \\ 
&&+ a_{103} {G_5}_{abeg}{}^{i,j} \
{G_5}_{cfhij,d} R^{abcd} R^{efgh}  + a_{104} {G_5}_{abe}{}^{ij}{}_{,g} {G_5}_{cfhij,d} R^{abcd} R^{efgh} \nonumber \\ 
&&+ a_{105} {G_5}_{abe}{}^{ij}{}_{,f} \
{G_5}_{cghij,d} R^{abcd} R^{efgh}  + a_{106} {G_5}_{abce}{}^{i,j} {G_5}_{dfghi,j} R^{abcd} \
R^{efgh} \nonumber \\ 
&&+ a_{107} {G_5}_{abce}{}^{i,j} {G_5}_{dfghj,i} R^{abcd} R^{efgh}  + a_{108} {G_5}_{abce}{}^{i,j} {G_5}_{dfgij,h} R^{abcd} \
R^{efgh} \nonumber \\ 
&&+ a_{109} {G_5}_{abc}{}^{ij}{}_{,e} \
{G_5}_{dfgij,h} R^{abcd} R^{efgh}  + a_{110} {G_5}_{abe}{}^{ij}{}_{,c} {G_5}_{dfgij,h} R^{abcd} R^{efgh} \nonumber \\ 
&&+ a_{111} {G_5}_{ace}{}^{ij}{}_{,b} \
{G_5}_{dfgij,h} R^{abcd} R^{efgh}  + a_{112} {G_5}_{abce}{}^{i,j} {G_5}_{dghij,f} R^{abcd} \
R^{efgh} \nonumber \\ 
&&+ a_{113} {G_5}_{abc}{}^{ij}{}_{,e} \
{G_5}_{dghij,f} R^{abcd} R^{efgh}  + a_{114} {G_5}_{abe}{}^{ij}{}_{,c} {G_5}_{dghij,f} R^{abcd} R^{efgh} \nonumber \\ 
&&+ a_{115} {G_5}_{abcd}{}^{i,j} \
{G_5}_{efghi,j} R^{abcd} R^{efgh}  + a_{116} {G_5}_{abcd}{}^{i,j} {G_5}_{efghj,i} R^{abcd} \
R^{efgh} \nonumber \\ 
&&+ a_{117} {G_5}_{abcd}{}^{i,j} {G_5}_{efgij,h} R^{abcd} R^{efgh}  + a_{118} {G_5}_{abc}{}^{ij}{}_{,d} {G_5}_{efgij,h} R^{abcd} R^{efgh} \nonumber \\ 
&&+ a_{119} {G_5}_{abce}{}^{i,j} \
{G_5}_{fghij,d} R^{abcd} R^{efgh}  + a_{120} {G_5}_{abc}{}^{ij}{}_{,e} {G_5}_{fghij,d} R^{abcd} R^{efgh},\label{G5G5RR}
\eea
where comma on the indices of field strengths refers to a partial derivative with respect to the index afterwards.

\subsection{The $ ({\partial {F_4}})^2 R^2 $ basis}\label{F42R2}

By imposing the linearised lowest-order equations of motion \cite{Peeters:2005tb}, one obtains 24 possible terms of the form $ (\pa {F_4})^2 R^2 $ in the action
\bea
&& b_1 {F_4}^{agh}{}_{i}{}^{,e} {F_4}^{bdfi,c} R_{abcd} \
R_{efgh} + b_2 {F_4}^{acg}{}_{i}{}^{,e} {F_4}^{bdfi,h} R_{abcd} R_{efgh} \nonumber \\ 
&& + b_3 {F_4}^{acg}{}_{i}{}^{,e} {F_4}^{bdhi,f} R_{abcd} \
R_{efgh} + b_4 {F_4}^{cdgh}{}_{,i} {F_4}^{iabe,f} R_{abcd} R_{efgh} \nonumber \\ 
&&+ b_5 {F_4}^{bc}{}_{hi}{}^{,a} \
{F_4}^{fghi,e} R_{abcd} R_{efg}{}^{d}+ b_6 {F_4}^{be}{}_{hi}{}^{,a} {F_4}^{fghi,c} R_{abcd} \
R_{efg}{}^{d}\nonumber \\ 
&& + b_7 {F_4}^{be}{}_{hi}{}^{,a} \
{F_4}^{cfhi,g} R_{abcd} R_{efg}{}^{d}+ b_8 {F_4}^{ce}{}_{hi}{}^{,a} \
{F_4}^{fghi,b} R_{abcd} R_{efg}{}^{d}\nonumber \\ 
&& + b_9 {F_4}^{bghi,f} {F_4}^{ce}{}_{hi}{}^{,a} R_{abcd} \
R_{efg}{}^{d}+ b_{10} {F_4}^{abfh,i} {F_4}^{ceg}{}_{h,i} R_{abcd} R_{efg}{}^{d} \nonumber \\ 
&&+ b_{11} {F_4}^{abf}{}_{i,h} {F_4}^{cegh,i} R_{abcd} R_{efg}{}^{d} + b_{12} {F_4}^{bahi,g} {F_4}^{ef}{}_{hi}{}^{,c} R_{abcd} R_{efg}{}^{d}  \nonumber \\ 
&&+ b_{13} {F_4}^{bf}{}_{gh,i} {F_4}^{degh,i} R_{abcd} R_{e}{}^{a}{}_{f}{}^{c}   + b_{14} {F_4}^{bd}{}_{gh,i} \
{F_4}^{efgh,i} R_{abcd} R_{e}{}^{a}{}_{f}{}^{c} \nonumber \\ 
&& + b_{15} {F_4}^{dghi,e} {F_4}^{f}{}_{ghi}{}^{,b} R_{abcd} R_{e}{}^{a}{}_{f}{}^{c} + b_{16} {F_4}^{d}{}_{ghi}{}^{,b} {F_4}^{fghi,e} R_{abcd} R_{e}{}^{a}{}_{f}{}^{c} \nonumber \\ 
&& + b_{17} \
{F_4}^{e}{}_{ghi}{}^{,b} {F_4}^{fghi,d} R_{abcd} R_{e}{}^{a}{}_{f}{}^{c}+ b_{18} {F_4}^{bghi,d} {F_4}^{ef}{}_{gh,i} R_{abcd} R_{e}{}^{a}{}_{f}{}^{c} \nonumber \\ 
&&+ b_{19} \
{F_4}^{ae}{}_{gh,i} {F_4}^{bfgh,i} R_{abcd} R_{ef}{}^{cd}  + b_{20} \
{F_4}^{e}{}_{ghi}{}^{,a} {F_4}^{fghi,b} R_{abcd} R_{ef}{}^{cd}\nonumber \\ 
&&+ b_{21} {F_4}^{bghi,f} {F_4}^{e}{}_{ghi}{}^{,a} R_{abcd} R_{ef}{}^{cd}+ b_{22} {F_4}_{fghi}{}^{,b} {F_4}^{fghi,e} R_{abcd} R_{e}{}^{acd} \nonumber \\ 
&&+ b_{23} {F_4}^{b}{}_{fgh,i} {F_4}^{efgh,i} R_{abcd} R_{e}{}^{acd}+b_{24} {F_4}^{eghi,f} {F_4}_{fghi,e} R_{abcd} R^{abcd}.\label{F4F4RR} 
\eea

\subsection{The $ ({\partial {F_4}})^4 $ basis}\label{F44}

The basis for the $ (\pa {F_4})^4 $ terms (at linearised on-shell level) \cite{Peeters:2005tb} is also given by 24 terms, that are
\bea
&& c_1 {F_4}^{aehj,i} {F_4}_{bcde,a} {F_4}^{c}{}_{fgh}{}^{,b} \
{F_4}^{dg}{}_{ij}{}^{,f} + c_2 {F_4}^{a}{}_{fgh}{}^{,b} {F_4}_{bcde,a} \
{F_4}^{cdf}{}_{j,i} {F_4}^{eghi,j} \nonumber \\ 
&&+ c_3 {F_4}_{bcde,a} \
{F_4}^{b}{}_{fgh}{}^{,a} {F_4}^{cdf}{}_{j,i} {F_4}^{eghj,i}  + c_4 {F_4}^{ad}{}_{ij}{}^{,f} {F_4}_{bcde,a} \
{F_4}^{c}{}_{fgh}{}^{,b} {F_4}^{eghj,i} \nonumber \\ 
&&+ c_5 {F_4}_{bcde,a} \
{F_4}^{b}{}_{fgh}{}^{,a} {F_4}^{df}{}_{ij}{}^{,c} {F_4}^{eghj,i} + c_6 \
{F_4}^{ac}{}_{fg}{}^{,b} {F_4}_{bcde,a} {F_4}^{df}{}_{ij,h} {F_4}^{eghj,i}  \nonumber \\ 
&&+ c_7 {F_4}_{bcde,a} {F_4}^{bc}{}_{fg}{}^{,a} {F_4}^{df}{}_{ij,h} \
{F_4}^{eghj,i} + c_8 {F_4}_{bcde,a} {F_4}^{cdf}{}_{j}{}^{,b} {F_4}^{ehij,g} \
{F_4}_{fghi}{}^{,a} \nonumber \\ 
&& + c_9 {F_4}^{aghi,j} {F_4}_{bcde,a} \
{F_4}^{cde}{}_{j}{}^{,f} {F_4}_{fghi}{}^{,b} + c_{10} {F_4}^{aehi,j} {F_4}_{bcde,a} {F_4}^{cdg}{}_{j}{}^{,f} \
{F_4}_{fghi}{}^{,b} \nonumber \\ 
&& + c_{11} {F_4}^{acd}{}_{j}{}^{,f} {F_4}_{bcde,a} \
{F_4}^{ehij,g} {F_4}_{fghi}{}^{,b} + c_{12} {F_4}_{bcde,a} \
{F_4}^{b}{}_{fgh}{}^{,a} {F_4}^{cde}{}_{j,i} {F_4}^{fghi,j} \nonumber \\ 
&& + c_{13} {F_4}_{bcde,a} {F_4}^{cdej,i} {F_4}_{fghi}{}^{,a} \
{F_4}^{fgh}{}_{j}{}^{,b} + c_{14} {F_4}^{ac}{}_{fg}{}^{,b} \
{F_4}_{bcde,a} {F_4}^{de}{}_{ij,h} {F_4}^{fghj,i} \nonumber \\ 
&&+ c_{15} {F_4}_{bcde,a} \
{F_4}^{b}{}_{fgh}{}^{,a} {F_4}^{ehij,d} {F_4}^{fg}{}_{ij}{}^{,c}  + c_{16} {F_4}^{aeij,h} {F_4}_{bcde,a} {F_4}^{c}{}_{fgh}{}^{,b} \
{F_4}^{fg}{}_{ij}{}^{,d} \nonumber \\ 
&&+ c_{17} {F_4}^{ac}{}_{fg}{}^{,b} \
{F_4}_{bcde,a} {F_4}^{e}{}_{hij}{}^{,d} {F_4}^{fgij,h}  + c_{18} {F_4}_{bcde,a} {F_4}^{bc}{}_{fg}{}^{,a} {F_4}^{ehij,g} \
{F_4}^{f}{}_{hij}{}^{,d} \nonumber \\ 
&&+ c_{19} {F_4}_{bcde,a} \
{F_4}^{bcd}{}_{f}{}^{,a} {F_4}^{e}{}_{hij,g} {F_4}^{fhij,g}  + c_{20} {F_4}^{acde,j} {F_4}_{bcde,a} {F_4}_{fghi}{}^{,b} \
{F_4}^{ghi}{}_{j}{}^{,f} \nonumber \\ 
&&+ c_{21} {F_4}_{bcde,a} \
{F_4}^{b}{}_{fgh}{}^{,a} {F_4}^{cde}{}_{j,i} {F_4}^{ghij,f} + c_{22} \
{F_4}_{bcde,a} {F_4}^{bc}{}_{fg}{}^{,a} {F_4}^{e}{}_{hij}{}^{,d} {F_4}^{ghij,f} \nonumber \\ 
&& + c_{23} {F_4}_{bcde,a} {F_4}^{bcd}{}_{f}{}^{,a} {F_4}_{ghij}{}^{,e} \
{F_4}^{ghij,f} + c_{24} {F_4}_{bcde,a} {F_4}^{bcde,a} {F_4}_{ghij,f} \
{F_4}^{ghij,f}.\label{F4F4F4F4}
\eea

\subsection{The $ ({\partial {G_5}})^4 $ basis}\label{G54}

The basis for the $ ({\partial {G_5}})^4 $ terms consists, at least at on-shell level, of 109 elements:
\bea
&& d_1 {G_5}_{abf}{}^{gh,i} {G_5}^{abcde,f} {G_5}_{cgi}{}^{jk,l}\
{G_5}_{dhjkl,e} + d_2 {G_5}_{ab}{}^{ghi,j} {G_5}^{abcde,f} \
{G_5}_{cfg}{}^{kl}{}_{,h} {G_5}_{dijkl,e} \nonumber \\ 
&&+ d_3 {G_5}_{ab}{}^{ghi,j}
{G_5}^{abcde,f} {G_5}_{cdgj}{}^{k,l} {G_5}_{efhik,l}  + d_4 {G_5}_{ab}{}^{ghi,j} {G_5}^{abcde,f} {G_5}_{cdgj}{}^{k,l}\
{G_5}_{efhil,k} \nonumber \\ 
&&+ d_5 {G_5}_{ab}{}^{ghi,j} {G_5}^{abcde,f}
{G_5}_{cdgj}{}^{k,l} {G_5}_{efhkl,i} + d_6 {G_5}_{ab}{}^{ghi,j} \
{G_5}^{abcde,f} {G_5}_{cdgh}{}^{k,l} {G_5}_{efijk,l} \nonumber \\ 
&& + d_7 {G_5}_{ab}{}^{ghi,j} {G_5}^{abcde,f} {G_5}_{cdgh}{}^{k,l} \
{G_5}_{efijl,k} + d_8 {G_5}_{abf}{}^{gh,i} {G_5}^{abcde,f} \
{G_5}_{cdi}{}^{jk,l} {G_5}_{eghjk,l} \nonumber \\ 
&&+ d_9 {G_5}_{abf}{}^{gh,i} \
{G_5}^{abcde,f} {G_5}_{cd}{}^{jkl}{}_{,i} {G_5}_{eghjk,l}  + d_{10} {G_5}_{abf}{}^{gh,i} {G_5}^{abcde,f} {G_5}_{cdi}{}^{jk,l} \
{G_5}_{eghjl,k} \nonumber \\ 
&&+ d_{11} {G_5}_{abc}{}^{gh,i} {G_5}^{abcde,f} \
{G_5}_{dfi}{}^{jk,l} {G_5}_{egjkl,h} + d_{12} {G_5}_{abc}{}^{gh,i} \
{G_5}^{abcde,f} {G_5}_{df}{}^{jkl}{}_{,i} {G_5}_{egjkl,h} \nonumber \\ 
&& + d_{13} {G_5}_{abf}{}^{gh,i} {G_5}^{abcde,f} {G_5}_{cdg}{}^{jk,l} \
{G_5}_{ehijk,l} + d_{14} {G_5}_{ab}{}^{ghi}{}_{,f} {G_5}^{abcde,f} \
{G_5}_{cdg}{}^{jk,l} {G_5}_{ehijk,l} \nonumber \\ 
&& + d_{15} {G_5}_{abf}{}^{gh,i} {G_5}^{abcde,f} \
{G_5}_{cd}{}^{jkl}{}_{,g} {G_5}_{ehijk,l} + d_{16} {G_5}_{abc}{}^{gh,i} \
{G_5}^{abcde,f} {G_5}_{dfg}{}^{jk,l} {G_5}_{ehijk,l} \nonumber \\ 
&&+ d_{17} \
{G_5}_{abcf}{}^{g,h} {G_5}^{abcde,f} {G_5}_{dg}{}^{ijk,l} {G_5}_{ehijk,l}  + d_{18} {G_5}_{abc}{}^{gh}{}_{,f} {G_5}^{abcde,f} \
{G_5}_{dg}{}^{ijk,l} {G_5}_{ehijk,l} \nonumber \\ 
&&+ d_{19} {G_5}_{abf}{}^{gh,i} \
{G_5}^{abcde,f} {G_5}_{cdg}{}^{jk,l} {G_5}_{ehijl,k}  + d_{20} {G_5}_{ab}{}^{ghi}{}_{,f} {G_5}^{abcde,f} \
{G_5}_{cdg}{}^{jk,l} {G_5}_{ehijl,k} \nonumber \\ 
&&+ d_{21} {G_5}_{abc}{}^{gh,i} \
{G_5}^{abcde,f} {G_5}_{dfg}{}^{jk,l} {G_5}_{ehijl,k} + d_{22} \
{G_5}_{abcf}{}^{g,h} {G_5}^{abcde,f} {G_5}_{dg}{}^{ijk,l} {G_5}_{ehijl,k} \nonumber \\ 
&& + d_{23} {G_5}_{abc}{}^{gh}{}_{,f} {G_5}^{abcde,f} \
{G_5}_{dg}{}^{ijk,l} {G_5}_{ehijl,k} + d_{24} {G_5}_{abf}{}^{gh}{}_{,c} \
{G_5}^{abcde,f} {G_5}_{dg}{}^{ijk,l} {G_5}_{ehijl,k} \nonumber \\ 
&& + d_{25} {G_5}_{ab}{}^{ghi,j} {G_5}^{abcde,f} {G_5}_{cdgj}{}^{k,l} \
{G_5}_{ehikl,f} + d_{26} {G_5}_{ab}{}^{ghi,j} {G_5}^{abcde,f} \
{G_5}_{cdg}{}^{kl}{}_{,j} {G_5}_{ehikl,f} \nonumber \\ 
&&+ d_{27} {G_5}_{abf}{}^{gh,i} \
{G_5}^{abcde,f} {G_5}_{cdg}{}^{jk,l} {G_5}_{ehjkl,i}  + d_{28} {G_5}_{ab}{}^{ghi,j} {G_5}^{abcde,f} \
{G_5}_{cfg}{}^{kl}{}_{,d} {G_5}_{ehjkl,i} \nonumber \\ 
&&+ d_{29} {G_5}_{abc}{}^{gh,i} \
{G_5}^{abcde,f} {G_5}_{dfg}{}^{jk,l} {G_5}_{ehjkl,i}  + d_{30} {G_5}_{abc}{}^{gh,i} {G_5}^{abcde,f} \
{G_5}_{df}{}^{jkl}{}_{,g} {G_5}_{ehjkl,i} \nonumber \\ 
&&+ d_{31} {G_5}_{abc}{}^{gh,i} \
{G_5}^{abcde,f} {G_5}_{dg}{}^{jkl}{}_{,f} {G_5}_{ehjkl,i}  + d_{32} {G_5}_{abf}{}^{gh,i} {G_5}^{abcde,f} {G_5}_{cdg}{}^{jk,l} \
{G_5}_{eijkl,h} \nonumber \\ 
&&+ d_{33} {G_5}_{abc}{}^{gh,i} {G_5}^{abcde,f} \
{G_5}_{dfg}{}^{jk,l} {G_5}_{eijkl,h} + d_{34} {G_5}_{abc}{}^{gh,i} \
{G_5}^{abcde,f} {G_5}_{df}{}^{jkl}{}_{,g} {G_5}_{eijkl,h} \nonumber \\ 
&& + d_{35} {G_5}_{abcf}{}^{g,h} {G_5}^{abcde,f} {G_5}_{dg}{}^{ijk,l} \
{G_5}_{eijkl,h} + d_{36} {G_5}_{abc}{}^{gh,i} {G_5}^{abcde,f} \
{G_5}_{dg}{}^{jkl}{}_{,f} {G_5}_{eijkl,h} \nonumber \\ 
&&+ d_{37} {G_5}_{abc}{}^{gh,i} \
{G_5}^{abcde,f} {G_5}_{dei}{}^{jk,l} {G_5}_{fghjk,l}  + d_{38} {G_5}_{abc}{}^{gh,i} {G_5}^{abcde,f} {G_5}_{dei}{}^{jk,l} \
{G_5}_{fghjl,k} \nonumber \\ 
&&+ d_{39} {G_5}_{abcd}{}^{g,h} {G_5}^{abcde,f} \
{G_5}_{eh}{}^{ijk,l} {G_5}_{fgijk,l} + d_{40} {G_5}_{abcd}{}^{g,h} \
{G_5}^{abcde,f} {G_5}_{eh}{}^{ijk,l} {G_5}_{fgijl,k} \nonumber \\ 
&& + d_{41} {G_5}_{abc}{}^{gh,i} {G_5}^{abcde,f} {G_5}_{dei}{}^{jk,l} \
{G_5}_{fgjkl,h} + d_{42} {G_5}_{abc}{}^{gh,i} {G_5}^{abcde,f} \
{G_5}_{deg}{}^{jk,l} {G_5}_{fhijk,l} \nonumber \\ 
&&+ d_{43} {G_5}_{abc}{}^{gh,i} \
{G_5}^{abcde,f} {G_5}_{de}{}^{jkl}{}_{,g} {G_5}_{fhijk,l}  + d_{44} {G_5}_{abcd}{}^{g,h} {G_5}^{abcde,f} {G_5}_{eg}{}^{ijk,l} \
{G_5}_{fhijk,l} \nonumber \\ 
&&+ d_{45} {G_5}_{abcd}{}^{g,h} {G_5}^{abcde,f} \
{G_5}_{e}{}^{ijkl}{}_{,g} {G_5}_{fhijk,l} + d_{46} {G_5}_{abc}{}^{gh,i} \
{G_5}^{abcde,f} {G_5}_{deg}{}^{jk,l} {G_5}_{fhijl,k} \nonumber \\ 
&& + d_{47} {G_5}_{abcd}{}^{g,h} {G_5}^{abcde,f} {G_5}_{eg}{}^{ijkl} \
{G_5}_{fhijl,k} + d_{48} {G_5}_{ab}{}^{ghi,j} {G_5}^{abcde,f} \
{G_5}_{cdgj}{}^{k,l} {G_5}_{fhikl,e} \nonumber \\ 
&&+ d_{49} {G_5}_{ab}{}^{ghi,j} \
{G_5}^{abcde,f} {G_5}_{cdg}{}^{kl}{}_{,j} {G_5}_{fhikl,e}  + d_{50} {G_5}_{ab}{}^{ghi,j} {G_5}^{abcde,f} \
{G_5}_{cdj}{}^{kl}{}_{,g} {G_5}_{fhikl,e} \nonumber \\ 
&&+ d_{51} {G_5}_{abc}{}^{gh,i} \
{G_5}^{abcde,f} {G_5}_{di}{}^{jkl}{}_{,g} {G_5}_{fhjkl,e}  + d_{52} {G_5}_{abc}{}^{gh,i} {G_5}^{abcde,f} {G_5}_{deg}{}^{jk,l} \
{G_5}_{fhjkl,i} \nonumber \\ 
&&+ d_{53} {G_5}_{abc}{}^{gh,i} {G_5}^{abcde,f} \
{G_5}_{de}{}^{jkl}{}_{,g} {G_5}_{fhjkl,i} + d_{54} {G_5}_{abcd}{}^{g,h} \
{G_5}^{abcde,f} {G_5}_{eh}{}^{ijk,l} {G_5}_{fijkl,g} \nonumber \\ 
&& + d_{55} {G_5}_{abc}{}^{gh,i} {G_5}^{abcde,f} {G_5}_{deg}{}^{jk,l} \
{G_5}_{fijkl,h} + d_{56} {G_5}_{abc}{}^{gh,i} {G_5}^{abcde,f} \
{G_5}_{de}{}^{jkl}{}_{,g} {G_5}_{fijkl,h} \nonumber \\ 
&&+ d_{57} {G_5}_{abcd}{}^{g,h} \
{G_5}^{abcde,f} {G_5}_{eg}{}^{ijk,l} {G_5}_{fijkl,h}  + d_{58} {G_5}_{abcd}{}^{g,h} {G_5}^{abcde,f} \
{G_5}_{e}{}^{ijkl}{}_{,g} {G_5}_{fijkl,h} \nonumber \\ 
&&+ d_{59} {G_5}_{abf}{}^{gh,i} \
{G_5}^{abcde,f} {G_5}_{cd}{}^{jkl}{}_{,e} {G_5}_{ghijk,l}  + d_{60} {G_5}_{abc}{}^{gh,i} {G_5}^{abcde,f} {G_5}_{def}{}^{jk,l} \
{G_5}_{ghijk,l} \nonumber \\ 
&&+ d_{61} {G_5}_{abcf}{}^{g,h} {G_5}^{abcde,f} \
{G_5}_{de}{}^{ijk,l} {G_5}_{ghijk,l} + d_{62} {G_5}_{abc}{}^{gh}{}_{,f} \
{G_5}^{abcde,f} {G_5}_{de}{}^{ijk,l} {G_5}_{ghijk,l} \nonumber \\ 
&& + d_{63} {G_5}_{abcf}{}^{g,h} {G_5}^{abcde,f} \
{G_5}_{d}{}^{ijkl}{}_{,e} {G_5}_{ghijk,l} + d_{64} {G_5}_{abcd}{}^{g,h} \
{G_5}^{abcde,f} {G_5}_{ef}{}^{ijk,l} {G_5}_{ghijk,l} \nonumber \\ 
&&+ d_{65} \
{G_5}_{abcdf}{}^{,g} {G_5}^{abcde,f} {G_5}_{e}{}^{hijk,l} {G_5}_{ghijk,l}  + d_{66} {G_5}_{abcd}{}^{g}{}_{,f} {G_5}^{abcde,f} \
{G_5}_{e}{}^{hijk,l} {G_5}_{ghijk,l} \nonumber \\ 
&&+ d_{67} {G_5}_{abcde}{}^{,g} \
{G_5}^{abcde,f} {G_5}_{f}{}^{hijk,l} {G_5}_{ghijk,l} + d_{68} \
{G_5}_{abc}{}^{gh,i} {G_5}^{abcde,f} {G_5}_{def}{}^{jk,l} {G_5}_{ghijl,k} \nonumber \\ 
&& + d_{69} {G_5}_{abcf}{}^{g,h} {G_5}^{abcde,f} {G_5}_{de}{}^{ijk,l} \
{G_5}_{ghijl,k} + d_{70} {G_5}_{abc}{}^{gh}{}_{,f} {G_5}^{abcde,f} \
{G_5}_{de}{}^{ijk,l} {G_5}_{ghijl,k} \nonumber \\ 
&& + d_{71} {G_5}_{abf}{}^{gh}{}_{,c} {G_5}^{abcde,f} \
{G_5}_{de}{}^{ijk,l} {G_5}_{ghijl,k} + d_{72} {G_5}_{abcd}{}^{g,h} \
{G_5}^{abcde,f} {G_5}_{ef}{}^{ijk,l} {G_5}_{ghijl,k} \nonumber \\ 
&&+ d_{73} \
{G_5}_{abcdf}{}^{,g} {G_5}^{abcde,f} {G_5}_{e}{}^{hijk,l} {G_5}_{ghijl,k} + d_{74} {G_5}_{abcd}{}^{g}{}_{,f} {G_5}^{abcde,f} \
{G_5}_{e}{}^{hijk,l} {G_5}_{ghijl,k} \nonumber \\ 
&& + d_{75} {G_5}_{abcf}{}^{g}{}_{,d} \
{G_5}^{abcde,f} {G_5}_{e}{}^{hijk,l} {G_5}_{ghijl,k}  + d_{76} {G_5}_{abcde}{}^{,g} {G_5}^{abcde,f} {G_5}_{f}{}^{hijk,l} \
{G_5}_{ghijl,k} \nonumber \\ 
&& + d_{77} {G_5}_{abf}{}^{gh,i} {G_5}^{abcde,f} \
{G_5}_{cdi}{}^{jk,l} {G_5}_{ghjkl,e} + d_{78} {G_5}_{abc}{}^{gh,i} \
{G_5}^{abcde,f} {G_5}_{dei}{}^{jk,l} {G_5}_{ghjkl,f} \nonumber \\ 
&& + d_{79} {G_5}_{abc}{}^{gh,i} {G_5}^{abcde,f} \
{G_5}_{de}{}^{jkl}{}_{,i} {G_5}_{ghjkl,f} + d_{80} {G_5}_{abc}{}^{gh,i} \
{G_5}^{abcde,f} {G_5}_{def}{}^{jk,l} {G_5}_{ghjkl,i} \nonumber \\ 
&& + d_{81} {G_5}_{abc}{}^{gh,i} {G_5}^{abcde,f} \
{G_5}_{de}{}^{jkl}{}_{,f} {G_5}_{ghjkl,i} + d_{82} {G_5}_{abcf}{}^{g,h} \
{G_5}^{abcde,f} {G_5}_{dh}{}^{ijk,l} {G_5}_{gijkl,e} \nonumber \\ 
&&+ d_{83} \
{G_5}_{abcd}{}^{g,h} {G_5}^{abcde,f} {G_5}_{eh}{}^{ijk,l} {G_5}_{gijkl,f}  + d_{84} {G_5}_{abcd}{}^{g,h} {G_5}^{abcde,f} \
{G_5}_{e}{}^{ijkl}{}_{,h} {G_5}_{gijkl,f} \nonumber \\ 
&&+ d_{85} {G_5}_{abc}{}^{gh,i} \
{G_5}^{abcde,f} {G_5}_{def}{}^{jk,l} {G_5}_{gijkl,h} + d_{86} \
{G_5}_{abcf}{}^{g,h} {G_5}^{abcde,f} {G_5}_{de}{}^{ijk,l} {G_5}_{gijkl,h} \nonumber \\ 
&& + d_{87} {G_5}_{abc}{}^{gh,i} {G_5}^{abcde,f} \
{G_5}_{de}{}^{jkl}{}_{,f} {G_5}_{gijkl,h} + d_{88} {G_5}_{abc}{}^{gh,i} \
{G_5}^{abcde,f} {G_5}_{df}{}^{jkl}{}_{,e} {G_5}_{gijkl,h} \nonumber \\ 
&& + d_{89} {G_5}_{abcd}{}^{g,h} {G_5}^{abcde,f} {G_5}_{ef}{}^{ijk,l} \
{G_5}_{gijkl,h} + d_{90} {G_5}_{abcd}{}^{g,h} {G_5}^{abcde,f} \
{G_5}_{e}{}^{ijkl}{}_{,f} {G_5}_{gijkl,h} \nonumber \\ 
&&+ d_{91} {G_5}_{abcde,f} \
{G_5}^{abcde,f} {G_5}_{ghijk,l} {G_5}^{ghijk,l}  + d_{92} {G_5}_{abcde,f} {G_5}^{abcde,f} {G_5}_{ghijl,k} {G_5}^{ghijk,l} \
\nonumber \\ 
&&+ d_{93} {G_5}_{abcdf,e} {G_5}^{abcde,f} {G_5}_{ghijl,k} {G_5}^{ghijk,l} + \
d_{94} {G_5}_{abf}{}^{gh,i} {G_5}^{abcde,f} {G_5}_{cdg}{}^{jk,l} \
{G_5}_{hijkl,e} \nonumber \\ 
&& + d_{95} {G_5}_{abc}{}^{gh,i} {G_5}^{abcde,f} {G_5}_{dfg}{}^{jk,l} \
{G_5}_{hijkl,e} + d_{96} {G_5}_{abc}{}^{gh,i} {G_5}^{abcde,f} \
{G_5}_{df}{}^{jkl}{}_{,g} {G_5}_{hijkl,e} \nonumber \\ 
&&+ d_{97} {G_5}_{abcf}{}^{g,h} \
{G_5}^{abcde,f} {G_5}_{dg}{}^{ijk,l} {G_5}_{hijkl,e}  + d_{98} {G_5}_{abcd}{}^{g,h} {G_5}^{abcde,f} \
{G_5}_{f}{}^{ijkl}{}_{,g} {G_5}_{hijkl,e} \nonumber \\ 
&&+ d_{99} {G_5}_{abcdf}{}^{,g} \
{G_5}^{abcde,f} {G_5}_{g}{}^{hijk,l} {G_5}_{hijkl,e} + d_{100} \
{G_5}_{abc}{}^{gh,i} {G_5}^{abcde,f} {G_5}_{deg}{}^{jk,l} {G_5}_{hijkl,f} \nonumber \\ 
&& + d_{101} {G_5}_{abc}{}^{gh,i} {G_5}^{abcde,f} \
{G_5}_{de}{}^{jkl}{}_{,g} {G_5}_{hijkl,f} + d_{102} \
{G_5}_{abcd}{}^{g,h} {G_5}^{abcde,f} {G_5}_{e}{}^{ijkl}{}_{,g} {G_5}_{hijkl,f} \nonumber \\ 
&& + d_{103} {G_5}_{abcf}{}^{g,h} {G_5}^{abcde,f} {G_5}_{de}{}^{ijk,l} \
{G_5}_{hijkl,g} + d_{104} {G_5}_{abcd}{}^{g,h} {G_5}^{abcde,f} \
{G_5}_{ef}{}^{ijk,l} {G_5}_{hijkl,g} \nonumber \\ 
&&+ d_{105} {G_5}_{abcdf}{}^{,g} \
{G_5}^{abcde,f} {G_5}_{e}{}^{hijk,l} {G_5}_{hijkl,g}  + d_{106} {G_5}_{abcd}{}^{g,h} {G_5}^{abcde,f} \
{G_5}_{e}{}^{ijkl}{}_{,f} {G_5}_{hijkl,g} \nonumber \\ 
&&+ d_{107} \
{G_5}_{abcde}{}^{,g} {G_5}^{abcde,f} {G_5}_{f}{}^{hijk,l} {G_5}_{hijkl,g}  + d_{108} {G_5}_{abcd}{}^{g,h} {G_5}^{abcde,f} \
{G_5}_{f}{}^{ijkl}{}_{,e} {G_5}_{hijkl,g} \nonumber \\ 
&&+ d_{109} \
{G_5}_{abcde}{}^{,g} {G_5}^{abcde,f} {G_5}_{hijkl,g} {G_5}^{hijkl}{}_{,f}.\label{G5G5G5G5}
\eea

\subsection{The $ ({\partial {G_5}})^2 ({\partial {F_4}})^2 $ basis}\label{G52F42}

There are 352 independent on-shell terms in $ ({\partial {G_5}})^2 ({\partial {F_4}})^2 $ basis which can be written as
\bea
&& e_1 {F_4}^{abcd,e} {F_4}^{fghi,j} {G_5}_{abefj}{}^{,k} {G_5}_{cdghk,i} + \
e_2 {F_4}^{abcd,e} {F_4}^{fghi,j} {G_5}_{abef}{}^{k}{}_{,j} {G_5}_{cdghk,i} \
\nonumber \\ 
&&+ e_3 {F_4}^{abcd,e} {F_4}^{fghi,j} {G_5}_{abej}{}^{k}{}_{,f} \
{G_5}_{cdghk,i}  + e_4 {F_4}^{abcd,e} {F_4}_{e}{}^{fgh,i} {G_5}_{abfi}{}^{j,k} \
{G_5}_{cdgjk,h} \nonumber \\ 
&&+ e_5 {F_4}^{abcd,e} {F_4}^{fghi,j} {G_5}_{abefg}{}^{,k} \
{G_5}_{cdhij,k} + e_6 {F_4}^{abcd,e} {F_4}_{e}{}^{fgh,i} \
{G_5}_{abfg}{}^{j,k} {G_5}_{cdhij,k} \nonumber \\ 
&& + e_7 {F_4}^{abcd,e} {F_4}^{fghi}{}_{,e} {G_5}_{abfg}{}^{j,k} \
{G_5}_{cdhij,k} + e_8 {F_4}^{abcd,e} {F_4}^{fghi,j} {G_5}_{abefg}{}^{,k} \
{G_5}_{cdhik,j} \nonumber \\ 
&&+ e_9 {F_4}^{abcd,e} {F_4}^{fghi,j} \
{G_5}_{abef}{}^{k}{}_{,g} {G_5}_{cdhik,j}  + e_{10} {F_4}^{abcd,e} {F_4}_{e}{}^{fgh,i} {G_5}_{abfg}{}^{j,k} \
{G_5}_{cdhik,j} \nonumber \\ 
&&+ e_{11} {F_4}^{abcd,e} {F_4}^{fghi}{}_{,e} \
{G_5}_{abfg}{}^{j,k} {G_5}_{cdhik,j} + e_{12} {F_4}^{abcd,e} {F_4}^{fghi,j} \
{G_5}_{abfg}{}^{k}{}_{,e} {G_5}_{cdhik,j} \nonumber \\ 
&& + e_{13} {F_4}^{abcd,e} {F_4}^{fghi,j} {G_5}_{abefg}{}^{,k} \
{G_5}_{cdhjk,i} + e_{14} {F_4}^{abcd,e} {F_4}^{fghi,j} \
{G_5}_{abef}{}^{k}{}_{,g} {G_5}_{cdhjk,i} \nonumber \\ 
&&+ e_{15} {F_4}^{abcd,e} \
{F_4}_{e}{}^{fgh,i} {G_5}_{abfg}{}^{j,k} {G_5}_{cdhjk,i}  + e_{16} {F_4}^{abcd,e} {F_4}^{fghi}{}_{,e} {G_5}_{abfg}{}^{j,k} \
{G_5}_{cdhjk,i} \nonumber \\ 
&&+ e_{17} {F_4}^{abcd,e} {F_4}^{fghi,j} \
{G_5}_{abfg}{}^{k}{}_{,e} {G_5}_{cdhjk,i} + e_{18} {F_4}^{abcd,e} \
{F_4}_{e}{}^{fgh,i} {G_5}_{abf}{}^{jk}{}_{,g} {G_5}_{cdhjk,i} \nonumber \\ 
&& + e_{19} {F_4}^{abcd,e} {F_4}^{fghi}{}_{,e} \
{G_5}_{abf}{}^{jk}{}_{,g} {G_5}_{cdhjk,i} + e_{20} {F_4}^{abcd,e} \
{F_4}_{e}{}^{fgh,i} {G_5}_{abfg}{}^{j,k} {G_5}_{cdijk,h} \nonumber \\ 
&&+ e_{21} \
{F_4}^{abcd,e} {F_4}_{e}{}^{fgh,i} {G_5}_{abf}{}^{jk}{}_{,g} {G_5}_{cdijk,h}  + e_{22} {F_4}^{abcd,e} {F_4}^{fghi,j} {G_5}_{abfj}{}^{k}{}_{,g} \
{G_5}_{cehik,d} \nonumber \\ 
&&+ e_{23} {F_4}^{abcd,e} {F_4}_{e}{}^{fgh,i} \
{G_5}_{abfi}{}^{j,k} {G_5}_{cghjk,d} + e_{24} {F_4}^{abcd,e} \
{F_4}_{e}{}^{fgh,i} {G_5}_{abf}{}^{jk}{}_{,i} {G_5}_{cghjk,d} \nonumber \\ 
&& + e_{25} {F_4}^{abcd,e} {F_4}_{e}{}^{fgh,i} \
{G_5}_{abi}{}^{jk}{}_{,f} {G_5}_{cghjk,d} + e_{26} {F_4}_{ae}{}^{fg,h} \
{F_4}^{abcd,e} {G_5}_{bfh}{}^{ij,k} {G_5}_{cgijk,d} \nonumber \\ 
&&+ e_{27} {F_4}^{abcd,e} \
{F_4}^{fghi,j} {G_5}_{abefg}{}^{,k} {G_5}_{chijk,d}  + e_{28} {F_4}^{abcd,e} {F_4}^{fghi,j} {G_5}_{abef}{}^{k}{}_{,g} \
{G_5}_{chijk,d} \nonumber \\ 
&&+ e_{29} {F_4}^{abcd,e} {F_4}_{e}{}^{fgh,i} \
{G_5}_{abfg}{}^{j,k} {G_5}_{chijk,d} + e_{30} {F_4}^{abcd,e} \
{F_4}_{e}{}^{fgh,i} {G_5}_{abf}{}^{jk}{}_{,g} {G_5}_{chijk,d} \nonumber \\ 
&& + e_{31} {F_4}^{abcd,e} {F_4}^{fghi}{}_{,e} \
{G_5}_{abf}{}^{jk}{}_{,g} {G_5}_{chijk,d} + e_{32} {F_4}^{abcd,e} \
{F_4}_{e}{}^{fgh,i} {G_5}_{afg}{}^{jk}{}_{,b} {G_5}_{chijk,d} \nonumber \\ 
&& + e_{33} {F_4}_{a}{}^{fgh,i} {F_4}^{abcd,e} \
{G_5}_{bef}{}^{jk}{}_{,g} {G_5}_{chijk,d} + e_{34} {F_4}_{ae}{}^{fg,h} \
{F_4}^{abcd,e} {G_5}_{bfg}{}^{ij,k} {G_5}_{chijk,d} \nonumber \\ 
&&+ e_{35} \
{F_4}_{ae}{}^{fg,h} {F_4}^{abcd,e} {G_5}_{bf}{}^{ijk}{}_{,g} {G_5}_{chijk,d}  + e_{36} {F_4}^{abcd,e} {F_4}^{fghi,j} {G_5}_{abcfj}{}^{,k} \
{G_5}_{deghi,k} \nonumber \\ 
&&+ e_{37} {F_4}_{a}{}^{fgh,i} {F_4}^{abcd,e} \
{G_5}_{bcfi}{}^{j,k} {G_5}_{deghj,k} + e_{38} {F_4}^{abcd,e} {F_4}^{fghi,j} \
{G_5}_{abcfj}{}^{,k} {G_5}_{deghk,i} \nonumber \\ 
&& + e_{39} {F_4}^{abcd,e} {F_4}^{fghi,j} {G_5}_{abcf}{}^{k}{}_{,j} \
{G_5}_{deghk,i} + e_{40} {F_4}^{abcd,e} {F_4}^{fghi,j} \
{G_5}_{abcj}{}^{k}{}_{,f} {G_5}_{deghk,i} \nonumber \\ 
&&+ e_{41} {F_4}^{abcd,e} \
{F_4}^{fghi,j} {G_5}_{abfj}{}^{k}{}_{,c} {G_5}_{deghk,i}  + e_{42} {F_4}_{a}{}^{fgh,i} {F_4}^{abcd,e} {G_5}_{bcfi}{}^{j,k} \
{G_5}_{deghk,j} \nonumber \\ 
&&+ e_{43} {F_4}_{a}{}^{fgh,i} {F_4}^{abcd,e} \
{G_5}_{bcfi}{}^{j,k} {G_5}_{degjk,h} + e_{44} {F_4}_{a}{}^{fgh,i} \
{F_4}^{abcd,e} {G_5}_{bcf}{}^{jk}{}_{,i} {G_5}_{degjk,h} \nonumber \\ 
&& + e_{45} {F_4}_{a}{}^{fgh,i} {F_4}^{abcd,e} \
{G_5}_{bci}{}^{jk}{}_{,f} {G_5}_{degjk,h} + e_{46} {F_4}^{abcd,e} \
{F_4}^{fghi,j} {G_5}_{abcfg}{}^{,k} {G_5}_{dehij,k} \nonumber \\ 
&&+ e_{47} {F_4}^{abcd,e} \
{F_4}^{fghi,j} {G_5}_{abcf}{}^{k}{}_{,g} {G_5}_{dehij,k}  + e_{48} {F_4}_{a}{}^{fgh,i} {F_4}^{abcd,e} {G_5}_{bcfg}{}^{j,k} \
{G_5}_{dehij,k} \nonumber \\ 
&&+ e_{49} {F_4}_{a}{}^{fgh,i} {F_4}^{abcd,e} \
{G_5}_{bcf}{}^{jk}{}_{,g} {G_5}_{dehij,k} + e_{50} {F_4}^{abcd,e} \
{F_4}^{fghi,j} {G_5}_{abcfg}{}^{,k} {G_5}_{dehik,j} \nonumber \\ 
&& + e_{51} {F_4}^{abcd,e} {F_4}^{fghi,j} {G_5}_{abcf}{}^{k}{}_{,g} \
{G_5}_{dehik,j} + e_{52} {F_4}^{abcd,e} {F_4}^{fghi,j} \
{G_5}_{abfg}{}^{k}{}_{,c} {G_5}_{dehik,j} \nonumber \\ 
&&+ e_{53} {F_4}_{a}{}^{fgh,i} \
{F_4}^{abcd,e} {G_5}_{bcfg}{}^{j,k} {G_5}_{dehik,j}  + e_{54} {F_4}^{abcd,e} {F_4}^{fghi,j} {G_5}_{abcfg}{}^{,k} \
{G_5}_{dehjk,i} \nonumber \\ 
&&+ e_{55} {F_4}^{abcd,e} {F_4}^{fghi,j} \
{G_5}_{abcf}{}^{k}{}_{,g} {G_5}_{dehjk,i} + e_{56} {F_4}^{abcd,e} \
{F_4}^{fghi,j} {G_5}_{abfg}{}^{k}{}_{,c} {G_5}_{dehjk,i} \nonumber \\ 
&& + e_{57} {F_4}_{a}{}^{fgh,i} {F_4}^{abcd,e} {G_5}_{bcfg}{}^{j,k} \
{G_5}_{dehjk,i} + e_{58} {F_4}_{a}{}^{fgh,i} {F_4}^{abcd,e} \
{G_5}_{bcf}{}^{jk}{}_{,g} {G_5}_{dehjk,i} \nonumber \\ 
&&+ e_{59} {F_4}_{a}{}^{fgh,i} \
{F_4}^{abcd,e} {G_5}_{bcfg}{}^{j,k} {G_5}_{deijk,h}  + e_{60} {F_4}_{a}{}^{fgh,i} {F_4}^{abcd,e} \
{G_5}_{bcf}{}^{jk}{}_{,g} {G_5}_{deijk,h} \nonumber \\ 
&&+ e_{61} {F_4}^{abcd,e} \
{F_4}_{e}{}^{fgh,i} {G_5}_{abci}{}^{j,k} {G_5}_{dfghj,k} + e_{62} \
{F_4}^{abcd,e} {F_4}_{e}{}^{fgh,i} {G_5}_{abc}{}^{jk}{}_{,i} {G_5}_{dfghj,k} \nonumber \\ 
&& + e_{63} {F_4}^{abcd,e} {F_4}_{e}{}^{fgh,i} \
{G_5}_{abi}{}^{jk}{}_{,c} {G_5}_{dfghj,k} + e_{64} {F_4}^{abcd,e} \
{F_4}^{fghi,j} {G_5}_{abcej}{}^{,k} {G_5}_{dfghk,i} \nonumber \\ 
&&+ e_{65} {F_4}^{abcd,e} \
{F_4}^{fghi,j} {G_5}_{abce}{}^{k}{}_{,j} {G_5}_{dfghk,i}  + e_{66} {F_4}^{abcd,e} {F_4}_{e}{}^{fgh,i} {G_5}_{abci}{}^{j,k} \
{G_5}_{dfghk,j} \nonumber \\ 
&&+ e_{67} {F_4}_{ae}{}^{fg,h} {F_4}^{abcd,e} \
{G_5}_{bch}{}^{ij,k} {G_5}_{dfgij,k} + e_{68} {F_4}_{ae}{}^{fg,h} \
{F_4}^{abcd,e} {G_5}_{bc}{}^{ijk}{}_{,h} {G_5}_{dfgij,k} \nonumber \\ 
&& + e_{69} {F_4}_{ae}{}^{fg,h} {F_4}^{abcd,e} {G_5}_{bch}{}^{ij,k} \
{G_5}_{dfgik,j} + e_{70} {F_4}^{abcd,e} {F_4}_{e}{}^{fgh,i} \
{G_5}_{abci}{}^{j,k} {G_5}_{dfgjk,h} \nonumber \\ 
&&+ e_{71} {F_4}^{abcd,e} \
{F_4}_{e}{}^{fgh,i} {G_5}_{abc}{}^{jk}{}_{,i} {G_5}_{dfgjk,h}  + e_{72} {F_4}^{abcd,e} {F_4}_{e}{}^{fgh,i} \
{G_5}_{abi}{}^{jk}{}_{,c} {G_5}_{dfgjk,h} \nonumber \\ 
&&+ e_{73} {F_4}_{a}{}^{fgh,i} \
{F_4}^{abcd,e} {G_5}_{bcei}{}^{j,k} {G_5}_{dfgjk,h} + e_{74} \
{F_4}_{a}{}^{fgh,i} {F_4}^{abcd,e} {G_5}_{bce}{}^{jk}{}_{,i} {G_5}_{dfgjk,h} \nonumber \\ 
&& + e_{75} {F_4}_{ae}{}^{fg,h} {F_4}^{abcd,e} {G_5}_{bch}{}^{ij,k} \
{G_5}_{dfijk,g} + e_{76} {F_4}_{ae}{}^{fg,h} {F_4}^{abcd,e} \
{G_5}_{bc}{}^{ijk}{}_{,h} {G_5}_{dfijk,g} \nonumber \\ 
&&+ e_{77} {F_4}_{ab}{}^{fg,h} \
{F_4}^{abcd,e} {G_5}_{ceh}{}^{ij,k} {G_5}_{dfijk,g}  + e_{78} {F_4}_{ab}{}^{fg,h} {F_4}^{abcd,e} \
{G_5}_{ce}{}^{ijk}{}_{,h} {G_5}_{dfijk,g} \nonumber \\ 
&&+ e_{79} {F_4}^{abcd,e} \
{F_4}^{fghi,j} {G_5}_{abcef}{}^{,k} {G_5}_{dghij,k} + e_{80} {F_4}^{abcd,e} \
{F_4}_{e}{}^{fgh,i} {G_5}_{abcf}{}^{j,k} {G_5}_{dghij,k} \nonumber \\ 
&& + e_{81} {F_4}^{abcd,e} {F_4}^{fghi}{}_{,e} {G_5}_{abcf}{}^{j,k} \
{G_5}_{dghij,k} + e_{82} {F_4}^{abcd,e} {F_4}_{e}{}^{fgh,i} \
{G_5}_{abc}{}^{jk}{}_{,f} {G_5}_{dghij,k} \nonumber \\ 
&&+ e_{83} {F_4}^{abcd,e} \
{F_4}_{e}{}^{fgh,i} {G_5}_{abf}{}^{jk}{}_{,c} {G_5}_{dghij,k}  + e_{84} {F_4}_{a}{}^{fgh,i} {F_4}^{abcd,e} {G_5}_{bcef}{}^{j,k} \
{G_5}_{dghij,k} \nonumber \\ 
&&+ e_{85} {F_4}_{ae}{}^{fg,h} {F_4}^{abcd,e} \
{G_5}_{bcf}{}^{ij,k} {G_5}_{dghij,k} + e_{86} {F_4}_{a}{}^{fgh}{}_{,e} \
{F_4}^{abcd,e} {G_5}_{bcf}{}^{ij,k} {G_5}_{dghij,k} \nonumber \\ 
&& + e_{87} {F_4}^{abcd,e} {F_4}_{e}{}^{fgh}{}_{,a} \
{G_5}_{bcf}{}^{ij,k} {G_5}_{dghij,k} + e_{88} {F_4}_{ae}{}^{fg,h} \
{F_4}^{abcd,e} {G_5}_{bc}{}^{ijk}{}_{,f} {G_5}_{dghij,k} \nonumber \\ 
&&+ e_{89} \
{F_4}_{ab}{}^{fg,h} {F_4}^{abcd,e} {G_5}_{cef}{}^{ij,k} {G_5}_{dghij,k}  + e_{90} {F_4}_{abe}{}^{f,g} {F_4}^{abcd,e} {G_5}_{cf}{}^{hij,k} \
{G_5}_{dghij,k} \nonumber \\ 
&&+ e_{91} {F_4}_{ab}{}^{fg}{}_{,e} {F_4}^{abcd,e} \
{G_5}_{cf}{}^{hij,k} {G_5}_{dghij,k} + e_{92} {F_4}_{ae}{}^{fg}{}_{,b} \
{F_4}^{abcd,e} {G_5}_{cf}{}^{hij,k} {G_5}_{dghij,k} \nonumber \\ 
&& + e_{93} {F_4}^{abcd,e} {F_4}^{fghi,j} {G_5}_{abcfj}{}^{,k} \
{G_5}_{dghik,e} + e_{94} {F_4}^{abcd,e} {F_4}^{fghi,j} \
{G_5}_{abcf}{}^{k}{}_{,j} {G_5}_{dghik,e} \nonumber \\ 
&&+ e_{95} {F_4}^{abcd,e} \
{F_4}^{fghi,j} {G_5}_{abcef}{}^{,k} {G_5}_{dghik,j}  + e_{96} {F_4}^{abcd,e} {F_4}^{fghi,j} {G_5}_{abce}{}^{k}{}_{,f} \
{G_5}_{dghik,j} \nonumber \\ 
&&+ e_{97} {F_4}^{abcd,e} {F_4}_{e}{}^{fgh,i} \
{G_5}_{abcf}{}^{j,k} {G_5}_{dghik,j} + e_{98} {F_4}^{abcd,e} \
{F_4}^{fghi}{}_{,e} {G_5}_{abcf}{}^{j,k} {G_5}_{dghik,j} \nonumber \\ 
&& + e_{99} {F_4}^{abcd,e} {F_4}^{fghi,j} {G_5}_{abcf}{}^{k}{}_{,e} \
{G_5}_{dghik,j} + e_{100} {F_4}_{a}{}^{fgh,i} {F_4}^{abcd,e} \
{G_5}_{bcef}{}^{j,k} {G_5}_{dghik,j} \nonumber \\ 
&&+ e_{101} {F_4}_{ae}{}^{fg,h} \
{F_4}^{abcd,e} {G_5}_{bcf}{}^{ij,k} {G_5}_{dghik,j}  + e_{102} {F_4}_{a}{}^{fgh}{}_{,e} {F_4}^{abcd,e} \
{G_5}_{bcf}{}^{ij,k} {G_5}_{dghik,j} \nonumber \\ 
&&+ e_{103} {F_4}^{abcd,e} \
{F_4}_{e}{}^{fgh}{}_{,a} {G_5}_{bcf}{}^{ij,k} {G_5}_{dghik,j}  + e_{104} {F_4}_{ab}{}^{fg,h} {F_4}^{abcd,e} {G_5}_{cef}{}^{ij,k} \
{G_5}_{dghik,j} \nonumber \\ 
&&+ e_{105} {F_4}_{abe}{}^{f,g} {F_4}^{abcd,e} \
{G_5}_{cf}{}^{hij,k} {G_5}_{dghik,j} + e_{106} {F_4}_{ab}{}^{fg}{}_{,e} \
{F_4}^{abcd,e} {G_5}_{cf}{}^{hij,k} {G_5}_{dghik,j} \nonumber \\ 
&& + e_{107} {F_4}_{ae}{}^{fg}{}_{,b} {F_4}^{abcd,e} \
{G_5}_{cf}{}^{hij,k} {G_5}_{dghik,j} + e_{108} {F_4}_{a}{}^{fgh,i} \
{F_4}^{abcd,e} {G_5}_{bcfi}{}^{j,k} {G_5}_{dghjk,e} \nonumber \\ 
&&+ e_{109} \
{F_4}_{a}{}^{fgh,i} {F_4}^{abcd,e} {G_5}_{bcf}{}^{jk}{}_{,i} {G_5}_{dghjk,e}  + e_{110} {F_4}^{abcd,e} {F_4}^{fghi,j} {G_5}_{abcef}{}^{,k} \
{G_5}_{dghjk,i} \nonumber \\ 
&&+ e_{111} {F_4}^{abcd,e} {F_4}^{fghi,j} \
{G_5}_{abce}{}^{k}{}_{,f} {G_5}_{dghjk,i} + e_{112} {F_4}^{abcd,e} \
{F_4}_{e}{}^{fgh,i} {G_5}_{abcf}{}^{j,k} {G_5}_{dghjk,i} \nonumber \\ 
&& + e_{113} {F_4}^{abcd,e} {F_4}^{fghi}{}_{,e} {G_5}_{abcf}{}^{j,k} \
{G_5}_{dghjk,i} + e_{114} {F_4}^{abcd,e} {F_4}^{fghi,j} \
{G_5}_{abcf}{}^{k}{}_{,e} {G_5}_{dghjk,i} \nonumber \\ 
&&+ e_{115} {F_4}^{abcd,e} \
{F_4}_{e}{}^{fgh,i} {G_5}_{abc}{}^{jk}{}_{,f} {G_5}_{dghjk,i}  + e_{116} {F_4}^{abcd,e} {F_4}^{fghi}{}_{,e} \
{G_5}_{abc}{}^{jk}{}_{,f} {G_5}_{dghjk,i} \nonumber \\ 
&&+ e_{117} {F_4}^{abcd,e} \
{F_4}^{fghi,j} {G_5}_{abef}{}^{k}{}_{,c} {G_5}_{dghjk,i} + e_{118} \
{F_4}^{abcd,e} {F_4}_{e}{}^{fgh,i} {G_5}_{abf}{}^{jk}{}_{,c} {G_5}_{dghjk,i} \nonumber \\ 
&& + e_{119} {F_4}^{abcd,e} {F_4}^{fghi}{}_{,e} \
{G_5}_{abf}{}^{jk}{}_{,c} {G_5}_{dghjk,i} + e_{120} {F_4}_{a}{}^{fgh,i} \
{F_4}^{abcd,e} {G_5}_{bcef}{}^{j,k} {G_5}_{dghjk,i} \nonumber \\ 
&&+ e_{121} \
{F_4}_{a}{}^{fgh,i} {F_4}^{abcd,e} {G_5}_{bce}{}^{jk}{}_{,f} {G_5}_{dghjk,i}  + e_{122} {F_4}_{a}{}^{fgh,i} {F_4}^{abcd,e} \
{G_5}_{bcf}{}^{jk}{}_{,e} {G_5}_{dghjk,i} \nonumber \\ 
&&+ e_{123} {F_4}^{abcd,e} \
{F_4}_{e}{}^{fgh,i} {G_5}_{abcf}{}^{j,k} {G_5}_{dgijk,h} + e_{124} \
{F_4}^{abcd,e} {F_4}_{e}{}^{fgh,i} {G_5}_{abc}{}^{jk}{}_{,f} {G_5}_{dgijk,h}  \nonumber \\ 
&&+ e_{125} {F_4}^{abcd,e} {F_4}_{e}{}^{fgh,i} \
{G_5}_{abf}{}^{jk}{}_{,c} {G_5}_{dgijk,h} + e_{126} {F_4}_{a}{}^{fgh,i} \
{F_4}^{abcd,e} {G_5}_{bcef}{}^{j,k} {G_5}_{dgijk,h} \nonumber \\ 
&&+ e_{127} \
{F_4}_{a}{}^{fgh,i} {F_4}^{abcd,e} {G_5}_{bce}{}^{jk}{}_{,f} {G_5}_{dgijk,h}  + e_{128} {F_4}_{ae}{}^{fg,h} {F_4}^{abcd,e} {G_5}_{bcf}{}^{ij,k} \
{G_5}_{dgijk,h} \nonumber \\ 
&&+ e_{129} {F_4}_{a}{}^{fgh}{}_{,e} {F_4}^{abcd,e} \
{G_5}_{bcf}{}^{ij,k} {G_5}_{dgijk,h} + e_{130} {F_4}^{abcd,e} \
{F_4}_{e}{}^{fgh}{}_{,a} {G_5}_{bcf}{}^{ij,k} {G_5}_{dgijk,h} \nonumber \\ 
&& + e_{131} {F_4}_{a}{}^{fgh,i} {F_4}^{abcd,e} \
{G_5}_{bcf}{}^{jk}{}_{,e} {G_5}_{dgijk,h} + e_{132} {F_4}_{ae}{}^{fg,h} \
{F_4}^{abcd,e} {G_5}_{bc}{}^{ijk}{}_{,f} {G_5}_{dgijk,h} \nonumber \\ 
&& + e_{133} {F_4}_{a}{}^{fgh}{}_{,e} {F_4}^{abcd,e} \
{G_5}_{bc}{}^{ijk}{}_{,f} {G_5}_{dgijk,h} + e_{134} {F_4}^{abcd,e} \
{F_4}_{e}{}^{fgh}{}_{,a} {G_5}_{bc}{}^{ijk}{}_{,f} {G_5}_{dgijk,h} \nonumber \\ 
&& + e_{135} {F_4}_{a}{}^{fgh,i} {F_4}^{abcd,e} \
{G_5}_{bef}{}^{jk}{}_{,c} {G_5}_{dgijk,h} + e_{136} {F_4}_{ae}{}^{fg,h} \
{F_4}^{abcd,e} {G_5}_{bf}{}^{ijk}{}_{,c} {G_5}_{dgijk,h} \nonumber \\ 
&& + e_{137} {F_4}_{a}{}^{fgh}{}_{,e} {F_4}^{abcd,e} \
{G_5}_{bf}{}^{ijk}{}_{,c} {G_5}_{dgijk,h} + e_{138} {F_4}^{abcd,e} \
{F_4}_{e}{}^{fgh}{}_{,a} {G_5}_{bf}{}^{ijk}{}_{,c} {G_5}_{dgijk,h} \nonumber \\ 
&& + e_{139} {F_4}_{ab}{}^{fg,h} {F_4}^{abcd,e} {G_5}_{cef}{}^{ij,k} \
{G_5}_{dgijk,h} + e_{140} {F_4}_{ab}{}^{fg,h} {F_4}^{abcd,e} \
{G_5}_{ce}{}^{ijk}{}_{,f} {G_5}_{dgijk,h} \nonumber \\ 
&&+ e_{141} {F_4}_{ab}{}^{fg,h} \
{F_4}^{abcd,e} {G_5}_{cf}{}^{ijk}{}_{,e} {G_5}_{dgijk,h}  + e_{142} {F_4}^{abcd,e} {F_4}^{fghi,j} {G_5}_{abcfg}{}^{,k} \
{G_5}_{dhijk,e} \nonumber \\ 
&&+ e_{143} {F_4}^{abcd,e} {F_4}^{fghi,j} \
{G_5}_{abcf}{}^{k}{}_{,g} {G_5}_{dhijk,e} + e_{144} {F_4}_{a}{}^{fgh,i} \
{F_4}^{abcd,e} {G_5}_{bcf}{}^{jk}{}_{,g} {G_5}_{dhijk,e} \nonumber \\ 
&& + e_{145} {F_4}_{abe}{}^{f,g} {F_4}^{abcd,e} {G_5}_{cg}{}^{hij,k} \
{G_5}_{dhijk,f} + e_{146} {F_4}_{ae}{}^{fg,h} {F_4}^{abcd,e} \
{G_5}_{bcf}{}^{ij,k} {G_5}_{dhijk,g} \nonumber \\ 
&&+ e_{147} {F_4}_{ae}{}^{fg,h} \
{F_4}^{abcd,e} {G_5}_{bc}{}^{ijk}{}_{,f} {G_5}_{dhijk,g}  + e_{148} {F_4}_{ae}{}^{fg,h} {F_4}^{abcd,e} \
{G_5}_{bf}{}^{ijk}{}_{,c} {G_5}_{dhijk,g} \nonumber \\ 
&&+ e_{149} {F_4}_{ab}{}^{fg,h} \
{F_4}^{abcd,e} {G_5}_{cef}{}^{ij,k} {G_5}_{dhijk,g} + e_{150} \
{F_4}_{ab}{}^{fg,h} {F_4}^{abcd,e} {G_5}_{ce}{}^{ijk}{}_{,f} {G_5}_{dhijk,g} \nonumber \\ 
&& + e_{151} {F_4}_{abe}{}^{f,g} {F_4}^{abcd,e} {G_5}_{cf}{}^{hij,k} \
{G_5}_{dhijk,g} + e_{152} {F_4}_{ab}{}^{fg}{}_{,e} {F_4}^{abcd,e} \
{G_5}_{cf}{}^{hij,k} {G_5}_{dhijk,g} \nonumber \\ 
&&+ e_{153} {F_4}_{ae}{}^{fg}{}_{,b} \
{F_4}^{abcd,e} {G_5}_{cf}{}^{hij,k} {G_5}_{dhijk,g}  + e_{154} {F_4}_{ab}{}^{fg,h} {F_4}^{abcd,e} \
{G_5}_{cf}{}^{ijk}{}_{,e} {G_5}_{dhijk,g} \nonumber \\ 
&&+ e_{155} {F_4}_{abe}{}^{f,g} \
{F_4}^{abcd,e} {G_5}_{c}{}^{hijk}{}_{,f} {G_5}_{dhijk,g}  + e_{156} {F_4}_{ab}{}^{fg}{}_{,e} {F_4}^{abcd,e} \
{G_5}_{c}{}^{hijk}{}_{,f} {G_5}_{dhijk,g} \nonumber \\ 
&&+ e_{157} \
{F_4}_{ae}{}^{fg}{}_{,b} {F_4}^{abcd,e} {G_5}_{c}{}^{hijk}{}_{,f} {G_5}_{dhijk,g}  + e_{158} {F_4}^{abcd,e} {F_4}^{fghi,j} {G_5}_{abcdj}{}^{,k} \
{G_5}_{efghi,k} \nonumber \\ 
&&+ e_{159} {F_4}_{a}{}^{fgh,i} {F_4}^{abcd,e} \
{G_5}_{bcdi}{}^{j,k} {G_5}_{efghj,k} + e_{160} {F_4}^{abcd,e} \
{F_4}^{fghi,j} {G_5}_{abcdj}{}^{,k} {G_5}_{efghk,i} \nonumber \\ 
&& + e_{161} {F_4}^{abcd,e} {F_4}^{fghi,j} {G_5}_{abcd}{}^{k}{}_{,j} \
{G_5}_{efghk,i} + e_{162} {F_4}^{abcd,e} {F_4}^{fghi,j} \
{G_5}_{abcj}{}^{k}{}_{,d} {G_5}_{efghk,i} \nonumber \\ 
&&+ e_{163} {F_4}_{a}{}^{fgh,i} \
{F_4}^{abcd,e} {G_5}_{bcdi}{}^{j,k} {G_5}_{efghk,j}  + e_{164} {F_4}_{ab}{}^{fg,h} {F_4}^{abcd,e} {G_5}_{cdh}{}^{ij,k} \
{G_5}_{efgij,k} \nonumber \\ 
&&+ e_{165} {F_4}_{ab}{}^{fg,h} {F_4}^{abcd,e} \
{G_5}_{cdh}{}^{ij,k} {G_5}_{efgik,j} + e_{166} {F_4}_{a}{}^{fgh,i} \
{F_4}^{abcd,e} {G_5}_{bcdi}{}^{j,k} {G_5}_{efgjk,h} \nonumber \\ 
&& + e_{167} {F_4}_{a}{}^{fgh,i} {F_4}^{abcd,e} \
{G_5}_{bcd}{}^{jk}{}_{,i} {G_5}_{efgjk,h} + e_{168} {F_4}_{a}{}^{fgh,i} \
{F_4}^{abcd,e} {G_5}_{bci}{}^{jk}{}_{,d} {G_5}_{efgjk,h} \nonumber \\ 
&& + e_{169} {F_4}_{abc}{}^{f,g} {F_4}^{abcd,e} {G_5}_{dg}{}^{hij,k} \
{G_5}_{efhij,k} + e_{170} {F_4}_{abc}{}^{f,g} {F_4}^{abcd,e} \
{G_5}_{dg}{}^{hij,k} {G_5}_{efhik,j} \nonumber \\ 
&&+ e_{171} {F_4}_{ab}{}^{fg,h} \
{F_4}^{abcd,e} {G_5}_{cdh}{}^{ij,k} {G_5}_{efijk,g}  + e_{172} {F_4}_{ab}{}^{fg,h} {F_4}^{abcd,e} \
{G_5}_{cd}{}^{ijk}{}_{,h} {G_5}_{efijk,g} \nonumber \\ 
&&+ e_{173} {F_4}_{ab}{}^{fg,h} \
{F_4}^{abcd,e} {G_5}_{ch}{}^{ijk}{}_{,d} {G_5}_{efijk,g} + e_{174} \
{F_4}^{abcd,e} {F_4}^{fghi,j} {G_5}_{abcdf}{}^{,k} {G_5}_{eghij,k} \nonumber \\ 
&& + e_{175} {F_4}^{abcd,e} {F_4}^{fghi,j} {G_5}_{abcd}{}^{k}{}_{,f} \
{G_5}_{eghij,k} + e_{176} {F_4}_{a}{}^{fgh,i} {F_4}^{abcd,e} \
{G_5}_{bcdf}{}^{j,k} {G_5}_{eghij,k} \nonumber \\ 
&&+ e_{177} {F_4}_{a}{}^{fgh,i} \
{F_4}^{abcd,e} {G_5}_{bcd}{}^{jk}{}_{,f} {G_5}_{eghij,k}  + e_{178} {F_4}_{ab}{}^{fg,h} {F_4}^{abcd,e} {G_5}_{cdf}{}^{ij,k} \
{G_5}_{eghij,k} \nonumber \\ 
&&+ e_{179} {F_4}_{ab}{}^{fg,h} {F_4}^{abcd,e} \
{G_5}_{cd}{}^{ijk}{}_{,f} {G_5}_{eghij,k} + e_{180} {F_4}_{abc}{}^{f,g} \
{F_4}^{abcd,e} {G_5}_{df}{}^{hij,k} {G_5}_{eghij,k} \nonumber \\ 
&& + e_{181} {F_4}_{abc}{}^{f,g} {F_4}^{abcd,e} \
{G_5}_{d}{}^{hijk}{}_{,f} {G_5}_{eghij,k} + e_{182} {F_4}^{abcd,e} \
{F_4}^{fghi,j} {G_5}_{abcfj}{}^{,k} {G_5}_{eghik,d} \nonumber \\ 
&&+ e_{183} \
{F_4}^{abcd,e} {F_4}^{fghi,j} {G_5}_{abcf}{}^{k}{}_{,j} {G_5}_{eghik,d}  + e_{184} {F_4}^{abcd,e} {F_4}^{fghi,j} {G_5}_{abcj}{}^{k}{}_{,f} \
{G_5}_{eghik,d} \nonumber \\ 
&&+ e_{185} {F_4}^{abcd,e} {F_4}^{fghi,j} \
{G_5}_{abcdf}{}^{,k} {G_5}_{eghik,j} + e_{186} {F_4}^{abcd,e} \
{F_4}^{fghi,j} {G_5}_{abcd}{}^{k}{}_{,f} {G_5}_{eghik,j} \nonumber \\ 
&& + e_{187} {F_4}^{abcd,e} {F_4}^{fghi,j} {G_5}_{abcf}{}^{k}{}_{,d} \
{G_5}_{eghik,j} + e_{188} {F_4}_{a}{}^{fgh,i} {F_4}^{abcd,e} \
{G_5}_{bcdf}{}^{j,k} {G_5}_{eghik,j} \nonumber \\ 
&&+ e_{189} {F_4}_{ab}{}^{fg,h} \
{F_4}^{abcd,e} {G_5}_{cdf}{}^{ij,k} {G_5}_{eghik,j}  + e_{190} {F_4}_{abc}{}^{f,g} {F_4}^{abcd,e} {G_5}_{df}{}^{hij,k} \
{G_5}_{eghik,j} \nonumber \\ 
&&+ e_{191} {F_4}_{a}{}^{fgh,i} {F_4}^{abcd,e} \
{G_5}_{bcfi}{}^{j,k} {G_5}_{eghjk,d} + e_{192} {F_4}_{a}{}^{fgh,i} \
{F_4}^{abcd,e} {G_5}_{bcf}{}^{jk}{}_{,i} {G_5}_{eghjk,d} \nonumber \\ 
&& + e_{193} {F_4}_{a}{}^{fgh,i} {F_4}^{abcd,e} \
{G_5}_{bci}{}^{jk}{}_{,f} {G_5}_{eghjk,d} + e_{194} {F_4}^{abcd,e} \
{F_4}^{fghi,j} {G_5}_{abcdf}{}^{,k} {G_5}_{eghjk,i} \nonumber \\ 
&&+ e_{195} \
{F_4}^{abcd,e} {F_4}^{fghi,j} {G_5}_{abcd}{}^{k}{}_{,f} {G_5}_{eghjk,i} + e_{196} {F_4}^{abcd,e} {F_4}^{fghi,j} {G_5}_{abcf}{}^{k}{}_{,d} \
{G_5}_{eghjk,i} \nonumber \\ 
&&+ e_{197} {F_4}_{a}{}^{fgh,i} {F_4}^{abcd,e} \
{G_5}_{bcdf}{}^{j,k} {G_5}_{eghjk,i} + e_{198} {F_4}_{a}{}^{fgh,i} \
{F_4}^{abcd,e} {G_5}_{bcd}{}^{jk}{}_{,f} {G_5}_{eghjk,i} \nonumber \\ 
&& + e_{199} {F_4}_{a}{}^{fgh,i} {F_4}^{abcd,e} \
{G_5}_{bcf}{}^{jk}{}_{,d} {G_5}_{eghjk,i} + e_{200} {F_4}_{ab}{}^{fg,h} \
{F_4}^{abcd,e} {G_5}_{ch}{}^{ijk}{}_{,f} {G_5}_{egijk,d} \nonumber \\ 
&& + e_{201} {F_4}_{a}{}^{fgh,i} {F_4}^{abcd,e} {G_5}_{bcdf}{}^{j,k} \
{G_5}_{egijk,h} + e_{202} {F_4}_{a}{}^{fgh,i} {F_4}^{abcd,e} \
{G_5}_{bcd}{}^{jk}{}_{,f} {G_5}_{egijk,h} \nonumber \\ 
&&+ e_{203} {F_4}_{a}{}^{fgh,i} \
{F_4}^{abcd,e} {G_5}_{bcf}{}^{jk}{}_{,d} {G_5}_{egijk,h}  + e_{204} {F_4}_{ab}{}^{fg,h} {F_4}^{abcd,e} {G_5}_{cdf}{}^{ij,k} \
{G_5}_{egijk,h} \nonumber \\ 
&&+ e_{205} {F_4}_{ab}{}^{fg,h} {F_4}^{abcd,e} \
{G_5}_{cd}{}^{ijk}{}_{,f} {G_5}_{egijk,h} + e_{206} {F_4}_{ab}{}^{fg,h} \
{F_4}^{abcd,e} {G_5}_{cf}{}^{ijk}{}_{,d} {G_5}_{egijk,h} \nonumber \\ 
&& + e_{207} {F_4}^{abcd,e} {F_4}^{fghi,j} {G_5}_{abcfg}{}^{,k} \
{G_5}_{ehijk,d} + e_{208} {F_4}^{abcd,e} {F_4}^{fghi,j} \
{G_5}_{abcf}{}^{k}{}_{,g} {G_5}_{ehijk,d} \nonumber \\ 
&&+ e_{209} {F_4}_{a}{}^{fgh,i} \
{F_4}^{abcd,e} {G_5}_{bcf}{}^{jk}{}_{,g} {G_5}_{ehijk,d}  + e_{210} {F_4}_{abc}{}^{f,g} {F_4}^{abcd,e} {G_5}_{dg}{}^{hij,k} \
{G_5}_{ehijk,f} \nonumber \\ 
&&+ e_{211} {F_4}_{abc}{}^{f,g} {F_4}^{abcd,e} \
{G_5}_{d}{}^{hijk}{}_{,g} {G_5}_{ehijk,f} + e_{212} {F_4}_{ab}{}^{fg,h} \
{F_4}^{abcd,e} {G_5}_{cdf}{}^{ij,k} {G_5}_{ehijk,g} \nonumber \\ 
&& + e_{213} {F_4}_{ab}{}^{fg,h} {F_4}^{abcd,e} \
{G_5}_{cd}{}^{ijk}{}_{,f} {G_5}_{ehijk,g} + e_{214} {F_4}_{ab}{}^{fg,h} \
{F_4}^{abcd,e} {G_5}_{cf}{}^{ijk}{}_{,d} {G_5}_{ehijk,g} \nonumber \\ 
&& + e_{215} {F_4}_{abc}{}^{f,g} {F_4}^{abcd,e} {G_5}_{df}{}^{hij,k} \
{G_5}_{ehijk,g} + e_{216} {F_4}_{abc}{}^{f,g} {F_4}^{abcd,e} \
{G_5}_{d}{}^{hijk}{}_{,f} {G_5}_{ehijk,g} \nonumber \\ 
&&+ e_{217} {F_4}^{abcd,e} \
{F_4}^{fghi,j} {G_5}_{abcde}{}^{,k} {G_5}_{fghij,k}  + e_{218} {F_4}^{abcd,e} {F_4}_{e}{}^{fgh,i} {G_5}_{abcd}{}^{j,k} \
{G_5}_{fghij,k} \nonumber \\ 
&&+ e_{219} {F_4}^{abcd,e} {F_4}^{fghi}{}_{,e} \
{G_5}_{abcd}{}^{j,k} {G_5}_{fghij,k} + e_{220} {F_4}^{abcd,e} \
{F_4}_{e}{}^{fgh,i} {G_5}_{abc}{}^{jk}{}_{,d} {G_5}_{fghij,k} \nonumber \\ 
&& + e_{221} {F_4}_{a}{}^{fgh,i} {F_4}^{abcd,e} {G_5}_{bcde}{}^{j,k} \
{G_5}_{fghij,k} + e_{222} {F_4}_{ae}{}^{fg,h} {F_4}^{abcd,e} \
{G_5}_{bcd}{}^{ij,k} {G_5}_{fghij,k} \nonumber \\ 
&&+ e_{223} {F_4}_{a}{}^{fgh}{}_{,e} \
{F_4}^{abcd,e} {G_5}_{bcd}{}^{ij,k} {G_5}_{fghij,k}  + e_{224} {F_4}^{abcd,e} {F_4}_{e}{}^{fgh}{}_{,a} \
{G_5}_{bcd}{}^{ij,k} {G_5}_{fghij,k} \nonumber \\ 
&&+ e_{225} {F_4}_{ae}{}^{fg,h} \
{F_4}^{abcd,e} {G_5}_{bc}{}^{ijk}{}_{,d} {G_5}_{fghij,k}  + e_{226} {F_4}_{ab}{}^{fg,h} {F_4}^{abcd,e} {G_5}_{cde}{}^{ij,k} \
{G_5}_{fghij,k} \nonumber \\ 
&&+ e_{227} {F_4}_{abe}{}^{f,g} {F_4}^{abcd,e} \
{G_5}_{cd}{}^{hij,k} {G_5}_{fghij,k} + e_{228} {F_4}_{ab}{}^{fg}{}_{,e} \
{F_4}^{abcd,e} {G_5}_{cd}{}^{hij,k} {G_5}_{fghij,k} \nonumber \\ 
&& + e_{229} {F_4}_{ae}{}^{fg}{}_{,b} {F_4}^{abcd,e} \
{G_5}_{cd}{}^{hij,k} {G_5}_{fghij,k} + e_{230} {F_4}_{abe}{}^{f,g} \
{F_4}^{abcd,e} {G_5}_{c}{}^{hijk}{}_{,d} {G_5}_{fghij,k} \nonumber \\ 
&& + e_{231} {F_4}_{abc}{}^{f,g} {F_4}^{abcd,e} {G_5}_{de}{}^{hij,k} \
{G_5}_{fghij,k} + e_{232} {F_4}_{abce}{}^{,f} {F_4}^{abcd,e} \
{G_5}_{d}{}^{ghij,k} {G_5}_{fghij,k} \nonumber \\ 
&&+ e_{233} {F_4}_{abc}{}^{f}{}_{,e} \
{F_4}^{abcd,e} {G_5}_{d}{}^{ghij,k} {G_5}_{fghij,k}  + e_{234} {F_4}_{abe}{}^{f}{}_{,c} {F_4}^{abcd,e} \
{G_5}_{d}{}^{ghij,k} {G_5}_{fghij,k} \nonumber \\ 
&&+ e_{235} {F_4}_{abcd}{}^{,f} \
{F_4}^{abcd,e} {G_5}_{e}{}^{ghij,k} {G_5}_{fghij,k} + e_{236} \
{F_4}^{abcd,e} {F_4}^{fghi,j} {G_5}_{abcdj}{}^{,k} {G_5}_{fghik,e} \nonumber \\ 
&& + e_{237} {F_4}^{abcd,e} {F_4}^{fghi,j} {G_5}_{abcd}{}^{k}{}_{,j} \
{G_5}_{fghik,e} + e_{238} {F_4}^{abcd,e} {F_4}^{fghi,j} \
{G_5}_{abcde}{}^{,k} {G_5}_{fghik,j} \nonumber \\ 
&&+ e_{239} {F_4}^{abcd,e} \
{F_4}_{e}{}^{fgh,i} {G_5}_{abcd}{}^{j,k} {G_5}_{fghik,j}  + e_{240} {F_4}^{abcd,e} {F_4}^{fghi}{}_{,e} {G_5}_{abcd}{}^{j,k} \
{G_5}_{fghik,j} \nonumber \\ 
&&+ e_{241} {F_4}^{abcd,e} {F_4}^{fghi,j} \
{G_5}_{abcd}{}^{k}{}_{,e} {G_5}_{fghik,j} + e_{242} {F_4}_{a}{}^{fgh,i} \
{F_4}^{abcd,e} {G_5}_{bcde}{}^{j,k} {G_5}_{fghik,j} \nonumber \\ 
&& + e_{243} {F_4}_{ae}{}^{fg,h} {F_4}^{abcd,e} {G_5}_{bcd}{}^{ij,k} \
{G_5}_{fghik,j} + e_{244} {F_4}_{a}{}^{fgh}{}_{,e} {F_4}^{abcd,e} \
{G_5}_{bcd}{}^{ij,k} {G_5}_{fghik,j} \nonumber \\ 
&&+ e_{245} {F_4}^{abcd,e} \
{F_4}_{e}{}^{fgh}{}_{,a} {G_5}_{bcd}{}^{ij,k} {G_5}_{fghik,j}  + e_{246} {F_4}_{ab}{}^{fg,h} {F_4}^{abcd,e} {G_5}_{cde}{}^{ij,k} \
{G_5}_{fghik,j} \nonumber \\ 
&&+ e_{247} {F_4}_{abe}{}^{f,g} {F_4}^{abcd,e} \
{G_5}_{cd}{}^{hij,k} {G_5}_{fghik,j} + e_{248} {F_4}_{ab}{}^{fg}{}_{,e} \
{F_4}^{abcd,e} {G_5}_{cd}{}^{hij,k} {G_5}_{fghik,j} \nonumber \\ 
&& + e_{249} {F_4}_{ae}{}^{fg}{}_{,b} {F_4}^{abcd,e} \
{G_5}_{cd}{}^{hij,k} {G_5}_{fghik,j} + e_{250} {F_4}_{abc}{}^{f,g} \
{F_4}^{abcd,e} {G_5}_{de}{}^{hij,k} {G_5}_{fghik,j} \nonumber \\ 
&&+ e_{251} \
{F_4}_{abce}{}^{,f} {F_4}^{abcd,e} {G_5}_{d}{}^{ghij,k} {G_5}_{fghik,j}  + e_{252} {F_4}_{abc}{}^{f}{}_{,e} {F_4}^{abcd,e} \
{G_5}_{d}{}^{ghij,k} {G_5}_{fghik,j} \nonumber \\ 
&&+ e_{253} {F_4}_{abe}{}^{f}{}_{,c} \
{F_4}^{abcd,e} {G_5}_{d}{}^{ghij,k} {G_5}_{fghik,j}  + e_{254} {F_4}_{abcd}{}^{,f} {F_4}^{abcd,e} {G_5}_{e}{}^{ghij,k} \
{G_5}_{fghik,j} \nonumber \\ 
&&+ e_{255} {F_4}^{abcd,e} {F_4}_{e}{}^{fgh,i} \
{G_5}_{abci}{}^{j,k} {G_5}_{fghjk,d} + e_{256} {F_4}^{abcd,e} \
{F_4}_{e}{}^{fgh,i} {G_5}_{abc}{}^{jk}{}_{,i} {G_5}_{fghjk,d} \nonumber \\ 
&& + e_{257} {F_4}^{abcd,e} {F_4}_{e}{}^{fgh,i} \
{G_5}_{abi}{}^{jk}{}_{,c} {G_5}_{fghjk,d} + e_{258} {F_4}_{a}{}^{fgh,i} \
{F_4}^{abcd,e} {G_5}_{bcdi}{}^{j,k} {G_5}_{fghjk,e} \nonumber \\ 
&&+ e_{259} \
{F_4}_{a}{}^{fgh,i} {F_4}^{abcd,e} {G_5}_{bcd}{}^{jk}{}_{,i} {G_5}_{fghjk,e}  + e_{260} {F_4}^{abcd,e} {F_4}^{fghi,j} {G_5}_{abcde}{}^{,k} \
{G_5}_{fghjk,i} \nonumber \\ 
&&+ e_{261} {F_4}^{abcd,e} {F_4}_{e}{}^{fgh,i} \
{G_5}_{abcd}{}^{j,k} {G_5}_{fghjk,i} + e_{262} {F_4}^{abcd,e} \
{F_4}^{fghi}{}_{,e} {G_5}_{abcd}{}^{j,k} {G_5}_{fghjk,i} \nonumber \\ 
&& + e_{263} {F_4}^{abcd,e} {F_4}^{fghi,j} {G_5}_{abcd}{}^{k}{}_{,e} \
{G_5}_{fghjk,i} + e_{264} {F_4}^{abcd,e} {F_4}^{fghi,j} \
{G_5}_{abce}{}^{k}{}_{,d} {G_5}_{fghjk,i} \nonumber \\ 
&&+ e_{265} {F_4}^{abcd,e} \
{F_4}_{e}{}^{fgh,i} {G_5}_{abc}{}^{jk}{}_{,d} {G_5}_{fghjk,i}  + e_{266} {F_4}^{abcd,e} {F_4}^{fghi}{}_{,e} \
{G_5}_{abc}{}^{jk}{}_{,d} {G_5}_{fghjk,i} \nonumber \\ 
&&+ e_{267} {F_4}_{a}{}^{fgh,i} \
{F_4}^{abcd,e} {G_5}_{bcde}{}^{j,k} {G_5}_{fghjk,i} + e_{268} \
{F_4}_{a}{}^{fgh,i} {F_4}^{abcd,e} {G_5}_{bcd}{}^{jk}{}_{,e} {G_5}_{fghjk,i} \nonumber \\ 
&&+ e_{269} {F_4}_{ae}{}^{fg,h} {F_4}^{abcd,e} {G_5}_{bch}{}^{ij,k} \
{G_5}_{fgijk,d} + e_{270} {F_4}_{ae}{}^{fg,h} {F_4}^{abcd,e} \
{G_5}_{bc}{}^{ijk}{}_{,h} {G_5}_{fgijk,d} \nonumber \\ 
&&+ e_{271} {F_4}_{ae}{}^{fg,h} \
{F_4}^{abcd,e} {G_5}_{bh}{}^{ijk}{}_{,c} {G_5}_{fgijk,d}  + e_{272} {F_4}_{ab}{}^{fg,h} {F_4}^{abcd,e} {G_5}_{cdh}{}^{ij,k} \
{G_5}_{fgijk,e} \nonumber \\ 
&&+ e_{273} {F_4}_{ab}{}^{fg,h} {F_4}^{abcd,e} \
{G_5}_{cd}{}^{ijk}{}_{,h} {G_5}_{fgijk,e} + e_{274} {F_4}^{abcd,e} \
{F_4}_{e}{}^{fgh,i} {G_5}_{abcd}{}^{j,k} {G_5}_{fgijk,h} \nonumber \\ 
&& + e_{275} {F_4}^{abcd,e} {F_4}_{e}{}^{fgh,i} \
{G_5}_{abc}{}^{jk}{}_{,d} {G_5}_{fgijk,h} + e_{276} {F_4}_{a}{}^{fgh,i} \
{F_4}^{abcd,e} {G_5}_{bcde}{}^{j,k} {G_5}_{fgijk,h} \nonumber \\ 
&&+ e_{277} \
{F_4}_{ae}{}^{fg,h} {F_4}^{abcd,e} {G_5}_{bcd}{}^{ij,k} {G_5}_{fgijk,h}  + e_{278} {F_4}_{a}{}^{fgh}{}_{,e} {F_4}^{abcd,e} \
{G_5}_{bcd}{}^{ij,k} {G_5}_{fgijk,h} \nonumber \\ 
&&+ e_{279} {F_4}^{abcd,e} \
{F_4}_{e}{}^{fgh}{}_{,a} {G_5}_{bcd}{}^{ij,k} {G_5}_{fgijk,h}  + e_{280} {F_4}_{a}{}^{fgh,i} {F_4}^{abcd,e} \
{G_5}_{bcd}{}^{jk}{}_{,e} {G_5}_{fgijk,h} \nonumber \\ 
&&+ e_{281} {F_4}_{a}{}^{fgh,i} \
{F_4}^{abcd,e} {G_5}_{bce}{}^{jk}{}_{,d} {G_5}_{fgijk,h}  + e_{282} {F_4}_{ae}{}^{fg,h} {F_4}^{abcd,e} \
{G_5}_{bc}{}^{ijk}{}_{,d} {G_5}_{fgijk,h} \nonumber \\ 
&&+ e_{283} \
{F_4}_{a}{}^{fgh}{}_{,e} {F_4}^{abcd,e} {G_5}_{bc}{}^{ijk}{}_{,d} {G_5}_{fgijk,h}  + e_{284} {F_4}^{abcd,e} {F_4}_{e}{}^{fgh}{}_{,a} \
{G_5}_{bc}{}^{ijk}{}_{,d} {G_5}_{fgijk,h} \nonumber \\ 
&&+ e_{285} {F_4}_{ab}{}^{fg,h} \
{F_4}^{abcd,e} {G_5}_{cde}{}^{ij,k} {G_5}_{fgijk,h} + e_{286} {F_4}_{ab}{}^{fg,h} {F_4}^{abcd,e} \
{G_5}_{cd}{}^{ijk}{}_{,e} {G_5}_{fgijk,h} \nonumber \\ 
&&+ e_{287} {F_4}_{abe}{}^{f,g} \
{F_4}^{abcd,e} {G_5}_{cg}{}^{hij,k} {G_5}_{fhijk,d} + e_{288} \
{F_4}_{abe}{}^{f,g} {F_4}^{abcd,e} {G_5}_{c}{}^{hijk}{}_{,g} {G_5}_{fhijk,d} \nonumber \\ 
&& + e_{289} {F_4}_{abc}{}^{f,g} {F_4}^{abcd,e} {G_5}_{dg}{}^{hij,k} \
{G_5}_{fhijk,e} + e_{290} {F_4}_{abc}{}^{f,g} {F_4}^{abcd,e} \
{G_5}_{d}{}^{hijk}{}_{,g} {G_5}_{fhijk,e} \nonumber \\ 
&&+ e_{291} {F_4}_{ae}{}^{fg,h} \
{F_4}^{abcd,e} {G_5}_{bcd}{}^{ij,k} {G_5}_{fhijk,g}  + e_{292} {F_4}_{ae}{}^{fg,h} {F_4}^{abcd,e} \
{G_5}_{bc}{}^{ijk}{}_{,d} {G_5}_{fhijk,g} \nonumber \\ 
&&+ e_{293} {F_4}_{ab}{}^{fg,h} \
{F_4}^{abcd,e} {G_5}_{cde}{}^{ij,k} {G_5}_{fhijk,g} + e_{294} \
{F_4}_{abe}{}^{f,g} {F_4}^{abcd,e} {G_5}_{cd}{}^{hij,k} {G_5}_{fhijk,g} \nonumber \\ 
&& + e_{295} {F_4}_{ab}{}^{fg}{}_{,e} {F_4}^{abcd,e} \
{G_5}_{cd}{}^{hij,k} {G_5}_{fhijk,g} + e_{296} {F_4}_{ae}{}^{fg}{}_{,b} \
{F_4}^{abcd,e} {G_5}_{cd}{}^{hij,k} {G_5}_{fhijk,g} \nonumber \\ 
&& + e_{297} {F_4}_{ab}{}^{fg,h} {F_4}^{abcd,e} \
{G_5}_{cd}{}^{ijk}{}_{,e} {G_5}_{fhijk,g} + e_{298} {F_4}_{ab}{}^{fg,h} \
{F_4}^{abcd,e} {G_5}_{ce}{}^{ijk}{}_{,d} {G_5}_{fhijk,g} \nonumber \\ 
&& + e_{299} {F_4}_{abe}{}^{f,g} {F_4}^{abcd,e} \
{G_5}_{c}{}^{hijk}{}_{,d} {G_5}_{fhijk,g} + e_{300} \
{F_4}_{ab}{}^{fg}{}_{,e} {F_4}^{abcd,e} {G_5}_{c}{}^{hijk}{}_{,d} {G_5}_{fhijk,g} \nonumber \\ 
&& + e_{301} {F_4}_{ae}{}^{fg}{}_{,b} {F_4}^{abcd,e} \
{G_5}_{c}{}^{hijk}{}_{,d} {G_5}_{fhijk,g} + e_{302} {F_4}_{abc}{}^{f,g} \
{F_4}^{abcd,e} {G_5}_{de}{}^{hij,k} {G_5}_{fhijk,g} \nonumber \\ 
&& + e_{303} {F_4}_{abc}{}^{f,g} {F_4}^{abcd,e} \
{G_5}_{d}{}^{hijk}{}_{,e} {G_5}_{fhijk,g} + e_{304} {F_4}_{abcd,e} \
{F_4}^{abcd,e} {G_5}_{fghij,k} {G_5}^{fghij,k} \nonumber \\ 
&&+ e_{305} {F_4}_{abce,d} \
{F_4}^{abcd,e} {G_5}_{fghij,k} {G_5}^{fghij,k}  + e_{306} {F_4}_{abcd,e} {F_4}^{abcd,e} {G_5}_{fghik,j} {G_5}^{fghij,k} \
\nonumber \\ 
&&+ e_{307} {F_4}_{abce,d} {F_4}^{abcd,e} {G_5}_{fghik,j} {G_5}^{fghij,k} + \
e_{308} {F_4}^{abcd,e} {F_4}^{fghi,j} {G_5}_{abcef}{}^{,k} {G_5}_{ghijk,d} \
\nonumber \\ 
&& + e_{309} {F_4}^{abcd,e} {F_4}^{fghi,j} {G_5}_{abce}{}^{k}{}_{,f} \
{G_5}_{ghijk,d} + e_{310} {F_4}^{abcd,e} {F_4}_{e}{}^{fgh,i} \
{G_5}_{abcf}{}^{j,k} {G_5}_{ghijk,d} \nonumber \\ 
&&+ e_{311} {F_4}^{abcd,e} \
{F_4}^{fghi}{}_{,e} {G_5}_{abcf}{}^{j,k} {G_5}_{ghijk,d}  + e_{312} {F_4}^{abcd,e} {F_4}_{e}{}^{fgh,i} \
{G_5}_{abc}{}^{jk}{}_{,f} {G_5}_{ghijk,d} \nonumber \\ 
&&+ e_{313} {F_4}^{abcd,e} \
{F_4}^{fghi}{}_{,e} {G_5}_{abc}{}^{jk}{}_{,f} {G_5}_{ghijk,d}  + e_{314} {F_4}^{abcd,e} {F_4}_{e}{}^{fgh,i} \
{G_5}_{abf}{}^{jk}{}_{,c} {G_5}_{ghijk,d} \nonumber \\ 
&&+ e_{315} {F_4}_{a}{}^{fgh,i} \
{F_4}^{abcd,e} {G_5}_{bcef}{}^{j,k} {G_5}_{ghijk,d} + e_{316} \
{F_4}_{a}{}^{fgh,i} {F_4}^{abcd,e} {G_5}_{bce}{}^{jk}{}_{,f} {G_5}_{ghijk,d} \nonumber \\ 
&& + e_{317} {F_4}_{ae}{}^{fg,h} {F_4}^{abcd,e} {G_5}_{bcf}{}^{ij,k} \
{G_5}_{ghijk,d} + e_{318} {F_4}_{a}{}^{fgh}{}_{,e} {F_4}^{abcd,e} \
{G_5}_{bcf}{}^{ij,k} {G_5}_{ghijk,d} \nonumber \\ 
&&+ e_{319} {F_4}^{abcd,e} \
{F_4}_{e}{}^{fgh}{}_{,a} {G_5}_{bcf}{}^{ij,k} {G_5}_{ghijk,d}  + e_{320} {F_4}_{ae}{}^{fg,h} {F_4}^{abcd,e} \
{G_5}_{bc}{}^{ijk}{}_{,f} {G_5}_{ghijk,d} \nonumber \\ 
&&+ e_{321} \
{F_4}_{a}{}^{fgh}{}_{,e} {F_4}^{abcd,e} {G_5}_{bc}{}^{ijk}{}_{,f} {G_5}_{ghijk,d}  + e_{322} {F_4}^{abcd,e} {F_4}_{e}{}^{fgh}{}_{,a} \
{G_5}_{bc}{}^{ijk}{}_{,f} {G_5}_{ghijk,d} \nonumber \\ 
&&+ e_{323} {F_4}_{ae}{}^{fg,h} \
{F_4}^{abcd,e} {G_5}_{bf}{}^{ijk}{}_{,c} {G_5}_{ghijk,d}  + e_{324} {F_4}_{ab}{}^{fg,h} {F_4}^{abcd,e} {G_5}_{cef}{}^{ij,k} \
{G_5}_{ghijk,d} \nonumber \\ 
&&+ e_{325} {F_4}_{ab}{}^{fg,h} {F_4}^{abcd,e} \
{G_5}_{ce}{}^{ijk}{}_{,f} {G_5}_{ghijk,d} + e_{326} {F_4}_{abe}{}^{f,g} \
{F_4}^{abcd,e} {G_5}_{cf}{}^{hij,k} {G_5}_{ghijk,d} \nonumber \\ 
&& + e_{327} {F_4}_{abe}{}^{f,g} {F_4}^{abcd,e} \
{G_5}_{c}{}^{hijk}{}_{,f} {G_5}_{ghijk,d} + e_{328} \
{F_4}_{ab}{}^{fg}{}_{,e} {F_4}^{abcd,e} {G_5}_{c}{}^{hijk}{}_{,f} {G_5}_{ghijk,d} \nonumber \\ 
&& + e_{329} {F_4}_{ae}{}^{fg}{}_{,b} {F_4}^{abcd,e} \
{G_5}_{c}{}^{hijk}{}_{,f} {G_5}_{ghijk,d} + e_{330} {F_4}_{abc}{}^{f,g} \
{F_4}^{abcd,e} {G_5}_{e}{}^{hijk}{}_{,f} {G_5}_{ghijk,d} \nonumber \\ 
&& + e_{331} {F_4}_{abce}{}^{,f} {F_4}^{abcd,e} {G_5}_{f}{}^{ghij,k} \
{G_5}_{ghijk,d} + e_{332} {F_4}_{abe}{}^{f,g} {F_4}^{abcd,e} \
{G_5}_{f}{}^{hijk}{}_{,c} {G_5}_{ghijk,d} \nonumber \\ 
&&+ e_{333} {F_4}^{abcd,e} \
{F_4}^{fghi,j} {G_5}_{abcdf}{}^{,k} {G_5}_{ghijk,e}  + e_{334} {F_4}^{abcd,e} {F_4}^{fghi,j} {G_5}_{abcd}{}^{k}{}_{,f} \
{G_5}_{ghijk,e} \nonumber \\ 
&&+ e_{335} {F_4}_{a}{}^{fgh,i} {F_4}^{abcd,e} \
{G_5}_{bcdf}{}^{j,k} {G_5}_{ghijk,e} + e_{336} {F_4}_{a}{}^{fgh,i} \
{F_4}^{abcd,e} {G_5}_{bcd}{}^{jk}{}_{,f} {G_5}_{ghijk,e} \nonumber \\ 
&& + e_{337} {F_4}_{ab}{}^{fg,h} {F_4}^{abcd,e} {G_5}_{cdf}{}^{ij,k} \
{G_5}_{ghijk,e} + e_{338} {F_4}_{ab}{}^{fg,h} {F_4}^{abcd,e} \
{G_5}_{cd}{}^{ijk}{}_{,f} {G_5}_{ghijk,e} \nonumber \\ 
&&+ e_{339} {F_4}_{abc}{}^{f,g} \
{F_4}^{abcd,e} {G_5}_{d}{}^{hijk}{}_{,f} {G_5}_{ghijk,e}  + e_{340} {F_4}_{abe}{}^{f,g} {F_4}^{abcd,e} {G_5}_{cd}{}^{hij,k} \
{G_5}_{ghijk,f} \nonumber \\ 
&&+ e_{341} {F_4}_{abe}{}^{f,g} {F_4}^{abcd,e} \
{G_5}_{c}{}^{hijk}{}_{,d} {G_5}_{ghijk,f} + e_{342} {F_4}_{abc}{}^{f,g} \
{F_4}^{abcd,e} {G_5}_{de}{}^{hij,k} {G_5}_{ghijk,f} \nonumber \\ 
&& + e_{343} {F_4}_{abce}{}^{,f} {F_4}^{abcd,e} {G_5}_{d}{}^{ghij,k} \
{G_5}_{ghijk,f} + e_{344} {F_4}_{abc}{}^{f}{}_{,e} {F_4}^{abcd,e} \
{G_5}_{d}{}^{ghij,k} {G_5}_{ghijk,f} \nonumber \\ 
&&+ e_{345} {F_4}_{abe}{}^{f}{}_{,c} \
{F_4}^{abcd,e} {G_5}_{d}{}^{ghij,k} {G_5}_{ghijk,f}  + e_{346} {F_4}_{abc}{}^{f,g} {F_4}^{abcd,e} \
{G_5}_{d}{}^{hijk}{}_{,e} {G_5}_{ghijk,f} \nonumber \\ 
&&+ e_{347} {F_4}_{abcd}{}^{,f} \
{F_4}^{abcd,e} {G_5}_{e}{}^{ghij,k} {G_5}_{ghijk,f} + e_{348} \
{F_4}_{abc}{}^{f,g} {F_4}^{abcd,e} {G_5}_{e}{}^{hijk}{}_{,d} {G_5}_{ghijk,f} \nonumber \\ 
&& + e_{349} {F_4}_{abce}{}^{,f} {F_4}^{abcd,e} {G_5}_{ghijk,f} \
{G_5}^{ghijk}{}_{,d} + e_{350} {F_4}_{abc}{}^{f}{}_{,e} {F_4}^{abcd,e} \
{G_5}_{ghijk,f} {G_5}^{ghijk}{}_{,d} \nonumber \\ 
&&+ e_{351} {F_4}_{abe}{}^{f}{}_{,c} \
{F_4}^{abcd,e} {G_5}_{ghijk,f} {G_5}^{ghijk}{}_{,d} + e_{352} {F_4}_{abcd}{}^{,f} {F_4}^{abcd,e} {G_5}_{ghijk,f} \
{G_5}^{ghijk}{}_{,e}.\label{G5G5F4F4}
\eea

\section*{Acknowledgement}\addcontentsline{toc}{section}{Acknowledgement}

We would like to thank the authors of Ref. \cite{Nutma:2013zea} for developing the excellent Mathematica package ``xTras" which we have used extensively for symbolic calculations. This work has been financially supported by the research deputy of Sirjan University of Technology.


\providecommand{\href}[2]{#2}\begingroup\raggedright
\endgroup
\end{document}